\documentclass[twocolumn,appendixfloats]{aastex63}

\usepackage{natbib}
\usepackage{amssymb}
\usepackage{amsmath}
\usepackage{verbatim}

\newcommand{\fiveyr}{NG5}        \newcommand{\nineyr}{NG9}        \newcommand{\elevenyr}{NG11}

\newcommand{\tempo}{{\sc Tempo}}
\newcommand{\tempotwo}{{\sc Tempo2}}
\newcommand{\pint}{{\sc PINT}}
\newcommand{\enterprise}{{\sc ENTERPRISE}}

\newcommand{\attn}[1]{\textbf{\textcolor{orange}{#1}}}

\NewPageAfterKeywords

\begin{document}

\title{The NANOGrav 12.5~yr Data Set: Observations and Narrowband Timing of 47 Millisecond Pulsars}
\shorttitle{The NANOGrav 12.5-year Data Set}
\shortauthors{Alam et al.}

\author[0000-0003-2449-5426]{Md F. Alam}
\affiliation{Department of Physics and Astronomy, Franklin \& Marshall College, P.O. Box 3003, Lancaster, PA 17604, USA}
\author{Zaven Arzoumanian}
\affiliation{X-Ray Astrophysics Laboratory, NASA Goddard Space Flight Center, Code 662, Greenbelt, MD 20771, USA}
\author[0000-0003-2745-753X]{Paul T. Baker}
\affiliation{Department of Physics and Astronomy, Widener University, One University Place, Chester, PA 19013, USA}
\author[0000-0003-4046-884X]{Harsha Blumer}
\affiliation{Department of Physics and Astronomy, West Virginia University, P.O. Box 6315, Morgantown, WV 26506, USA}
\affiliation{Center for Gravitational Waves and Cosmology, West Virginia University, Chestnut Ridge Research Building, Morgantown, WV 26505, USA}
\author{Keith E. Bohler}
\affiliation{Center for Advanced Radio Astronomy, University of Texas-Rio Grande Valley, Brownsville, TX 78520, USA}
\author{Adam Brazier}
\affiliation{Cornell Center for Astrophysics and Planetary Science and Department of Astronomy, Cornell University, Ithaca, NY 14853, USA}
\author[0000-0003-3053-6538]{Paul R. Brook}
\affiliation{Department of Physics and Astronomy, West Virginia University, P.O. Box 6315, Morgantown, WV 26506, USA}
\affiliation{Center for Gravitational Waves and Cosmology, West Virginia University, Chestnut Ridge Research Building, Morgantown, WV 26505, USA}
\author[0000-0003-4052-7838]{Sarah Burke-Spolaor}
\affiliation{Department of Physics and Astronomy, West Virginia University, P.O. Box 6315, Morgantown, WV 26506, USA}
\affiliation{Center for Gravitational Waves and Cosmology, West Virginia University, Chestnut Ridge Research Building, Morgantown, WV 26505, USA}
\author{Keeisi Caballero}
\affiliation{Center for Advanced Radio Astronomy, University of Texas-Rio Grande Valley, Brownsville, TX 78520, USA}
\author{Richard S. Camuccio}
\affiliation{Center for Advanced Radio Astronomy, University of Texas-Rio Grande Valley, Brownsville, TX 78520, USA}
\author{Rachel L. Chamberlain}
\affiliation{Department of Physics and Astronomy, Franklin \& Marshall College, P.O. Box 3003, Lancaster, PA 17604, USA}
\author[0000-0002-2878-1502]{Shami Chatterjee}
\affiliation{Cornell Center for Astrophysics and Planetary Science and Department of Astronomy, Cornell University, Ithaca, NY 14853, USA}
\author[0000-0002-4049-1882]{James M. Cordes}
\affiliation{Cornell Center for Astrophysics and Planetary Science and Department of Astronomy, Cornell University, Ithaca, NY 14853, USA}
\author[0000-0002-7435-0869]{Neil J. Cornish}
\affiliation{Department of Physics, Montana State University, Bozeman, MT 59717, USA}
\author[0000-0002-2578-0360]{Fronefield Crawford}
\affiliation{Department of Physics and Astronomy, Franklin \& Marshall College, P.O. Box 3003, Lancaster, PA 17604, USA}
\author[0000-0002-6039-692X]{H. Thankful Cromartie}
\affiliation{Cornell Center for Astrophysics and Planetary Science and Department of Astronomy, Cornell University, Ithaca, NY 14853, USA}
\author[0000-0002-2185-1790]{Megan E. DeCesar}
\affiliation{Department of Physics, Lafayette College, Easton, PA 18042, USA}
\affiliation{George Mason University, Fairfax, VA 22030, resident at U.S. Naval Research Laboratory, Washington, D.C. 20375, USA}
\author[0000-0002-6664-965X]{Paul B. Demorest}
\affiliation{National Radio Astronomy Observatory, 1003 Lopezville Rd., Socorro, NM 87801, USA}
\author[0000-0001-8885-6388]{Timothy Dolch}
\affiliation{Department of Physics, Hillsdale College, 33 E. College Street, Hillsdale, Michigan 49242, USA}
\author{Justin A. Ellis}
\affiliation{Infinia ML, 202 Rigsbee Avenue, Durham NC, 27701}
\author[0000-0002-2223-1235]{Robert D. Ferdman}
\affiliation{School of Chemistry, University of East Anglia, Norwich, NR4 7TJ, United Kingdom}
\author{Elizabeth C. Ferrara}
\affiliation{NASA Goddard Space Flight Center, Greenbelt, MD 20771, USA}
\author[0000-0001-5645-5336]{William Fiore}
\affiliation{Center for Gravitation, Cosmology and Astrophysics, Department of Physics, University of Wisconsin-Milwaukee,\\ P.O. Box 413, Milwaukee, WI 53201, USA}
\affiliation{Department of Physics and Astronomy, West Virginia University, P.O. Box 6315, Morgantown, WV 26506, USA}
\affiliation{Center for Gravitational Waves and Cosmology, West Virginia University, Chestnut Ridge Research Building, Morgantown, WV 26505, USA}
\author[0000-0001-8384-5049]{Emmanuel Fonseca}
\affiliation{Department of Physics, McGill University, 3600  University St., Montreal, QC H3A 2T8, Canada}
\author{Yhamil Garcia}
\affiliation{Center for Advanced Radio Astronomy, University of Texas-Rio Grande Valley, Brownsville, TX 78520, USA}
\author[0000-0001-6166-9646]{Nathan Garver-Daniels}
\affiliation{Department of Physics and Astronomy, West Virginia University, P.O. Box 6315, Morgantown, WV 26506, USA}
\affiliation{Center for Gravitational Waves and Cosmology, West Virginia University, Chestnut Ridge Research Building, Morgantown, WV 26505, USA}
\author[0000-0001-8158-638X]{Peter A. Gentile}
\affiliation{Department of Physics and Astronomy, West Virginia University, P.O. Box 6315, Morgantown, WV 26506, USA}
\affiliation{Center for Gravitational Waves and Cosmology, West Virginia University, Chestnut Ridge Research Building, Morgantown, WV 26505, USA}
\author[0000-0003-1884-348X]{Deborah C. Good}
\affiliation{Department of Physics and Astronomy, University of British Columbia, 6224 Agricultural Road, Vancouver, BC V6T 1Z1, Canada}
\author[0000-0001-6609-2997]{Jordan A. Gusdorff}
\affiliation{Department of Physics, Lafayette College, Easton, PA 18042, USA}
\author[0000-0003-2289-2575]{Daniel Halmrast}
\affiliation{Department of Physics, Hillsdale College, 33 E. College Street, Hillsdale, Michigan 49242, USA}
\affiliation{Department of Mathematics, University of California, Santa Barbara, CA 93106, USA}
\author[0000-0003-2742-3321]{Jeffrey S. Hazboun}
\affiliation{University of Washington Bothell, 18115 Campus Way NE, Bothell, WA 98011, USA}
\author{Kristina Islo}
\affiliation{Center for Gravitation, Cosmology and Astrophysics, Department of Physics, University of Wisconsin-Milwaukee,\\ P.O. Box 413, Milwaukee, WI 53201, USA}
\author[0000-0003-1082-2342]{Ross J. Jennings}
\affiliation{Cornell Center for Astrophysics and Planetary Science and Department of Astronomy, Cornell University, Ithaca, NY 14853, USA}
\author[0000-0002-4188-6827]{Cody Jessup}
\affiliation{Department of Physics, Hillsdale College, 33 E. College Street, Hillsdale, Michigan 49242, USA}
\affiliation{Department of Physics, Montana State University, Bozeman, MT 59717, USA}
\author[0000-0001-6607-3710]{Megan L. Jones}
\affiliation{Center for Gravitation, Cosmology and Astrophysics, Department of Physics, University of Wisconsin-Milwaukee,\\ P.O. Box 413, Milwaukee, WI 53201, USA}
\author[0000-0002-3654-980X]{Andrew R. Kaiser}
\affiliation{Department of Physics and Astronomy, West Virginia University, P.O. Box 6315, Morgantown, WV 26506, USA}
\affiliation{Center for Gravitational Waves and Cosmology, West Virginia University, Chestnut Ridge Research Building, Morgantown, WV 26505, USA}
\author[0000-0001-6295-2881]{David L. Kaplan}
\affiliation{Center for Gravitation, Cosmology and Astrophysics, Department of Physics, University of Wisconsin-Milwaukee,\\ P.O. Box 413, Milwaukee, WI 53201, USA}
\author[0000-0002-6625-6450]{Luke Zoltan Kelley}
\affiliation{Center for Interdisciplinary Exploration and Research in Astrophysics (CIERA), Northwestern University, Evanston, IL 60208}
\author[0000-0003-0123-7600]{Joey Shapiro Key}
\affiliation{University of Washington Bothell, 18115 Campus Way NE, Bothell, WA 98011, USA}
\author[0000-0003-0721-651X]{Michael T. Lam}
\affiliation{School of Physics and Astronomy, Rochester Institute of Technology, Rochester, NY 14623, USA}
\affiliation{Laboratory for Multiwavelength Astrophysics, Rochester Institute of Technology, Rochester, NY 14623, USA}
\author{T. Joseph W. Lazio}
\affiliation{Jet Propulsion Laboratory, California Institute of Technology, 4800 Oak Grove Drive, Pasadena, CA 91109, USA}
\author[0000-0003-1301-966X]{Duncan R. Lorimer}
\affiliation{Department of Physics and Astronomy, West Virginia University, P.O. Box 6315, Morgantown, WV 26506, USA}
\affiliation{Center for Gravitational Waves and Cosmology, West Virginia University, Chestnut Ridge Research Building, Morgantown, WV 26505, USA}
\author{Jing Luo}
\affiliation{Department of Astronomy \& Astrophysics, University of Toronto, 50 Saint George Street, Toronto, ON M5S 3H4, Canada}
\author[0000-0001-5229-7430]{Ryan S. Lynch}
\affiliation{Green Bank Observatory, P.O. Box 2, Green Bank, WV 24944, USA}
\author[0000-0003-2285-0404]{Dustin R. Madison}
\altaffiliation{NANOGrav Physics Frontiers Center Postdoctoral Fellow}
\affiliation{Department of Physics and Astronomy, West Virginia University, P.O. Box 6315, Morgantown, WV 26506, USA}
\affiliation{Center for Gravitational Waves and Cosmology, West Virginia University, Chestnut Ridge Research Building, Morgantown, WV 26505, USA}
\author{Kaleb Maraccini}
\affiliation{Center for Gravitation, Cosmology and Astrophysics, Department of Physics, University of Wisconsin-Milwaukee,\\ P.O. Box 413, Milwaukee, WI 53201, USA}
\author[0000-0001-7697-7422]{Maura A. McLaughlin}
\affiliation{Department of Physics and Astronomy, West Virginia University, P.O. Box 6315, Morgantown, WV 26506, USA}
\affiliation{Center for Gravitational Waves and Cosmology, West Virginia University, Chestnut Ridge Research Building, Morgantown, WV 26505, USA}
\author[0000-0002-4307-1322]{Chiara M. F. Mingarelli}
\affiliation{Center for Computational Astrophysics, Flatiron Institute, 162 5th Avenue, New York, New York, 10010, USA}
\affiliation{Department of Physics, University of Connecticut, 196 Auditorium Road, U-3046, Storrs, CT 06269-3046, USA}
\author[0000-0002-3616-5160]{Cherry Ng}
\affiliation{Dunlap Institute for Astronomy and Astrophysics, University of Toronto, 50 St. George St., Toronto, ON M5S 3H4, Canada}
\author{Benjamin M. X. Nguyen}
\affiliation{Department of Physics and Astronomy, Franklin \& Marshall College, P.O. Box 3003, Lancaster, PA 17604, USA}
\author[0000-0002-6709-2566]{David J. Nice}
\affiliation{Department of Physics, Lafayette College, Easton, PA 18042, USA}
\author[0000-0001-5465-2889]{Timothy T. Pennucci}
\altaffiliation{NANOGrav Physics Frontiers Center Postdoctoral Fellow}
\affiliation{National Radio Astronomy Observatory, 520 Edgemont Road, Charlottesville, VA 22903, USA}
\affiliation{Institute of Physics, E\"{o}tv\"{o}s Lor\'{a}nd University, P\'{a}zm\'{a}ny P. s. 1/A, 1117 Budapest, Hungary}
\author[0000-0002-8826-1285]{Nihan S. Pol}
\affiliation{Department of Physics and Astronomy, West Virginia University, P.O. Box 6315, Morgantown, WV 26506, USA}
\affiliation{Center for Gravitational Waves and Cosmology, West Virginia University, Chestnut Ridge Research Building, Morgantown, WV 26505, USA}
\author[0000-0002-4709-6236]{Joshua Ramette}
\affiliation{Department of Physics, Hillsdale College, 33 E. College Street, Hillsdale, Michigan 49242, USA}
\affiliation{Department of Physics, Massachusetts Institute of Technology, 77 Massachusetts Avenue, Cambridge, MA 02139-4307}
\author[0000-0001-5799-9714]{Scott M. Ransom}
\affiliation{National Radio Astronomy Observatory, 520 Edgemont Road, Charlottesville, VA 22903, USA}
\author[0000-0002-5297-5278]{Paul S. Ray}
\affiliation{Space Science Division, Naval Research Laboratory, Washington, DC 20375-5352, USA}
\author[0000-0002-7283-1124]{Brent J. Shapiro-Albert}
\affiliation{Department of Physics and Astronomy, West Virginia University, P.O. Box 6315, Morgantown, WV 26506, USA}
\affiliation{Center for Gravitational Waves and Cosmology, West Virginia University, Chestnut Ridge Research Building, Morgantown, WV 26505, USA}
\author[0000-0002-7778-2990]{Xavier Siemens}
\affiliation{Department of Physics, Oregon State University, Corvallis, OR 97331, USA}
\affiliation{Center for Gravitation, Cosmology and Astrophysics, Department of Physics, University of Wisconsin-Milwaukee,\\ P.O. Box 413, Milwaukee, WI 53201, USA}
\author[0000-0003-1407-6607]{Joseph Simon}
\affiliation{Jet Propulsion Laboratory, California Institute of Technology, 4800 Oak Grove Drive, Pasadena, CA 91109, USA}
\author[0000-0002-6730-3298]{Ren\'{e}e Spiewak}
\affiliation{Centre for Astrophysics and Supercomputing, Swinburne University of Technology, P.O. Box 218, Hawthorn, Victoria 3122, Australia}
\author[0000-0001-9784-8670]{Ingrid H. Stairs}
\affiliation{Department of Physics and Astronomy, University of British Columbia, 6224 Agricultural Road, Vancouver, BC V6T 1Z1, Canada}
\author[0000-0002-1797-3277]{Daniel R. Stinebring}
\affiliation{Department of Physics and Astronomy, Oberlin College, Oberlin, OH 44074, USA}
\author[0000-0002-7261-594X]{Kevin Stovall}
\affiliation{National Radio Astronomy Observatory, 1003 Lopezville Rd., Socorro, NM 87801, USA}
\author[0000-0002-1075-3837]{Joseph K. Swiggum}
\altaffiliation{NANOGrav Physics Frontiers Center Postdoctoral Fellow}
\affiliation{Department of Physics, Lafayette College, Easton, PA 18042, USA}
\author[0000-0003-0264-1453]{Stephen R. Taylor}
\affiliation{Department of Physics and Astronomy, Vanderbilt University, 2301 Vanderbilt Place, Nashville, TN 37235, USA}
\author[0000-0002-4657-9826]{Michael Tripepi}
\affiliation{Department of Physics, Hillsdale College, 33 E. College Street, Hillsdale, Michigan 49242, USA}
\affiliation{Department of Physics, The Ohio State University, Columbus, OH 43210, USA}
\author[0000-0002-4162-0033]{Michele Vallisneri}
\affiliation{Jet Propulsion Laboratory, California Institute of Technology, 4800 Oak Grove Drive, Pasadena, CA 91109, USA}
\author[0000-0003-4700-9072]{Sarah J. Vigeland}
\affiliation{Center for Gravitation, Cosmology and Astrophysics, Department of Physics, University of Wisconsin-Milwaukee,\\ P.O. Box 413, Milwaukee, WI 53201, USA}
\author[0000-0002-6020-9274]{Caitlin A. Witt}
\affiliation{Department of Physics and Astronomy, West Virginia University, P.O. Box 6315, Morgantown, WV 26506, USA}
\affiliation{Center for Gravitational Waves and Cosmology, West Virginia University, Chestnut Ridge Research Building, Morgantown, WV 26505, USA}
\author[0000-0001-5105-4058]{Weiwei Zhu}
\affiliation{National Astronomical Observatories, Chinese Academy of Science, 20A Datun Road, Chaoyang District, Beijing 100012, China}
 
\collaboration{1000}{The NANOGrav Collaboration\\%\attn{Please check your name, affiliation, and ORCID!}\\\attn{Please check the contributions \& funding acknowledgements sections at the end of the draft!}}
  }
\noaffiliation

\correspondingauthor{Megan E. DeCesar}
\email{megan.decesar@nanograv.org}

\begin{abstract}

We present time-of-arrival (TOA) measurements and timing models of 47
millisecond pulsars (MSPs) observed from 2004 to 2017 at the Arecibo
Observatory and the Green Bank Telescope by the North American
Nanohertz Observatory for Gravitational Waves (NANOGrav).  The observing
cadence was three to four weeks for most pulsars over most of this
time span, with weekly observations of six sources.  These data were
collected for use in low-frequency gravitational wave searches and
for other astrophysical purposes.  We detail our observational
methods and present a set of TOA measurements,
based on ``narrowband'' analysis, in which many TOAs are calculated
within narrow radio-frequency bands for data collected simultaneously across a
wide bandwidth.  A separate set of ``wideband'' TOAs will be presented
in a companion paper.  We detail a number of methodological changes,
compared to our previous work, which yield a cleaner and more uniformly
processed data set.  Our timing models include several new astrometric
and binary pulsar measurements, including previously unpublished
values for the parallaxes of PSRs~J1832$-$0836 and J2322$+$2057, the
secular derivatives of the projected semi-major orbital axes of
PSRs~J0613$-$0200 and J2229$+$2643, and the first detection of the
Shapiro delay in PSR~J2145$-$0750.  We report detectable levels of red
noise in the time series for 14 pulsars.  As a check on timing
model reliability, we investigate the stability of astrometric
parameters across data sets of different lengths.  We also report flux
density measurements for all pulsars observed.  Searches for
stochastic and continuous gravitational waves using these data will be
subjects of forthcoming publications.
 
\end{abstract}


\begin{comment}

\setcounter{section}{-1}
\section{To-do list}

Things to do:

\begin{itemize}

\item Write abstract and summary/conclusion

\item Address questions/comments in orange text

\item Remove J1022+1001 from figures, tables, text

\end{itemize}

\end{comment}

\section{Introduction}
\label{sec:intro}
High-precision timing of millisecond
pulsars (MSPs) produces a wealth of both astrophysics and basic
physics, including strong constraints on the dense matter equation of
state \citep[e.g.,][]{Lattimer2019}, unique tests of theories of
gravity \citep[e.g.,][]{Renevey2019}, and the potential to soon detect
nHz-frequency gravitational waves \citep[e.g.,][]{taylor2016,ps18}.
The North American Nanohertz Observatory for Gravitational Waves
\citep[NANOGrav;][]{Ransom2019} is a collaboration pursuing long-term
goals of detecting and characterizing gravitational waves using the timing data from an
array of high-precision MSPs (a.k.a.~a pulsar timing array or PTA).
Such efforts promise a wide variety of astrophysical results at
virtually all scales from the solar system to the
cosmological \citep[e.g.,][]{bs+19}.

This paper describes the current public release of NANOGrav data, the
``12.5-year Data Set,'' which we have collected over 12.5~years (July 2004 to June 2017) using
the Arecibo Observatory and the Green Bank Telescope.  The data and
analyses described here are built on and extend those found in our
previous data releases for our
5-year \citep[][herein \fiveyr]{Demorest2013},
9-year \citep[][herein \nineyr]{Arzoumanian2015b}, and
11-year \citep[][herein \elevenyr]{Arzoumanian2018a} data sets.
The present release includes data from 47~MSPs.

We have taken two approaches to measuring pulse arrival times
in the 12.5-year data set, which we report in two separate papers.
In the present paper, we follow the procedures of our earlier data sets: 
we divide our
observations made across wide radio frequency bands into narrow
frequency subbands and determine pulse times of arrival (TOAs) for
each subband, resulting in a large number of measurements (``narrowband TOAs'') for each
observation.  An alternative approach, wideband
timing \citep{Liu14,PDR14,Pennucci19}, extracts a single TOA and 
dispersion measure (DM) for each observation, resulting in a more
compact data set of ``wideband TOAs.''  We analyze the 12.5-year data set 
using wideband timing in \citet{Alam_2020_wideband}.

Analyses to search the 12.5-year data set for signals indicative of
gravitational waves will be presented elsewhere.
Analyses of our previous-generation data set, \elevenyr, for
stochastic, continuous, and bursting gravitational waves can be found in 
\cite{11yrGWBG},
\cite{11yrGWCW},
and
\cite{11yrBWM},
respectively.

NANOGrav is part of the International Pulsar Timing
Array \citep[IPTA;][]{IPTA2010}, and the 12.5-year data set will
become part of a future IPTA data release \citep{IPTADR2} along with
data from the European Pulsar Timing
Array \citep[EPTA;][]{Desvignes2016} and the Parkes Pulsar Timing
Array \citep[PPTA;][]{Kerr2020}.

The plan for this paper is as follows.
In Section~\ref{sec:obs}, we describe the observations and data
reduction.  In Section~\ref{sec:timing}, we describe timing models fit
to the TOAs for each pulsar, including both deterministic
astrophysical phenomena and stochastic noise terms.  In
Section~\ref{sec:pint}, we compare timing models generated with the
longstanding \tempo\footnote{\url{https://github.com/nanograv/tempo}}
pulsar timing software package
with those generated using the
new \pint\footnote{\url{https://github.com/nanograv/PINT}} package
\citep{pint,Luo21}.
We list astrometric and binary parameters that have been newly measured with NANOGrav data in Section~\ref{sec:newsigpars}. In Section~\ref{sec:astrometry}, we
compare astrometric measurements between the present paper and
previous data sets.  In Section~\ref{sec:binary}, we highlight five binary pulsars for which new post-Keplerian parameters have been measured, or for which extensive testing was needed to obtain their timing solution.  In Section~\ref{sec:flux}, we
present flux density measurements for each pulsar at two or more radio
frequencies.  In Section~\ref{sec:conclusion}, we summarize the work.
In Appendix~\ref{sec:resid}, we present the timing residuals and
DM variations for all pulsars in this data set.

The NANOGrav 12.5 yr data set files include narrowband TOAs developed in the present paper, wideband TOAs developed in \citet{Alam_2020_wideband}, parameterized timing models for all pulsars for each of the TOA sets, and supporting files such as telescope clock offset measurements. The data set presented in the present paper has been preserved to Zenodo at \url{doi:10.5281/zenodo.4312297}\footnote{All of NANOGrav's data sets are available at \url{http://data.nanograv.org}, including the data set presented here, which is the ``v4'' version of the 12.5 yr data set. Raw telescope data products are also available from the same website. Version ``v4'' of the 12.5y data set has also been preserved in Zenodo at \url{doi:10.5281/zenodo.4312297.}}.
 
\section{Observations, Data Reduction, and Times-of-Arrival}
\label{sec:obs}
Here we describe the telescope observations and data reduction used
to produce our ``narrowband'' TOA data set.
The procedures we used
are nearly identical to those in \nineyr\ and \elevenyr.  We
therefore provide only a brief overview of analysis details that were
fully presented in \nineyr\ and \elevenyr, noting changes from those procedures where
applicable.
The ``wideband'' TOAs contained within our data set
use intermediate data products resulting from the
procedures described in the
subsections
below, but use a different TOA calculation algorithm
as described in \citet{Alam_2020_wideband}.

\subsection{Data Collection}
\label{subsec:data_collection}

\begin{figure*}
\begin{center}
\includegraphics[width=6.0in]{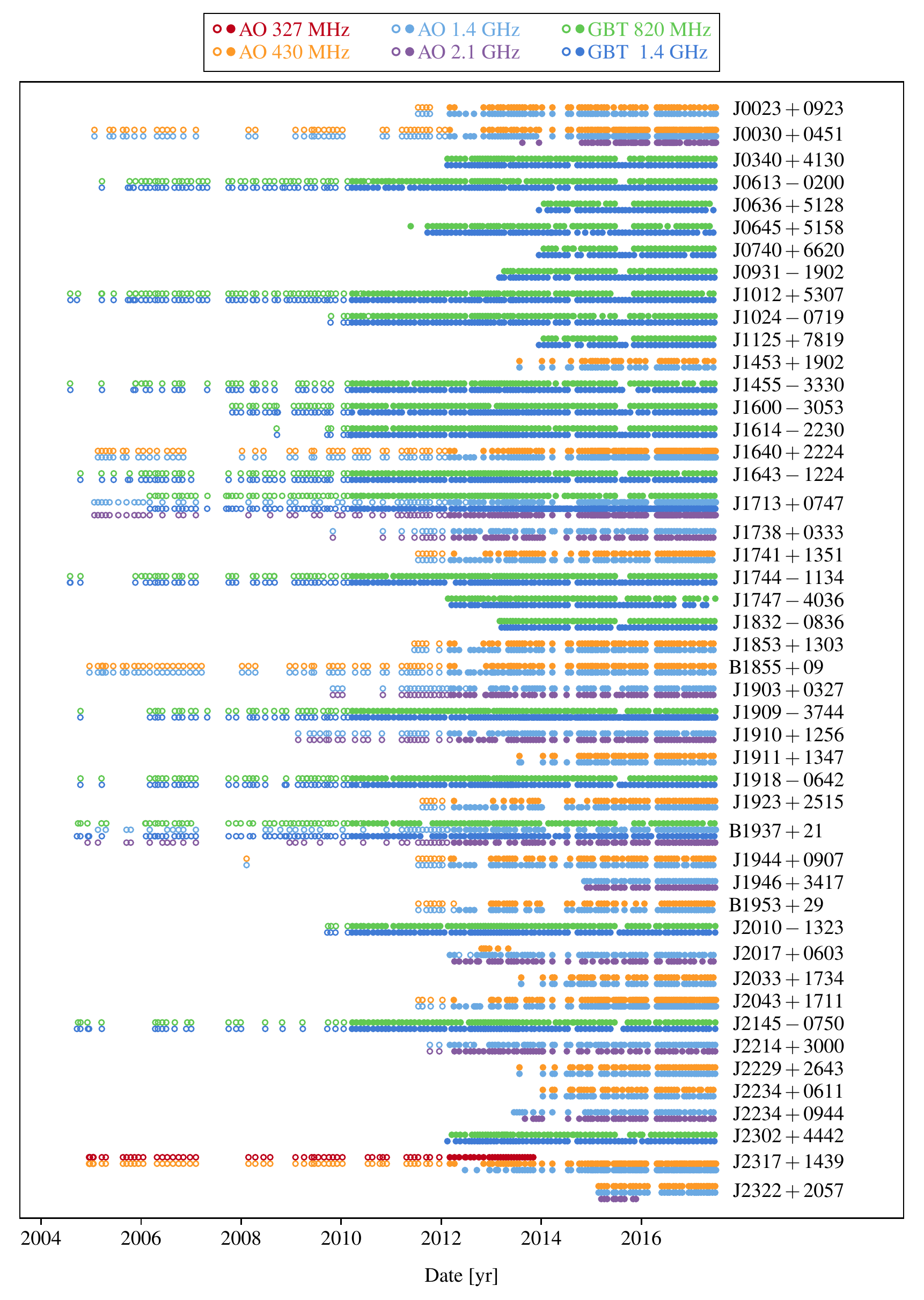}
\caption{\label{fig:obs:epochs} 
Epochs of all observations in the data set.  
The observatory and observing frequency are indicated by color:
Arecibo observations are red (327~MHz), orange (430~MHz), light
blue (1.4~GHz), and purple (2.1~GHz).  GBT observations
are green (820~MHz) and dark blue (1.4~GHz). 
The data acquisition system is indicated by symbol:
open circles are ASP or GASP, and filled circles are PUPPI or GUPPI.  
}
\end{center}
\end{figure*}

The data presented here were collected between 2004 July through 2017
June.  Timing baselines of individual pulsars range from 2.3 to 12.9
years.  Compared to \elevenyr, this data set adds 1.5 years of data
and two MSPs: J1946+3417 and J2322+2057.
The sources and observing epochs are summarized in Figure~\ref{fig:obs:epochs}.

Data were collected at the 305-m Arecibo Observatory (Arecibo or AO)
and the 100-m Robert C. Byrd Green Bank Telescope (GBT).  Twenty-six
pulsars were observed with Arecibo.  These include all pulsars in our
program within the Arecibo declination range of
$0^\circ<\delta<+39^\circ$.  Twenty-three pulsars were observed with
the GBT.  This includes all pulsars in our program outside the Arecibo
declination range, along with two pulsars also observed with
Arecibo, PSRs~J1713+0747 and B1937+21, for
which we have continuous data sets at both telescopes for the length
of the observing program.
All sources were observed with an approximately 3-week cadence at
Arecibo or 4-week cadence at the GBT (herein referred to as
``monthly'' observations).  In addition, six sources were observed
weekly to increase sensitivity to continuous waves from individual
foreground GW sources \citep{Arzoumanian2014}: PSRs~J0030+0451,
J1640+2224, J1713+0747, J1909$-$3744, J2043+1711, and J2317+1439
(herein referred to as ``high cadence'' observations). For each pulsar in the high cadence program, the observations were taken at the same telescope using the same methodologies as for the monthly observations, with the exception of the weekly GBT data, which covered only the 1.4~GHz frequency band.

Some interruptions in the data sets are evident in Figure~\ref{fig:obs:epochs}.
The most prominent of these were caused by telescope painting at Arecibo (2007), earthquake damage
at Arecibo (2014) and azimuth track refurbishment at the GBT (2007).

With few exceptions, each pulsar at each epoch was observed with at least two
receivers widely separated in observing frequency in order to measure
and remove interstellar propagation effects, including variations
in DM (Section~\ref{subsec:dm_variation_msmts}).  Such multi-receiver
observations were made on the same day at Arecibo or within a few days
at the GBT.
Exceptions to the two-receiver convention were made for
the high-cadence GBT observations, and for occasions at either
telescope when a receiver was not available for technical reasons.
Our criteria for using such data are described in Section~\ref{subsec:cleaningdata}.

Telescope receivers and data collection systems employed for this project are described in Table 1 of \nineyr.
At the GBT, we
used both the 820~MHz and 1.4~GHz receivers for monthly observations,
but only the 1.4~GHz receiver for the high-cadence observations.
At Arecibo, all sources were observed with
the 1.4~GHz receiver and a second receiver, either 430~MHz or 2.1~GHz, with choice of
receiver based on the pulsar's spectral index and timing precision in each frequency band.
Some pulsars that were initially observed with 430~MHz were later moved to 2.1~GHz, or vice versa, due to additional evaluation finding that a given pulsar is better timed at one frequency or the other.
One pulsar, PSR~J2317+1439, was initially observed at 327~MHz
and 430~MHz, but it is now observed at 430~MHz and 1.4~GHz, and no other use of the
327~MHz receiver has been made.

Two generations of backend instrumentation were used for data
collection.  The ASP and GASP systems \citep[64~MHz of
bandwidth;][]{Demorest2007} at Arecibo and the GBT, respectively, were
used for approximately the first six years of NANOGrav data
acquisition.  We transitioned to PUPPI (at Arecibo) and GUPPI (at
Green Bank) in 2012 and 2010, respectively.  PUPPI and GUPPI have
been used for all subsequent data collection, including all new
data in the present paper.  They can process up to
800-MHz bandwidths \citep{DuPlain2008, Ford2010} and significantly
improved our timing precision relative to ASP and GASP.  During the
transition from ASP/GASP to PUPPI/GUPPI, we made precise measurements
of time offsets between the instruments (Appendix A of \nineyr).  We
continue to use the offset measurements from \nineyr.

These instruments divide the telescope passband into narrow spectral
channels, undertake coherent dedispersion of the signals within each
channel, evaluate self- and cross-products to enable recovery of four
Stokes parameters, and fold the resulting time series in real time
using a nominal pulsar timing model.  Thus, the raw data are in the
form of folded pulse profiles as a function of time, radio frequency,
and polarization. The raw profiles have 2048 phase bins, a frequency
resolution of 4~MHz (ASP/GASP) or 1.5~MHz (GUPPI/PUPPI), and
subintegrations of
1~second (PUPPI at 1.4 and 2.1~GHz) or 10~seconds (all other
receiver/backend combinations).

Observations were calibrated in two steps.  Prior to each pulsar observation,
we inject a pulsed noise signal into the receiver path for use in calibrating
the signal amplitudes.  
The pulsed noise signals, in turn,
are calibrated approximately monthly via a series of on- and off-source observations
on an unpolarized continuum radio source of known flux density.  Details
of the continuum source are given in Section~\ref{sec:flux}.

\subsection{ASP/GASP Times-of-Arrival}\label{sec:aspgasp}

We collected data using the ASP and GASP instruments through 2012 and 2010,
respectively.  There are no new ASP/GASP data in this data release.  For these
data, we used TOAs generated in \nineyr\ without modification.
These TOAs were computed using procedures similar to those for PUPPI/GUPPI
data described below.  The ASP/GASP TOAs incorporate time offsets relative
to the PUPPI/GUPPI instruments as described in \nineyr.

During the transition between the ASP/GASP and PUPPI/GUPPI
instruments, parallel data were collected on two instruments,
resulting in two (redundant) sets of TOAs.  In these situations, we
use only the PUPPI/GUPPI TOAs, but we retain the commented-out
ASP/GASP TOAs in the data set, flagging these TOAs in a method similar
to the cut TOAs described in section~\ref{subsec:cleaningdata}.

\subsection{PUPPI/GUPPI Data Reduction}
\label{subsec:data_reduction}

In this section, we describe the processing of PUPPI/GUPPI 
folded pulse profiles described above to remove various artifacts,
to calibrate the data amplitudes, and to produce more compact
data sets which were then used for TOA generation.

\subsubsection{Artifact Removal}

\begin{figure*}
\begin{center}
\includegraphics[width=6.0in]{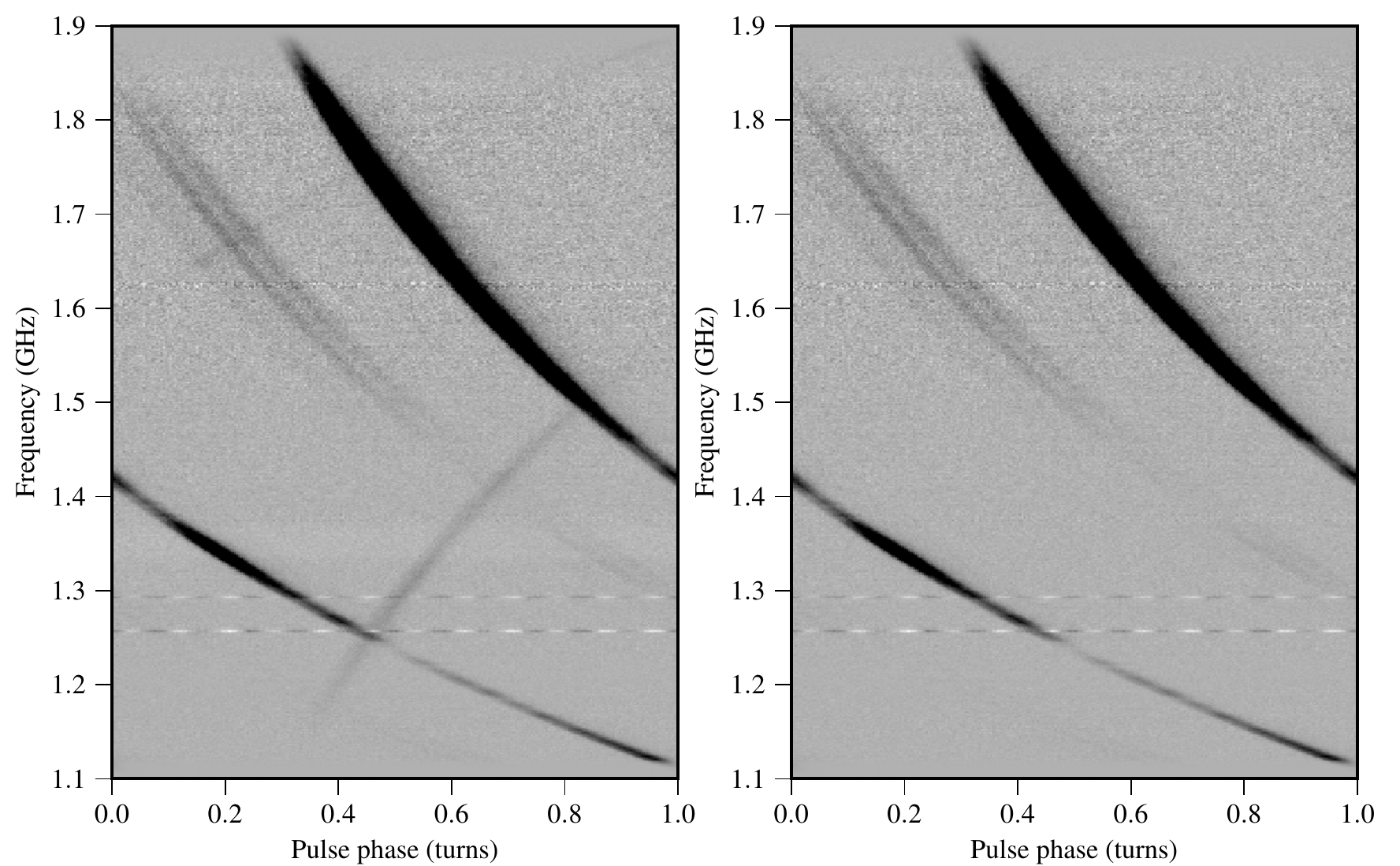}
\caption{\label{fig:obs:image}
Observation of J1744+1134 illustrating artifacts from GUPPI's
interleaved ADCs.  This is one of the lowest-DM pulsars in our sample,
therefore the effect is easily visible.  Dispersion has not been
removed, so the true pulsar signal arrives earlier at higher radio
frequencies.  The image artifact can be seen as the faint, apparently
negatively-dispersed, signal ``reflected'' about the band center
frequency of 1.5~GHz.  No interference excision has been applied to
these data; the spurious narrowband signals visible between 1.2 and
1.3~GHz are RFI.  The plot color scale has been saturated at 10\% of the
maximum data value.  The left panel shows the raw data, while the right
panel shows the same data after the correction procedure has been
applied.
}
\end{center}
\end{figure*}

GUPPI and PUPPI employ an interleaved analog-to-digital conversion
(ADC) scheme to achieve their wide bandwidths.  Rather than a single
ADC running at the Nyquist sampling rate of $2B$ (for a bandwidth
$B$), two converters are running in parallel at rate $B$, offset in
time from each other by half a cycle.  If the gain of the two
converters is not identical, or if there is a timing skew such that
the time offset is not exactly $(2B)^{-1}$, an image rejection
artifact will appear in the data.  This looks like a copy of the input
signal, frequency reversed about the center of the sampled band
(Fig.~\ref{fig:obs:image}, left panel).  For pulsar data this artifact
appears negatively dispersed and therefore can be distinguished from most typical
RFI.  The amplitude of the image signal depends on the magnitude of
the gain mismatch or timing skew.  Gain mismatch results in a constant
image ratio versus frequency, while the ratio from timing skew
increases with frequency within the sampled band.
\cite{Kurosawa} present a detailed analysis of this effect in
interleaved sampler systems, and derive analytic expressions for the
ratio of image to true signal as a function of the mismatch
parameters.

If not corrected, the presence of these artifacts will result in a
frequency-dependent systematic TOA bias.  The effect is largest at
those points in the band where the image pulse crosses the true pulse.
Low-DM, slow-spinning pulsars with wide pulse profile shapes are the
most affected, while for high-DM sources dispersion will smear out the
image relative to the pulsar signal within each channel, reducing its
impact.  For J2145$-$0750, a 1\% image ratio could shift some
individual channel TOAs by up to 1~$\mu$s, while for J1909$-$3744 the
effect is 40~ns and confined to smaller parts of the band.  On average
this effect will cancel out when averaged over the full wide band,
with equal amounts of positively and negatively shifted channels,
however this depends on the details of bandpass shape and
scintillation pattern.  In NG11 and previous data sets, there was no
effort to mitigate these images.  In the present work, we introduce a
new procedure to remove the images.

Using data from several bright pulsars in our sample, we measured the
relative amplitudes of the pulsar and image signals as a function of
frequency, bandwidth, and observing epoch.  We find that the observed
effect is consistent with a pure timing skew, with a typical value of
$\sim$30~ps.  This results in image ratios ranging from 0.5\% to 4\%
across the band.  The timing skew values are consistent between pulsars
for a given backend setup.  The values vary with time, showing
occasional step-like changes at dates corresponding to known maintenance
procedures such as replacement of a faulty ADC board or synthesizer.
Based on this observed behavior, we developed a piecewise constant in time model for the
skew value that can be used to correct the data.

With the known skew values, we calculate image ratio as a function of
frequency following \cite{Kurosawa}.  For all input data, we apply this
ratio to a frequency-reversed copy of the data, and subtract the result,
giving a corrected data set (Fig.~\ref{fig:obs:image}, right-hand
panel).  Based on the scatter of the measured skew values we
conservatively estimate this correction is good to at least the 10\%
level, therefore reducing image artifacts by one order of magnitude.

This procedure relies on having data measured continuously across the
full sampled bandwidth, as normally was the case in our observations.
However, in the PUPPI data set occasional instrumental failures resulted
in portions of the band not being recorded.  In these cases, it is not
possible to apply the correction to the corresponding subband
``mirrored'' about the band center.  Nevertheless, we elected to include
such data in the data set.  In the TOA-flagging system described below,
TOAs generated from such data are marked using the \texttt{-img uncorr}
flag.

\subsubsection{Calibration and Integration in Time and Frequency}

After removing artifact images, we performed standard data
reduction procedures as described in \elevenyr, with one additional
step of excising radio frequency interference (RFI) from the
calibration files as well as from the data files; this additional step
led to improvements in TOA measurements, especially at 2.1~GHz with
PUPPI.
The remaining steps of calibrating, reducing, and excising RFI from
the data were the same as in \elevenyr.  Data were
frequency-averaged into channels with bandwidths between
1.5 and 12.5\,MHz, depending on the receiver.  We time-averaged the
calibrated and cleaned profiles into subintegrations up to 30~minutes
in length, except in the few cases of binary pulsars with very short
orbital periods; in those cases we averaged the data into
subintegrations no longer in duration than 2.5\% of the orbital period
in order to maintain time resolution over the orbit.

\subsection{PUPPI/GUPPI Time-of-Arrival Generation}
\label{subsec:toa_gen}

We generally followed the methods described in \nineyr\ and \elevenyr\
to calculate narrowband TOAs, but with an improved algorithm to
calculate TOA uncertainties.  The uncertainties were calculated by
numerical integration of the TOA probability distribution presented in
Eqn.~12 of \nineyr\ Appendix~B.  This mitigates underestimation of
uncertainties calculated by conventional methods in the
low-signal-to-noise regime.

We used previously-generated template pulse
profiles for the 45 pulsars from \elevenyr, generating new templates
only for the two pulsars newly added to this data set.
To make the template profiles, we iteratively aligned and averaged together the
reduced data profiles, and applied wavelet smoothing to the final average profile.  With
these templates, we measured TOAs from the reduced GUPPI and PUPPI profiles,
and collated them with the existing GASP and ASP TOAs from \nineyr.

\begin{comment}

\attn{I think the following paragraph can be deleted --DJN}
The wideband data set uses an augmented Fourier phase gradient shift
algorithm to simultaneously measure a single TOA and dispersion
measure (DM)
\citep{Liu14,PDR14}.  Additionally, instead of using a constant template
profile in this measurement, the wideband data set uses a
frequency-dependent model that accounts for profile evolution, which
obviates the need for ``FD'' parameters to be included in the timing
model.  The exact profile data set from which these TOAs and DMs are
extracted, the profile modeling and TOA measurement procedures, the
curation of the data set, and the new wideband timing analysis
performed on it are all described in depth in our separate, parallel
publication, Alam et al. (\textit{submitted to ApJS}).
Most of the steps described in the remainder of this paper have
counterparts in the wideband analysis.  One main difference is that we
did not use PINT in the analysis of the wideband data set; that
development is ongoing at present.  The results from the analysis of
the wideband data set are presented in the parallel paper largely in
comparison to the bulk of findings from the present analysis, with the
conclusion that we arrive at very similar results.  This is
encouraging for the future development, implementation, and deployment
of wideband techniques for our experiment.

\end{comment}

\begin{deluxetable*}{lcp{5.8cm}p{7cm}}
\centerwidetable
\tabletypesize{\scriptsize}
\tablewidth{0pt}
\tablecaption{TOA Removal Methods\label{tab:toa_flags}}
\tablehead{\colhead{Flag} & \colhead{No.\ of TOAs} & \colhead{Reason for TOA Removal} & \colhead{Notes (including differences with the} \\[-4pt]
 & \multicolumn{1}{c}{Removed} & & \multicolumn{1}{c}{wideband data set procedures)}
}
  \startdata
  \texttt{-cut snr} & 92,290 & (Section~\ref{sec:snrcut}) Profile data used to generate TOA does not meet signal-to-noise ratio threshold & S/N $<$ 8 for narrowband TOAs, S/N $<$ 25 for wideband TOAs  \\
  \texttt{-cut badepoch} & 11,650 & (Sections~\ref{sec:outliermanualcut}, \ref{sec:calcut}, \ref{sec:fluxcut}) Observation is significantly corrupted by instrumentation issues or RFI & Identified by human inspection; these observations are not included in the wideband data set          \\
  \texttt{-cut dmx} & 10,874 & (Section~\ref{sec:baddmxcut}) Ratio of maximum to minimum frequency in an observing epoch (in a single DMX bin) $\nu_{\rm{max}}/\nu_{\rm{min}} < 1.1$ & $\nu_{\rm{max}}$ and $\nu_{\rm{min}}$ are TOA reference frequencies in the narrowband data set and are separately calculated for each TOA in the wideband data set  \\
  \texttt{-cut simul} & 5,194 & (Section~\ref{sec:aspgasp}) ASP/GASP TOA acquired at the same time as a PUPPI/GUPPI TOA & Removed at the last stage of analysis from both narrowband and wideband data sets \\
  \texttt{-cut epochdrop} & 2,384 & (Section~\ref{sec:epochcut}) Entire epoch removed based on $F$-test $p < 10^{-6}$ & Epochs identified in the narrowband data set analysis are also removed from the wideband data set  \\
  \texttt{-cut outlier10} & 1,022 & (Section~\ref{sec:outliertoa}) TOA has outlier probability $p_{i,{\rm out}} > 0.1$ & This particular outlier analysis applies only to the narrowband data set  \\
  \texttt{-cut orphaned} & 490 & (Section~\ref{sec:orphancut})  Insignificant data volume & A small number of TOAs originate from test observations in different receiver bands; these observations are not included in the wideband data set  \\
  \texttt{-cut manual} & 70 & (Section~\ref{sec:manualcut}) The TOAs corresponding to individual pulse profile are corrupted by instrumentation or RFI, but were not identified and removed via S/N threshold or outlier probability cuts & Identified by human inspection; both narrowband and wideband data sets have a small volume of manually excised TOAs, but they were determined independently  \\
  \enddata
  \tablecomments{The flags are listed here in order of the number of narrowband TOAs that were removed from the data set via each method.
        All cut TOAs are provided as commented-out TOAs in the ASCII-text TOA files; excluding these, there are 415,173 narrowband TOAs in the data set.}   \end{deluxetable*}
 
\subsection{Cleaning the Data Set for Improved Data Quality}  \label{subsec:cleaningdata}

Calculated TOAs can be biased by a variety of observational problems,
including imperfections in instrumentation, flawed calibration, RFI,
or other non-astrophysical influences.
In past data releases,
we have ensured a high level of data quality by systematically removing
RFI, excluding low signal-to-noise (S/N) TOAs
(see details in \nineyr), removing outliers identified by Bayesian
analysis of residuals (see
details in \elevenyr), and manual inspection of the data sets.  
These same procedures were carried out for the
present data set, along with a series of new cleaning techniques
described below.  

The remainder of this section details these quality-control measures, which
led to the removal of uninformative or suspect TOAs, 
as well as entire observations in some cases. The data quality
analysis steps were typically iterative in nature.  For example, the ``bad
DMX range'' criterion described below was re-checked after any change in the
data set made for other reasons.

All TOAs removed by these procedures are included in the data set files.  They
are marked as comments in the TOA files, and each excluded TOA includes a
flag indicating the reason for its exclusion.  The exclusion methods, flags,
and statistics of removed TOAs are summarized in Table~\ref{tab:toa_flags}.
After all TOA cutting was
complete, the final narrowband data set had 415,173 TOAs.

\subsubsection{Signal-to-noise ratio cut}\label{sec:snrcut}
As in \nineyr\ and \elevenyr, we removed TOAs that were generated
from pulse profiles with S/N $<$ 8. These TOAs contain little information
and, at the very lowest S/N values, can be miscalculated due to
the dominance of noise over any pulsar signal that is present.
We maintained this S/N cutoff in the present work.

\subsubsection{Bad DM range cut}\label{sec:baddmxcut}
As detailed in Section~\ref{subsec:dm_variation_msmts}, our timing models
include a step-wise model for variation in DM, in which DM is allowed to have
independently varying values in time intervals.  The time intervals range in length 
from 0.5 to 15~days, depending on the telescope and instrumentation.
In order to achieve DM measurements of reasonable precision, we
require that the maximum-to-minimum frequency ratio of TOAs with
each of these time ranges satisfies
$\nu_{\mathrm{max}}/\nu_{\mathrm{min}} > 1.1$.
Data within any DM time
range that did not meet this criterion were removed.

As with \elevenyr, this criterion lead to the removal of some data that had been incorporated in \nineyr, in particular lengthy subsets of single-receiver data from a few pulsars initially observed as part of a non-NANOGrav timing program and later merged into our data set.

\subsubsection{Outlier TOA Cut}\label{sec:outliertoa}
We used the automated outlier-identification algorithm
of \cite{Vallisneri2017} to estimate the probability $p_{i,{\rm out}}$
that each individual TOA is an outlier, based on the initial 12.5-year
timing models. The $p_{i,{\rm out}}$ are defined in Equation~6 of
Vallisneri \& van Haasteren, where Equations~1--5 provide definitions
of the parameters in Equation 6.  The $p_{i,{\rm out}}$ estimate is fully
consistent with the Bayesian inference of all noise parameters of the
pulsar being analyzed.  As in \elevenyr, we removed all TOAs with
$p_{i,{\rm out}} > 0.1$, resulting in a total of 1,022 TOAs being
removed from the full 12.5-year data set\footnote{We note that this
outlier analysis step was incorporated as an additional step toward
the end of the pipeline in \elevenyr, after the TOA sets had already
been manually edited.  For the current data set, we followed the S/N
thresholding step with this outlier analysis, and only later manually
edited the TOA data set.}. This outlier probability threshold was chosen based on inspection of a subset of pulsars, from which we found that $p_{i,{\rm out}}$ has an extremely bimodal distribution: nearly all TOAs had $p_{i,{\rm out}} < 0.1$ or $p_{i,{\rm out}} > 0.9$. The $p_{i,{\rm out}} > 0.1$ threshold for TOA removal is thus empirically motivated, and additionally remains consistent with the threshold used in \elevenyr.

We note that the outlier analysis was not run
iteratively as timing models were updated. Such an iterative procedure
was tested on several pulsars, and it was found that the updated
outlier results had very minor, if any, differences from the original outlier
results. This finding is likely explained by the fact that the results of the outlier runs are already marginalized over the linearized timing model parameters, with very large priors. Thus a different set of outliers from a subsequent analysis would only be expected if the original outliers had taken the timing model outside its linear range, or if so many data points were excluded after the initial analysis that the noise profile changed and new, previously-obscured outliers were revealed. Based on our empirical findings from the iterative outlier analysis runs on a subset of pulsars, and on this theoretical explanation, we chose to only run the outlier analysis
once at the beginning of the timing analysis for the full data set.

\subsubsection{Manual Removal of Individual Observations Guided by Outlier TOAs}\label{sec:outliermanualcut}
The outlier analysis was used as a guide in identifying observing
epochs in which most or all TOAs may have been corrupted by
instrumentation issues or by excessive RFI. If a single observation had
more than five TOAs with $p_{i,{\rm out}} > 0.1$, we reviewed the corresponding
observing log and examined the data manually to determine if there were
instrumental problems or RFI that rendered the
observation unusable.  In such cases, all TOAs from the affected observations were
removed.

\subsubsection{Corrupt Calibration Cut}\label{sec:calcut}
As described above, pulsar observations were preceded by measurements using
an artificial pulsed noise signal injected in the telescope signal path,
and the same pulsed-noise-signal method was used during continuum calibrator observations.
We searched for anomalies in the pulsed-noise-signal measurements; these could
result from instrumentation failures or from use of incorrect noise signals
in an individual observation.

The following are specific anomalies that were identified in the calibration files.
(1) A pulsed-noise amplitude that was unusually high or low (88 affected observations).
(2) Cross-polarization flux calibration amplitudes deviating
significantly from the mean locus of amplitudes for a given
receiver/backend combination (65 observations affected and removed, due to corrupted continuum source
observations on MJDs~57229 and 57249).
(3) Pulsed calibrator phase not smoothly varying with frequency
(one observation
was removed solely for this
reason, but this was also seen in observations flagged for othe reasons).
(4) Polarization fraction, $f_{UV} =
 [(U^2+V^2)/I^2]^{1/2}$, deviating from expected values. 
For pulsars with high intrinsic
 polarization ($f_{UV} \sim 1$), a small number of observations had
 significantly lower estimated $\hat{f}_{UV}$ values (typically $\hat{f}_{UV} < 0.3$) that
 signified corrupted data, and a small number
 of profiles had $\hat{f}_{UV} > 1$, suggesting a problem with the
 digitization levels (23 affected observations).

\subsubsection{Flux Measurement Cut}\label{sec:fluxcut}
Extremely high or low apparent flux density values can result from
incorrect calibration or digitization levels.  We therefore searched for
outlier flux densities as a proxy for calibration errors 
not detected through the means listed above.
We manually examined such observations 
and removed them as needed.
Only one observation was identified and removed by this method, but this analysis
informed the development of our other quality-check methods.
A detailed flux density analysis is
in Section~\ref{sec:flux}.

\subsubsection{Epoch $F$-Test Cut}\label{sec:epochcut}
We tested for the presence of otherwise-undetected bad data for
each pulsar by removing 
data from one observing epoch at a time and examining its impact on
the timing residuals.  
This method is effectively an outlier analysis for full observing
epochs, rather than for individual TOAs.
We compared the chi-square of the timing residuals before and after
data removal, $\chi^2_0$ and $\chi^2$, respectively, using an $F$-statistic,
\begin{equation}
    F = \frac{(\chi^2_0 - \chi^2) / (n_0 - n)}{(\chi^2 / n)},
\end{equation}
where $n_0$ and $n$ are the number of degrees of freedom in the original
and epoch-removed analyses.  We removed data for epochs for which the
$F$-test reported a chance probability $p < 10^{-6}$ ($\sim$~5$\sigma$).
This process was run iteratively.
We examined the profiles and calibration files for a subset of the observations that were flagged in this way and found that a majority of
these observations were faulty in obvious ways (calibration errors, extreme
RFI, etc.).
\subsubsection{Manual TOA Cut}\label{sec:manualcut}
After completing the above data quality checks, uninformative or
outlying TOAs were still present in the data sets of some 
pulsars. A total of 70 additional TOAs were removed 
after visual inspection.  These were TOAs whose timing
residuals appeared to be outliers but were not flagged by the outlier
analysis (typically a small number of pulse profiles
had been corrupted by narrow-frequency RFI, such that TOAs in a subset
of the full bandwidth had to be removed but the rest of the profile or
epoch was not adversely affected) or TOAs with very large
uncertainties and large timing residuals (usually resulting
from a low-S/N TOA measurement, just above our cutoff threshold, or RFI).

\smallskip

\subsubsection{Orphan Data Cut}\label{sec:orphancut}
For a few pulsars, in addition to the receivers normally used for
observations, a small number of observations were made with a
different receiver, typically for testing purposes near the start
of observations of this source.  We cut such TOAs in the 
same manner as data that were cut for other reasons.

\subsubsection{Wideband TOA Residual Check}

We note that for the methods described above, in some cases it was
difficult to determine if data were corrupted (e.g., the residuals
from a given epoch may have been larger than expected, but no evidence
of an issue with the calibration or profile data was found upon
inspection).  In those cases, we also examined the residual profiles
generated from fitting wideband TOAs, which use high-fidelity,
evolving profile models in the matched-template algorithm. For
example, the residual profile may reveal that the evolving profile
model did not adequately represent the profile during that observation,
suggesting a problem with the data that was not found using other
methods. Inspection of the wideband residual profiles thus aided in
identifying and removing more corrupted data in the narrowband data
set.  Using the evolving profile models as a means to identify
corrupted data profiles is an ongoing development.

\section{Timing Analysis}
\label{sec:timing}
\begin{table*}[th!]
\centering
\caption{Basic Pulsar Parameters and TOA Statistics\label{tab:timingpars}}
{\footnotesize\begin{tabular}{crrrrrlrlrlrlrlr}
\hline
\hline
Source & \multicolumn{1}{c}{$P$}  & \multicolumn{1}{c}{$dP/dt$}      & \multicolumn{1}{c}{DM}           & \multicolumn{1}{c}{$P_b$}
         & \multicolumn{10}{c}{Median scaled TOA uncertainty$^a$ ($\mu$s) / Number of epochs}
         & \multicolumn{1}{c}{Span} \\ \cline{6-15}
       & \multicolumn{1}{c}{(ms)} & \multicolumn{1}{c}{($10^{-20}$)} & \multicolumn{1}{c}{(pc~cm$^{-3}$)} & \multicolumn{1}{c}{(d)}
         & \multicolumn{2}{c}{327~MHz}
         & \multicolumn{2}{c}{430~MHz}
         & \multicolumn{2}{c}{820~MHz}
         & \multicolumn{2}{c}{1.4~GHz}
         & \multicolumn{2}{c}{2.1~GHz}
         & \multicolumn{1}{c}{(yr)} \\
\hline
J0023$+$0923 & 3.05 & 1.14 & 14.3 & 0.1& \multicolumn{2}{c}{\nodata} & 0.063 & 62 & \multicolumn{2}{c}{\nodata} & 0.556 & 68 & \multicolumn{2}{c}{\nodata} & 6.0\\
J0030$+$0451 & 4.87 & 1.02 & 4.3 & -& \multicolumn{2}{c}{\nodata} & 0.214 & 175 & \multicolumn{2}{c}{\nodata} & 0.424 & 187 & 1.558 & 71 & 12.4\\
J0340$+$4130 & 3.30 & 0.70 & 49.6 & -& \multicolumn{2}{c}{\nodata} & \multicolumn{2}{c}{\nodata} & 0.868 & 68 & 2.108 & 71 & \multicolumn{2}{c}{\nodata} & 5.3\\
J0613$-$0200 & 3.06 & 0.96 & 38.8 & 1.2& \multicolumn{2}{c}{\nodata} & \multicolumn{2}{c}{\nodata} & 0.109 & 134 & 0.582 & 135 & \multicolumn{2}{c}{\nodata} & 12.2\\
J0636$+$5128 & 2.87 & 0.34 & 11.1 & 0.1& \multicolumn{2}{c}{\nodata} & \multicolumn{2}{c}{\nodata} & 0.279 & 39 & 0.579 & 42 & \multicolumn{2}{c}{\nodata} & 3.5\\
J0645$+$5158 & 8.85 & 0.49 & 18.2 & -& \multicolumn{2}{c}{\nodata} & \multicolumn{2}{c}{\nodata} & 0.297 & 67 & 0.836 & 74 & \multicolumn{2}{c}{\nodata} & 6.1\\
J0740$+$6620 & 2.89 & 1.22 & 15.0 & 4.8& \multicolumn{2}{c}{\nodata} & \multicolumn{2}{c}{\nodata} & 0.445 & 38 & 0.651 & 40 & \multicolumn{2}{c}{\nodata} & 3.5\\
J0931$-$1902 & 4.64 & 0.36 & 41.5 & -& \multicolumn{2}{c}{\nodata} & \multicolumn{2}{c}{\nodata} & 1.030 & 51 & 1.777 & 53 & \multicolumn{2}{c}{\nodata} & 4.3\\
J1012$+$5307 & 5.26 & 1.71 & 9.0 & 0.6& \multicolumn{2}{c}{\nodata} & \multicolumn{2}{c}{\nodata} & 0.403 & 135 & 0.725 & 143 & \multicolumn{2}{c}{\nodata} & 12.9\\
J1024$-$0719 & 5.16 & 1.86 & 6.5 & -& \multicolumn{2}{c}{\nodata} & \multicolumn{2}{c}{\nodata} & 0.520 & 90 & 0.981 & 94 & \multicolumn{2}{c}{\nodata} & 7.7\\
J1125$+$7819 & 4.20 & 0.69 & 12.0 & 15.4& \multicolumn{2}{c}{\nodata} & \multicolumn{2}{c}{\nodata} & 0.974 & 40 & 2.024 & 42 & \multicolumn{2}{c}{\nodata} & 3.5\\
J1453$+$1902 & 5.79 & 1.17 & 14.1 & -& \multicolumn{2}{c}{\nodata} & 1.141 & 35 & \multicolumn{2}{c}{\nodata} & 2.120 & 40 & \multicolumn{2}{c}{\nodata} & 3.9\\
J1455$-$3330 & 7.99 & 2.43 & 13.6 & 76.2& \multicolumn{2}{c}{\nodata} & \multicolumn{2}{c}{\nodata} & 1.100 & 115 & 1.937 & 117 & \multicolumn{2}{c}{\nodata} & 12.9\\
J1600$-$3053 & 3.60 & 0.95 & 52.3 & 14.3& \multicolumn{2}{c}{\nodata} & \multicolumn{2}{c}{\nodata} & 0.271 & 113 & 0.227 & 115 & \multicolumn{2}{c}{\nodata} & 9.6\\
J1614$-$2230 & 3.15 & 0.96 & 34.5 & 8.7& \multicolumn{2}{c}{\nodata} & \multicolumn{2}{c}{\nodata} & 0.374 & 96 & 0.593 & 107 & \multicolumn{2}{c}{\nodata} & 8.8\\
J1640$+$2224 & 3.16 & 0.28 & 18.5 & 175.5& \multicolumn{2}{c}{\nodata} & 0.048 & 180 & \multicolumn{2}{c}{\nodata} & 0.375 & 189 & \multicolumn{2}{c}{\nodata} & 12.3\\
J1643$-$1224 & 4.62 & 1.85 & 62.3 & 147.0& \multicolumn{2}{c}{\nodata} & \multicolumn{2}{c}{\nodata} & 0.288 & 131 & 0.499 & 131 & \multicolumn{2}{c}{\nodata} & 12.7\\
J1713$+$0747 & 4.57 & 0.85 & 15.9 & 67.8& \multicolumn{2}{c}{\nodata} & \multicolumn{2}{c}{\nodata} & 0.188 & 129 & 0.077 & 451 & 0.061 & 186 & 12.4\\
J1738$+$0333 & 5.85 & 2.41 & 33.8 & 0.4& \multicolumn{2}{c}{\nodata} & \multicolumn{2}{c}{\nodata} & \multicolumn{2}{c}{\nodata} & 0.520 & 71 & 0.901 & 68 & 7.6\\
J1741$+$1351 & 3.75 & 3.02 & 24.2 & 16.3& \multicolumn{2}{c}{\nodata} & 0.142 & 63 & \multicolumn{2}{c}{\nodata} & 0.352 & 73 & \multicolumn{2}{c}{\nodata} & 5.9\\
J1744$-$1134 & 4.07 & 0.89 & 3.1 & -& \multicolumn{2}{c}{\nodata} & \multicolumn{2}{c}{\nodata} & 0.155 & 130 & 0.237 & 128 & \multicolumn{2}{c}{\nodata} & 12.9\\
J1747$-$4036 & 1.65 & 1.31 & 153.0 & -& \multicolumn{2}{c}{\nodata} & \multicolumn{2}{c}{\nodata} & 1.033 & 61 & 1.160 & 65 & \multicolumn{2}{c}{\nodata} & 5.3\\
J1832$-$0836 & 2.72 & 0.83 & 28.2 & -& \multicolumn{2}{c}{\nodata} & \multicolumn{2}{c}{\nodata} & 0.596 & 53 & 0.524 & 53 & \multicolumn{2}{c}{\nodata} & 4.3\\
J1853$+$1303 & 4.09 & 0.87 & 30.6 & 115.7& \multicolumn{2}{c}{\nodata} & 0.353 & 67 & \multicolumn{2}{c}{\nodata} & 0.593 & 72 & \multicolumn{2}{c}{\nodata} & 6.0\\
B1855$+$09\phantom{....} & 5.36 & 1.78 & 13.3 & 12.3& \multicolumn{2}{c}{\nodata} & 0.208 & 117 & \multicolumn{2}{c}{\nodata} & 0.211 & 124 & \multicolumn{2}{c}{\nodata} & 12.5\\
J1903$+$0327 & 2.15 & 1.88 & 297.5 & 95.2& \multicolumn{2}{c}{\nodata} & \multicolumn{2}{c}{\nodata} & \multicolumn{2}{c}{\nodata} & 0.443 & 75 & 0.511 & 78 & 7.6\\
J1909$-$3744 & 2.95 & 1.40 & 10.4 & 1.5& \multicolumn{2}{c}{\nodata} & \multicolumn{2}{c}{\nodata} & 0.066 & 126 & 0.124 & 269 & \multicolumn{2}{c}{\nodata} & 12.7\\
J1910$+$1256 & 4.98 & 0.97 & 38.1 & 58.5& \multicolumn{2}{c}{\nodata} & \multicolumn{2}{c}{\nodata} & \multicolumn{2}{c}{\nodata} & 0.338 & 82 & 0.767 & 83 & 8.3\\
J1911$+$1347 & 4.63 & 1.69 & 31.0 & -& \multicolumn{2}{c}{\nodata} & 0.590 & 42 & \multicolumn{2}{c}{\nodata} & 0.157 & 46 & \multicolumn{2}{c}{\nodata} & 3.9\\
J1918$-$0642 & 7.65 & 2.57 & 26.5 & 10.9& \multicolumn{2}{c}{\nodata} & \multicolumn{2}{c}{\nodata} & 0.518 & 126 & 0.901 & 128 & \multicolumn{2}{c}{\nodata} & 12.7\\
J1923$+$2515 & 3.79 & 0.96 & 18.9 & -& \multicolumn{2}{c}{\nodata} & 0.259 & 55 & \multicolumn{2}{c}{\nodata} & 1.023 & 67 & \multicolumn{2}{c}{\nodata} & 5.8\\
B1937$+$21\phantom{....} & 1.56 & 10.51 & 71.1 & -& \multicolumn{2}{c}{\nodata} & \multicolumn{2}{c}{\nodata} & 0.007 & 127 & 0.014 & 220 & 0.018 & 86 & 12.8\\
J1944$+$0907 & 5.19 & 1.73 & 24.4 & -& \multicolumn{2}{c}{\nodata} & 0.278 & 63 & \multicolumn{2}{c}{\nodata} & 0.825 & 73 & \multicolumn{2}{c}{\nodata} & 9.3\\
J1946$+$3417 & 3.17 & 0.32 & 110.2 & 27.0& \multicolumn{2}{c}{\nodata} & \multicolumn{2}{c}{\nodata} & \multicolumn{2}{c}{\nodata} & 0.414 & 40 & 0.547 & 39 & 2.6\\
B1953$+$29\phantom{....} & 6.13 & 2.97 & 104.5 & 117.3& \multicolumn{2}{c}{\nodata} & 0.255 & 54 & \multicolumn{2}{c}{\nodata} & 0.815 & 65 & \multicolumn{2}{c}{\nodata} & 5.9\\
J2010$-$1323 & 5.22 & 0.48 & 22.2 & -& \multicolumn{2}{c}{\nodata} & \multicolumn{2}{c}{\nodata} & 0.412 & 94 & 0.983 & 96 & \multicolumn{2}{c}{\nodata} & 7.8\\
J2017$+$0603 & 2.90 & 0.80 & 23.9 & 2.2& \multicolumn{2}{c}{\nodata} & 0.195 & 6 & \multicolumn{2}{c}{\nodata} & 0.425 & 67 & 0.537 & 50 & 5.3\\
J2033$+$1734 & 5.95 & 1.11 & 25.1 & 56.3& \multicolumn{2}{c}{\nodata} & 0.194 & 40 & \multicolumn{2}{c}{\nodata} & 1.163 & 46 & \multicolumn{2}{c}{\nodata} & 3.8\\
J2043$+$1711 & 2.38 & 0.52 & 20.8 & 1.5& \multicolumn{2}{c}{\nodata} & 0.079 & 137 & \multicolumn{2}{c}{\nodata} & 0.281 & 151 & \multicolumn{2}{c}{\nodata} & 5.9\\
J2145$-$0750 & 16.05 & 2.98 & 9.0 & 6.8& \multicolumn{2}{c}{\nodata} & \multicolumn{2}{c}{\nodata} & 0.289 & 111 & 0.650 & 116 & \multicolumn{2}{c}{\nodata} & 12.8\\
J2214$+$3000 & 3.12 & 1.47 & 22.5 & 0.4& \multicolumn{2}{c}{\nodata} & \multicolumn{2}{c}{\nodata} & \multicolumn{2}{c}{\nodata} & 0.743 & 72 & 1.059 & 57 & 5.7\\
J2229$+$2643 & 2.98 & 0.15 & 22.7 & 93.0& \multicolumn{2}{c}{\nodata} & 0.324 & 45 & \multicolumn{2}{c}{\nodata} & 1.096 & 47 & \multicolumn{2}{c}{\nodata} & 3.9\\
J2234$+$0611 & 3.58 & 1.20 & 10.8 & 32.0& \multicolumn{2}{c}{\nodata} & 0.429 & 41 & \multicolumn{2}{c}{\nodata} & 0.221 & 45 & \multicolumn{2}{c}{\nodata} & 3.4\\
J2234$+$0944 & 3.63 & 2.01 & 17.8 & 0.4& \multicolumn{2}{c}{\nodata} & \multicolumn{2}{c}{\nodata} & \multicolumn{2}{c}{\nodata} & 0.314 & 45 & 0.746 & 44 & 4.0\\
J2302$+$4442 & 5.19 & 1.39 & 13.8 & 125.9& \multicolumn{2}{c}{\nodata} & \multicolumn{2}{c}{\nodata} & 1.200 & 69 & 2.413 & 68 & \multicolumn{2}{c}{\nodata} & 5.3\\
J2317$+$1439 & 3.45 & 0.24 & 21.9 & 2.5& 0.085 & 79 & 0.068 & 188 & \multicolumn{2}{c}{\nodata} & 0.642 & 141 & \multicolumn{2}{c}{\nodata} & 12.5\\
J2322$+$2057 & 4.81 & 0.97 & 13.4 & -& \multicolumn{2}{c}{\nodata} & 0.291 & 35 & \multicolumn{2}{c}{\nodata} & 1.021 & 34 & 1.431 & 10 & 2.3\\
\hline
\multicolumn{5}{r}{Nominal scaling factor$^b$ (ASP/GASP)}
  & \multicolumn{2}{c}{0.6}
  & \multicolumn{2}{c}{0.4}
  & \multicolumn{2}{c}{0.8}
  & \multicolumn{2}{c}{0.8}
  & \multicolumn{2}{c}{0.8}
  & \\
\multicolumn{5}{r}{Nominal scaling factor$^b$ (GUPPI/PUPPI)}
  & \multicolumn{2}{c}{0.7}
  & \multicolumn{2}{c}{0.5}
  & \multicolumn{2}{c}{1.4}
  & \multicolumn{2}{c}{2.5}
  & \multicolumn{2}{c}{2.1}
  & \\
\hline
\end{tabular}

\vspace{0.5em}

{$^a$ For this table, the original TOA uncertainties were scaled by their
bandwidth-time product $\left( \frac{\Delta \nu}{100~\mathrm{MHz}}
\frac{\tau}{1800~\mathrm{s}} \right)^{1/2}$ to remove variation due to
different instrument bandwidths and integration time. We note that in \elevenyr, we
incorrectly calculated the tabulated TOA uncertainties due to a
scripting error.  This generally led to overestimates of the uncertainty
at lower frequencies and underestimates at higher frequencies.  The
error only applied to values shown in Table 1 of \elevenyr, and did not affect the
released data or any other results in the paper.  We have corrected this
error for the present work.}

\vspace{0.5em}

{$^b$ TOA uncertainties can be rescaled to the nominal full instrumental
bandwidth as listed in Table~1 of \cite{Arzoumanian2015b} by dividing by the
scaling factors given here.}

}

 \end{table*}

\begin{table*}[th!]
\centering
\caption{Summary of Timing Model Fits\label{tab:rms_rednoise}}
{\footnotesize
\begin{tabular}{@{\extracolsep{6pt}}crrrrrrrccccrc}
\hline
\hline
Source
  & Number
  & \multicolumn{6}{c}{Number of Fit Parameters$^a$}
  & \multicolumn{2}{c}{RMS$^b$ ($\mu$s)}
  & \multicolumn{3}{c}{Red Noise$^c$}
  & Figure \\ \cline{3-8} \cline{9-10} \cline{11-13}

  & of TOAs
  & S
  & A
  & B
  & DM
  & FD
  & J
  & Full
  & White
  & $A_{\mathrm{red}}$
  & $\gamma_{\mathrm{red}}$
  & log$_{10}B$
  & Number\\
\hline
J0023$+$0923 & 12516 & 3 & 5 & 9 & 67 & 4 & 1 & 0.285  & \nodata & \nodata & \nodata & 1.21  & \ref{fig:summary-J0023+0923} \\
J0030$+$0451 & 12543 & 3 & 5 & 0 & 190 & 4 & 2 & 25.157  & 0.200  & 0.003 & $-$6.3 & $>$2  & \ref{fig:summary-J0030+0451} \\
J0340$+$4130 & 8069 & 3 & 5 & 0 & 74 & 4 & 1 & 0.446  & \nodata & \nodata & \nodata & $-$0.21  & \ref{fig:summary-J0340+4130} \\
J0613$-$0200 & 13201 & 3 & 5 & 8 & 139 & 2 & 1 & 0.486  & 0.178  & 0.123 & $-$2.1 & $>$2  & \ref{fig:summary-J0613-0200} \\
J0636$+$5128 & 21374 & 3 & 5 & 6 & 44 & 1 & 1 & 0.640  & \nodata & \nodata & \nodata & $-$0.09  & \ref{fig:summary-J0636+5128} \\
J0645$+$5158 & 7893 & 3 & 5 & 0 & 79 & 2 & 1 & 0.207  & \nodata & \nodata & \nodata & $-$0.20  & \ref{fig:summary-J0645+5158} \\
J0740$+$6620 & 3328 & 3 & 5 & 7 & 44 & 1 & 1 & 0.132  & \nodata & \nodata & \nodata & $-$0.17  & \ref{fig:summary-J0740+6620} \\
J0931$-$1902 & 3712 & 3 & 5 & 0 & 57 & 0 & 1 & 0.452  & \nodata & \nodata & \nodata & $-$0.15  & \ref{fig:summary-J0931-1902} \\
J1012$+$5307 & 19307 & 3 & 5 & 6 & 142 & 4 & 1 & 0.999  & 0.272  & 0.406 & $-$1.6 & $>$2  & \ref{fig:summary-J1012+5307} \\
J1024$-$0719 & 9792 & 4 & 5 & 0 & 100 & 2 & 1 & 0.334  & \nodata & \nodata & \nodata & $-$0.08  & \ref{fig:summary-J1024-0719} \\
J1125$+$7819 & 4821 & 3 & 5 & 5 & 43 & 3 & 1 & 0.862  & \nodata & \nodata & \nodata & 0.09  & \ref{fig:summary-J1125+7819} \\
J1453$+$1902 & 1555 & 3 & 5 & 0 & 39 & 0 & 1 & 0.606  & \nodata & \nodata & \nodata & $-$0.13  & \ref{fig:summary-J1453+1902} \\
J1455$-$3330 & 8408 & 3 & 5 & 6 & 122 & 2 & 1 & 0.656  & \nodata & \nodata & \nodata & $-$0.14  & \ref{fig:summary-J1455-3330} \\
J1600$-$3053 & 14374 & 3 & 5 & 8 & 128 & 2 & 1 & 0.245  & \nodata & \nodata & \nodata & 0.55  & \ref{fig:summary-J1600-3053} \\
J1614$-$2230 & 12775 & 3 & 5 & 8 & 114 & 2 & 1 & 0.177  & \nodata & \nodata & \nodata & $-$0.24  & \ref{fig:summary-J1614-2230} \\
J1640$+$2224 & 9256 & 3 & 5 & 8 & 188 & 4 & 1 & 0.177  & \nodata & \nodata & \nodata & $-$0.20  & \ref{fig:summary-J1640+2224} \\
J1643$-$1224 & 12798 & 3 & 5 & 6 & 141 & 2 & 1 & 2.645  & 0.534  & 1.498 & $-$1.4 & $>$2  & \ref{fig:summary-J1643-1224} \\
J1713$+$0747 & 37698 & 3 & 5 & 8 & 325 & 5 & 3 & 0.101  & 0.069  & 0.030 & $-$1.3 & $>$2  & \ref{fig:summary-J1713+0747} \\
J1738$+$0333 & 6977 & 3 & 5 & 5 & 78 & 1 & 1 & 0.276  & \nodata & \nodata & \nodata & $-$0.24  & \ref{fig:summary-J1738+0333} \\
J1741$+$1351 & 3845 & 3 & 5 & 8 & 73 & 2 & 1 & 0.156  & \nodata & \nodata & \nodata & $-$0.08  & \ref{fig:summary-J1741+1351} \\
J1744$-$1134 & 13380 & 3 & 5 & 0 & 136 & 4 & 1 & 0.832  & 0.307  & 0.155 & $-$2.2 & $>$2  & \ref{fig:summary-J1744-1134} \\
J1747$-$4036 & 7572 & 3 & 5 & 0 & 71 & 1 & 1 & 6.343  & 1.414  & 0.709 & $-$3.3 & $>$2  & \ref{fig:summary-J1747-4036} \\
J1832$-$0836 & 5364 & 3 & 5 & 0 & 58 & 0 & 1 & 0.187  & \nodata & \nodata & \nodata & $-$0.05  & \ref{fig:summary-J1832-0836} \\
J1853$+$1303 & 3544 & 3 & 5 & 8 & 72 & 0 & 1 & 0.392  & 0.110  & 0.140 & $-$2.2 & $>$2  & \ref{fig:summary-J1853+1303} \\
B1855$+$09\phantom{....} & 6464 & 3 & 5 & 7 & 125 & 3 & 1 & 1.757  & 0.357  & 0.054 & $-$3.4 & $>$2  & \ref{fig:summary-B1855+09} \\
J1903$+$0327 & 4854 & 3 & 5 & 8 & 82 & 1 & 1 & 2.668  & 0.315  & 1.482 & $-$1.6 & $>$2  & \ref{fig:summary-J1903+0327} \\
J1909$-$3744 & 22633 & 3 & 5 & 9 & 223 & 1 & 1 & 0.334  & 0.061  & 0.028 & $-$2.7 & $>$2  & \ref{fig:summary-J1909-3744} \\
J1910$+$1256 & 5012 & 3 & 5 & 6 & 88 & 1 & 1 & 0.187  & \nodata & \nodata & \nodata & $-$0.06  & \ref{fig:summary-J1910+1256} \\
J1911$+$1347 & 2625 & 3 & 5 & 0 & 46 & 2 & 1 & 0.118  & \nodata & \nodata & \nodata & 0.20  & \ref{fig:summary-J1911+1347} \\
J1918$-$0642 & 13675 & 3 & 5 & 7 & 133 & 5 & 1 & 0.299  & \nodata & \nodata & \nodata & 0.02  & \ref{fig:summary-J1918-0642} \\
J1923$+$2515 & 3009 & 3 & 5 & 0 & 67 & 1 & 1 & 0.269  & \nodata & \nodata & \nodata & $-$0.15  & \ref{fig:summary-J1923+2515} \\
B1937$+$21\phantom{....} & 17024 & 3 & 5 & 0 & 204 & 5 & 3 & 2.277  & 0.103  & 0.099 & $-$3.3 & $>$2  & \ref{fig:summary-B1937+21} \\
J1944$+$0907 & 3931 & 3 & 5 & 0 & 73 & 2 & 1 & 0.365  & \nodata & \nodata & \nodata & 0.12  & \ref{fig:summary-J1944+0907} \\
J1946$+$3417 & 3016 & 3 & 5 & 8 & 41 & 1 & 1 & 0.468  & \nodata & \nodata & \nodata & 1.77  & \ref{fig:summary-J1946+3417} \\
B1953$+$29\phantom{....} & 3421 & 3 & 5 & 6 & 65 & 2 & 1 & 0.475  & \nodata & \nodata & \nodata & 1.05  & \ref{fig:summary-B1953+29} \\
J2010$-$1323 & 13306 & 3 & 5 & 0 & 108 & 1 & 1 & 0.244  & \nodata & \nodata & \nodata & $-$0.22  & \ref{fig:summary-J2010-1323} \\
J2017$+$0603 & 2986 & 3 & 5 & 7 & 73 & 0 & 2 & 0.076  & \nodata & \nodata & \nodata & $-$0.22  & \ref{fig:summary-J2017+0603} \\
J2033$+$1734 & 2691 & 3 & 5 & 5 & 46 & 2 & 1 & 0.561  & \nodata & \nodata & \nodata & $-$0.12  & \ref{fig:summary-J2033+1734} \\
J2043$+$1711 & 5624 & 3 & 5 & 7 & 151 & 4 & 1 & 0.151  & \nodata & \nodata & \nodata & 1.41  & \ref{fig:summary-J2043+1711} \\
J2145$-$0750 & 13961 & 3 & 5 & 7 & 123 & 2 & 1 & 1.467  & 0.328  & 0.347 & $-$2.1 & $>$2  & \ref{fig:summary-J2145-0750} \\
J2214$+$3000 & 6269 & 3 & 5 & 5 & 77 & 1 & 1 & 0.402  & \nodata & \nodata & \nodata & $-$0.17  & \ref{fig:summary-J2214+3000} \\
J2229$+$2643 & 2442 & 3 & 5 & 6 & 47 & 2 & 1 & 0.194  & \nodata & \nodata & \nodata & $-$0.18  & \ref{fig:summary-J2229+2643} \\
J2234$+$0611 & 2475 & 3 & 5 & 7 & 45 & 2 & 1 & 0.061  & \nodata & \nodata & \nodata & 0.60  & \ref{fig:summary-J2234+0611} \\
J2234$+$0944 & 5892 & 3 & 5 & 5 & 51 & 2 & 1 & 0.160  & \nodata & \nodata & \nodata & $-$0.13  & \ref{fig:summary-J2234+0944} \\
J2302$+$4442 & 7833 & 3 & 5 & 7 & 75 & 1 & 1 & 0.716  & \nodata & \nodata & \nodata & $-$0.15  & \ref{fig:summary-J2302+4442} \\
J2317$+$1439 & 9835 & 3 & 5 & 6 & 210 & 3 & 2 & 8.798  & 0.253  & 0.007 & $-$6.4 & $>$2  & \ref{fig:summary-J2317+1439} \\
J2322$+$2057 & 2093 & 3 & 5 & 0 & 35 & 4 & 2 & 0.235  & \nodata & \nodata & \nodata & $-$0.13  & \ref{fig:summary-J2322+2057} \\
\hline
\end{tabular}

\vspace{0.5em}

{$^a$ Fit parameters: S=spin; A=astrometry; B=binary; DM=dispersion measure;
FD=frequency dependence; J=jump}

\vspace{0.5em}

{$^b$ Weighted root-mean-square of epoch-averaged post-fit timing residuals,
calculated using the procedure described in Appendix D of \nineyr.
For sources with red noise, the ``Full'' RMS value includes the red noise
contribution, while the ``White'' RMS does not.}

\vspace{0.5em}

{$^c$ Red noise parameters: $A_{\mathrm{red}}$ = amplitude of red noise
spectrum at $f$=1~yr$^{-1}$ measured in $\mu$s yr$^{1/2}$;
$\gamma_{\mathrm{red}}$ = spectral index; $B$ = Bayes factor (``$>$2'' indicates a Bayes factor larger than our threshold log$_{10}$B~$>$~2, but which could not be estimated using the Savage-Dickey ratio).  See
Eqn.~\ref{eqn:rn_spec} and Appendix~C of \nineyr\ for details.}

}

 \end{table*}

The cleaned TOA data set for each pulsar was fit with a physical
timing model, with the \tempo\ timing software used for the primary
analysis.  The timing models were checked using the
\tempotwo\footnote{\url{https://bitbucket.org/psrsoft/tempo2}} and
\pint\footnote{See \url{https://github.com/nanograv/PINT} and \citet{pint}.} packages.  \pint\ \citep{pint,Luo21} was
developed independently of \tempo\ and \tempotwo\ and thus provides a
particularly robust independent check of the timing models
(Section~\ref{sec:pint}).  Our expectation is to transition to
\pint\ as the primary timing software for future data sets due to its
modularity and its use of modern programming tools, including coding
in Python.

An overarching development in the current release is our use of
standardized and automated timing procedures. In previous NANOGrav
data releases, two core portions of the data analysis were already
automated: data reduction (calibration, RFI removal, time- and
frequency-averaging) and TOA generation, all of which was done using
\texttt{nanopipe}\footnote{\url{https://github.com/demorest/nanopipe}}
\citep{nanopipe}; and checking for timing parameter significance
(e.g., as in \elevenyr). For the 12.5-year data set, we standardized
and automated the timing procedure using Jupyter notebooks.
These notebooks did not replace the often iterative nature of pulsar timing. Rather, once a reasonable timing solution was found for a pulsar, it was input into the notebook, which ran through an entire standard analysis that included checking for parameter significance (Section~\ref{subsec:timing:models_and_parameters}) and performing noise modeling (Section~\ref{subsec:noise}). This process allowed for systematic and transparent addition or removal of timing and noise parameters, and ensured that the final timing models were assembled in a standardized way. Additional benefits of automating the timing analysis in this way are that it makes NANOGrav timing analysis more accessible to new students or researchers, and much of the automated process can also be applied to other (non-PTA) pulsar timing work.

\subsection{Timing Models and Parameters}
\label{subsec:timing:models_and_parameters}

Timing fits were done using
the JPL DE436 solar system ephemeris and the TT(BIPM2017)
timescale. 
As in previous releases, we used a standard procedure to
determine which parameters are included in each pulsar's timing
model. Always included as free parameters were the intrinsic spin and
spin-down rate, and five astrometric parameters (two position
parameters, two proper motion parameters, and parallax), regardless of
measurement significance. We used ecliptic coordinates for all
astrometric parameters to minimize parameter covariances.  For binary
pulsars, five Keplerian parameters were also always fit: (i) the orbital
period, ($P_b$) or orbital frequency ($F_b$); (ii) the projected semi-major axis ($x$); and (iii-v)
either the eccentricity ($e$), longitude of periastron ($\omega$), and epoch of
periastron passage ($T_0$); or two Laplace-Lagrange parameters
($\epsilon_1$, $\epsilon_2$) and the epoch of the ascending node
($T_{\mathrm{asc}}$).

The particular binary model chosen was based on
orbital characteristics, including the presence of post-Keplerian parameters. For
low-eccentricity orbits, we used the ELL1 model,
which approximates the orbit using the Laplace-Lagrange parameterization of the
eccentricity with $\epsilon_1$ and $\epsilon_2$
\citep{Lange2001}.  In all cases in which we used ELL1,
the model deviated from a more precise timing model by at most 25~ns 
at any point in the orbit.
The pulsars in this data set that satisfy this criterion for the use of the ELL1 model all have $e<10^{-5}$, although we did not apply an explicit eccentricity criterion for this binary model.
If Shapiro delay was marginally present in a
low-eccentricity system, we used ELL1H, which incorporates the
orthometric parameterization of the Shapiro delay \citep{Freire2010}
into the ELL1 model; note that the ELL1H model employs the $h_3$ and
$h_4$ parameters, as opposed to $h_3$ and $\varsigma=h_4/h_3$ of the
DDFWHE model \citep{Freire2010, Weisberg2016}, which is for
high-eccentricity systems and is not used in any timing models in this
data set. For pulsars with higher eccentricity, we used the DD
binary model \citep{Damour85, Damour86, Damour1992}; and for
PSR~J1713+0747, we used DDK \citep{Kopeikin1995, Kopeikin1996}, which
allows us to measure annual-orbital parallax.
For PSR~J1713+0747, a \tempotwo-compatible timing model that uses
the T2 binary model instead of DDK is also included in the data
release. For some short-period binaries ($P_b \lesssim 0.5$\,d), we used orbital frequency and one or more orbital frequency
derivatives, rather than period and period derivative, to better
describe the orbit and allow for simple testing of additional
orbital frequency derivatives.

We determined parameter significance via an
$F$-test, with the requirement that $p < 0.0027$ ($\sim$~3$\sigma$)
for a parameter to be included in the timing model. This requirement
does not apply to the five astrometric and five Keplerian binary
parameters that are always included in the fit (for very
low-eccentricity binaries, the eccentricity parameters $\epsilon_1$
and $\epsilon_2$ may not be measured at a significant level for many
years). We specifically tested for the significance of additional
frequency-dependent pulse shape or evolution parameters (``FD''
parameters; see \nineyr). We allowed FD1 through FD5 to be fit, and require that all FD parameters up to the highest-order significant FD parameter be included in the fit, even if the lower-order parameters are not found to be significant. For example, if FD4 is significant but FD3 is not, then FD3 would still be included in the timing model.

For binary pulsars, we tested for the
secular evolution of binary parameters (e.g., $\dot{x}$,
$\dot{\omega}$, or $\dot{P_b}$), higher-order orbital frequency
derivatives if using orbital frequency rather than period, and Shapiro
delay parameters. For binaries modeled by ELL1 without
previously-measured Shapiro delay parameters, we converted the binary
model to ELL1H and tested the significance of $h_3$ and $h_4$.
If both $h_3$ and $h_4$ were significant, it
raised the possibility of measuring the traditional Shapiro delay
parameters (orbital inclination $i$ and companion mass $m_{\mathrm
  c}$) directly from the timing model fit. Thus, for pulsars with
significant detections of $h_3$ and $h_4$, we also tested the use of
the traditional Shapiro delay parameters with the ELL1 model: if $i$
and $m_{\mathrm c}$ converged to physically meaningful and
significantly-measured values, and if the use of these parameters
significantly improved the fit according to a $\Delta\chi^2$ test,
then we included $i$ and $m_{\mathrm c}$ in the timing model; 
otherwise, we continued to use $h_3$ and $h_4$. Compared with \elevenyr,
these significance tests resulted in the inclusion of one or more new
binary parameters for 19 pulsars, and the exclusion of
previously-included parameters for 3 pulsars
(Section~\ref{sec:newsigpars}).

Constant phase ``jumps'' were included as fit parameters to account for unknown offsets between data subsets collected with different receivers and/or telescopes. For data subsets
collected with the same receiver and telescope but different back end instruments, the measured offsets between GASP and GUPPI, and ASP and PUPPI, from \nineyr\ are included in the TOA data set (with flag ``\texttt{-to}'' on the TOA lines) rather than in the timing model. 

We included white and red noise models as described in Section~\ref{subsec:noise}.  We derived best-fit timing model parameter values using a generalized-least-squares fit that uses the noise-model covariance. It is important that the noise model be included when testing for parameter significance, especially if a pulsar shows significant red noise; for several pulsars, one or more parameters were found to be significant when \tempo\ was run without generalized-least-squares fitting, but were no longer significant when the noise model was included. Thus, the $F$-test significance tests described above were always performed with generalized-least-squares fitting.

A summary of TOA statistics, basic timing parameters, noise parameters, and other statistics are provided in Tables~\ref{tab:timingpars} and \ref{tab:rms_rednoise}.

\subsection{Dispersion Measure Variations}
\label{subsec:dm_variation_msmts}

Variations in dispersion measure are caused by the relative motion of
the Earth-pulsar sightline through the ionized interstellar medium (IISM)
as well as the Earth's motion through the ionized solar wind, and lead to
variations in pulse arrival times.
It is therefore necessary to include short-timescale DM variations in the timing model \citep{Jones2017}.

We used the piecewise-constant model called \texttt{DMX} in both
\tempo\ and \pint\ to measure the short-timescale DM variations in our
data set.  All Arecibo data were grouped into DMX windows of 0.5\,d,
because observations of any given pulsar normally use two receivers 
back-to-back.  For the GBT, observations with separate
receivers are made on different days; we grouped GASP data into
15\,d time ranges, and we grouped GUPPI data into 6.5\,d time ranges
in order to include data from multiple receivers in most DMX windows.
We imposed a minimal frequency range criteria for each DMX window;
this is described in section~\ref{sec:baddmxcut}.

If within these DMX time ranges we found that the expected solar wind
contribution to the epoch-specific DM induced a timing variation of
more than $100$\,ns, those time ranges were further divided into
0.5\,d windows (thus effectively measuring the DMX for a single
observing day). We used a toy model as in \elevenyr\ to estimate the
expected solar wind-induced time delays: the solar wind electron
density is modeled as $n_e = n_0(r/r_0)^{-2}$ (where $r$ is the
distance from the Sun and $n_0$ is the electron density at $r_0 =
1$\,AU), and use a representative value of $n_0 =
  5$\,cm$^{-3}$ \citep[e.g.,][]{Splaver2005}.  \citep[A similar value
    of 7.9 cm$^{-3}$ was found by][with NG11 data.]{Madison2019}
Additionally, for PSR~J1713+0747, it was necessary to break up the DMX
time range surrounding its second chromatic event \citep{Lam2018}: the
DM changes so rapidly that using only a single DMX value over the full
length of the event introduces significant noise into the data
set. The original DMX time range spanned MJD~57508.36--57512.3; we
divided this time range into two ranges, spanning
MJD~57508.36--57510.36 and 57510.36--57512.3.

\subsection{Noise Modeling}
\label{subsec:noise}

The noise model used in this analysis is nearly identical to that of \nineyr\ and \elevenyr.  The primary difference is that in this work we used the new PTA analysis software \enterprise\footnote{\url{https://github.com/nanograv/enterprise}} \citep{enterprise}. In all cases, the final noise model assumes Gaussian noise after all outlier TOAs and otherwise corrupted TOA data have been removed from the data set.

Noise in the timing residuals is modeled as additive Gaussian noise with three white-noise components and, if significantly detected, one red-noise component.
For convenience, here we provide a qualitative description of the noise model; for more details, we refer the reader to \nineyr\ and \elevenyr.
The four noise components are:

\begin{enumerate}

\item EFAC, $E_k$: Measured TOA uncertainties $\sigma_i$ may be underestimated.
A separate EFAC parameter, $E_k$, is therefore used for each combination of pulsar, backend, and receiver, indexed by $k$, to account for any systematics in TOA measurement uncertainties; hence $\sigma_i$ becomes $\sigma_{i,k}$.  For the majority of NANOGrav pulsars, $E_k \sim 1$, suggesting that our observing and analysis procedures are resulting in near-true TOA uncertainty estimates.

\item EQUAD, $Q_k$: Within \enterprise, the EQUAD term is added in quadrature to the EFAC-scaled TOA uncertainty, i.e., $\sigma^\prime_{i,k} = (E_k^2\sigma_{i,k}^2+Q_k^2)^{1/2}$. This term accounts for any uncorrelated systematic white noise that is present in addition to the statistical uncertainties in the TOA calculations. As with EFAC, we use a separate EQUAD parameter, $Q_k$, for each combination of pulsar, backend, and receiver, indexed by $k$. \tempo\ uses a different white noise formulation such that the maximum likelihood EQUAD values contained within the timing models in this data release were obtained via the conversion $Q_{k,\textrm{\tempo}} = Q_{k} / E_{k}$.

\item ECORR: This parameter describes a short-timescale noise process that has no correlation between observing epochs, but is completely correlated between TOAs that were obtained simultaneously at different observing frequencies (see Appendix C of \nineyr\ for details). Wideband noise processes such as pulse jitter \citep{Lam2016_mar16,Shannon14,Oslowski11} are accounted for by ECORR.

\item Red noise: Any steep-spectrum noise components are modeled as a single stationary Gaussian process, whose spectrum we parameterize by a power-law,
  \begin{equation}
  \label{eqn:rn_spec}
  P(f) = A^2_{\mathrm{red}} \left(\frac{f}{1\,\textrm{yr}^{-1}}\right)^{\gamma_{\mathrm{red}}},
  \end{equation}

where $f$ is a given Fourier frequency in the power spectrum and $A_{\mathrm{red}}$ is the amplitude of the red noise at reference frequency 1\,yr$^{-1}$.

\end{enumerate}

For each pulsar, we incorporated all noise components and timing model parameters into a joint likelihood using \enterprise\ and sampled the posterior distribution using the sampler PTMCMC \citep{ptmcmc}.  The red noise prior distribution was log-uniform, while all other prior distributions were uniform.
Since the model without red noise is nested within the general model (corresponding to a red noise amplitude of zero), we used the Savage-Dickey ratio to estimate the Bayes factor favoring the presence of red noise \citep{dickey1971}.
For pulsars with red noise Bayes factor $B$ above a threshold of $B > 100$, we included the red noise parameters in the final timing models; for the rest of the pulsars, we re-ran their analyses without red noise. This exercise typically did not affect the detectability of other parameters, but in a small number of cases the presence or absence of red noise did affect marginal timing parameters like $\dot{x}$, $\epsilon_1$, or $\epsilon_2$ (see the discussion of the PSR\,J1909$-$3744 binary model in section~\ref{sec:binary}).

The Savage-Dickey ratio fails to estimate a finite Bayes factor for heavily preferred red noise models with a finite-length chain of samples.  Indeed, for all of the pulsars with above-threshold red noise, $B$ was large enough that it was not robustly
estimated.  As such, we simply report the log$_{10}B$ for those pulsars in Table~\ref{tab:timingpars} as $>$2. Further details on the red noise characterization are provided in \nineyr\ and \elevenyr, and Appendix~C of \nineyr\ provides a complete description of the Bayesian inference model.

Fourteen pulsars were found to have red noise with $B > 100$; this includes all eleven sources with detected red noise in \elevenyr.  Of the three MSPs with newly detected red noise (J1744$-$1134, J1853$+$1303, and J2317$+$1439), two of them are among the longest-observed pulsars in the data set.  

The Bayes factors for the pulsars in which red noise
was not detected are listed in Table~\ref{tab:rms_rednoise}. Several pulsars have sub-threshold Bayes factors sufficiently larger than 1, such that we may expect those pulsars to display red noise above our defined threshold with several more years of data. A more detailed noise analysis of each pulsar is beyond the scope of this work, but will be performed as part of the gravitational wave analyses for the data set in a forthcoming work.
\begin{table*}[th!]
\centering
\caption{Noise Parameter Comparison for 11-Year Data Subset\label{tab:noise_ng11yr}
  }
{\footnotesize\begin{tabular}{cccc}
\hline
\hline
Noise Parameter 	& \# Decreased$^a$ 	&   \# Increased$^a$ 	&  Mean Difference$^b$  \\
\hline
EFAC 			& $109$ 			&  $30$ 			& $-0.044$ \\
EQUAD 			& $60$ 			&  $79$ 			& $-0.013\;\mu$s \\
ECORR 			& $79$ 			&  $60$ 			& $-0.017\;\mu$s \\ [0.5ex]
\hline
\end{tabular}

\vspace{0.5em}

{$^a$The number of noise parameters whose values decreased or increased in the 11-year ``slice'' of the 12.5-year data set, compared with their values from the 11-year data set (Section~\ref{subsec:noise_improvements}).}

{$^b$Weighted mean difference in noise parameters. These are computed as the difference between the 11-year data set values and the 11-year ``slice'' values, weighted by the errors on the 11-year data set values.}
}
 \end{table*}

\subsection{Improved Noise Parameters over the 11-year Data Set}
\label{subsec:noise_improvements}

Our data reduction methods (Section~\ref{subsec:data_reduction}) and data cleaning methods (Section~\ref{subsec:cleaningdata}) are designed to minimize non-astrophysical noise sources in the data set.  Minimizing these noise sources is important because both white noise and red noise have important consequences for the detection of nanohertz gravitational waves \citep{Hazboun20,Lam18b,Lam2016_mar16,Siemens2013}.  

Here we test the methods used in the present work with our previous-generation \elevenyr\ data set to see whether the refined methods result in a reduction of noise.  To make this comparison with \elevenyr, we use a subset of the present data set that corresponds to the pulsars and date range of \elevenyr, i.e., a data set equivalent to ``generating a NG11 data set with procedures of the present work.''  A full noise analysis was done on this data subset, consistent with the analyses done in \elevenyr. This sliced analysis has the advantage of using the same timespan of data, which avoids biases from searching for different frequencies of a steep spectral-index red process. Using the same timespan also keeps the number of TOAs in a receiver+backend combination similar so there is no bias in determining white noise parameters with largely differing numbers of TOAs. While a refit to the timing model is out of the scope of such a comparison, the data sets are otherwise similar except for the various data pipeline improvements referenced above. For the pulsars considered for this analysis, there were $139$ pulsar-receiver-backend combinations analyzed, and the changes in their white noise parameters are shown in Table~\ref{tab:noise_ng11yr}. The most dramatic change is seen in the EFAC parameters, where $109$ parameters had EFAC values smaller than in \elevenyr, with a mean difference of $-0.044$. Both the mean EQUAD and ECORR also decrease, but by smaller amounts. The changes in red noise are more subtle, with some pulsars showing mildly increased support for steep spectral index noise processes. These will be discussed in detail in the forthcoming paper presenting our results from GW analyses.

\section{Comparison of \tempo\ and \pint\ Timing Models}
\label{sec:pint}
The NANOGrav 12.5-year data analysis results are cross-checked by the new timing package \pint~(version 0.5.7), which has a completely independent code base from \tempo\ and \tempotwo.
In this section, we present a comparison between the \pint\ and \tempo\ timing results for the narrowband TOA data set.
This comparison focuses on the discrepancies of the post-fit parameter values.
We used timing models produced by \tempo\ as the initial input models for \pint, and refit the TOAs with \pint's general least-squares fitter.
Then, the best-fit parameters from these two packages were compared against each other.
The noise parameters obtained by the \enterprise\ analysis (Section~\ref{subsec:noise}) were not altered by \pint, thus we do not compare the noise parameters\footnote{We ran \enterprise\ using both \tempo\ and \pint\ on a small subset of pulsars from this data release in order to obtain posterior distributions of the noise parameters independently using both timing packages. Using a K-S test to compare the resulting distributions for a given pulsar, we found that all noise parameter posterior distributions were statistically consistent between the \pint- and \tempo-mode runs of \enterprise.}.
To describe the changes of parameter values with an intuitive, standardized quantity, we divide the parameter value differences by the \tempo\ uncertainties.
We have compared 5,417 parameters in total; 3,929 best-fit parameter values from \pint\ deviate from the \tempo\ values by less than 5\% of their \tempo\ uncertainties. Among the rest, 1,442 parameters' \pint\ results changed by less than 50\% of their \tempo\ uncertainties (for example, changes of the order $10^{-14}$--$10^{-10}$\,Hz for spin frequencies, or $10^{-11}$--$10^{-3}$\,d for orbital periods),
and 46 parameters' discrepancies are more significant than 50\%.

A majority of the outstanding discrepancies ($>$50\% of \tempo\ uncertainty) can be explained by different implementations in the two software packages.
Thirty-two of these outlier parameters belong to the two pulsars that have the largest amplitude of red noise, J1643$-$1224 and J1903$+$0327.
\tempo\ effectively uses a lower cutoff frequency in the spectrum describing red noise, which will make the uncertainties on the spin frequency and its derivative larger.
This implementation difference also causes the discrepancies seen in the spin frequency and spin frequency derivative parameters and uncertainties for other pulsars with red noise. For instance, PSR J2317$+$1439's spin frequency has the largest difference, 61\% of its \tempo\ uncertainty (corresponding to a change of $\approx 2\times10^{-11}$\,Hz);
it also has the steepest red noise index in the data set, $\gamma_{\textrm{red}} = -6.5$.
Another implementation difference is that \pint\ uses a different definition for the longitude of ascending node (``KOM'' parameter) in the DDK binary model.
Thus, ten parameters from PSR J1713$+$0747, which uses the DDK binary model, showed discrepancies greater than 50\% of their original \tempo\ uncertainty.
Other known implementation differences, which can induce small systematic offsets or differences on the order of $\sim$~10~ns, are summarized in \citep{Luo21}.
However, the reasons for the remaining three parameters with larger differences---$h_4$ of PSR J1853+1303, $T_{\textrm{asc}}$ of PSR J1918-0642, and the ecliptic longitude $\lambda$ of PSR J1640+2224---are still under investigation.
 
\section{Newly Measured Timing Parameters in the NANOGrav Data Set}
\label{sec:newsigpars}

\begin{table*}[th!]
\centering
\caption{New NANOGrav 12.5-year Parallax Measurements\label{tab:parallax}}
{\footnotesize\begin{tabular}{ccccccc}
\hline
\hline
PSR & Parallax & Previous Measurement &  Technique &  Reference \\
& (mas) & (mas) \\
\hline
J0636+5128   & $1.37\pm0.23$ &  $4.9\pm0.6$   &  Timing & \citet{Stovall2014} \\
J1012+5307   & $1.13\pm0.35$ & $1.21^{+0.03}_{-0.08}$ & VLBI & \citet{Ding2020} \\
J1832$-$0836 & $0.48\pm0.13$ & $\cdots$ & $\cdots$ & $\cdots$ \\
J1853+1303   & $0.48\pm0.14$ &  $1.0\pm0.6$   & Timing & \citet{Gonzalez2011} \\
B1937+21 & $0.28\pm0.05$ & $0.40\pm0.16^a$ & Timing & \citet{Reardon2016} \\
J2010$-$1323 & $0.41\pm0.12$ &  $0.48^{+0.17}_{-0.12}$ & VLBI (VLBA) & \citet{Deller2019} \\ [0.5ex]
J2322+2057 & $0.98\pm0.26$ & $<4.8$ & Timing & \citet{NiceTaylor95} \\ [0.5ex]
\hline
\end{tabular}

\vspace{0.5em}

{$^a$For PSR~B1937+21, we quote the timing parallax measurement from \citep{Reardon2016} that was corrected for the Lutz-Kelker bias \citep{Verbiest2012}.}

}
 \end{table*}

Comparing the present data set with \elevenyr, we find a number of astrometric and binary timing parameters that were not previously measured at a significant level (as defined by the $F$-test in \S\ref{sec:timing}) with NANOGrav data. Here we highlight those parameters and compare with any previously-published values from other teams.

\subsection{Newly Significant Astrometric Parameters}
\label{subsec:newastrometricparams}

The NANOGrav 12.5-year data release includes seven new measurements of annual trigonometric parallax compared to \elevenyr\ (Table~\ref{tab:parallax}), although in contrast to \elevenyr, the parallax measurements for PSRs~J0740+6620 and J2234+0944 are no longer significant \citep[but see][for the former]{cfr+20}. In addition, both components of proper motion are newly measured for
PSR~J1747$-$4036, along with one component for PSRs~J0023+0923, B1937+21, and
J2017+0603. (For the latter three MSPs, the other proper motion component was already
measured in previous data sets.)

In Table~\ref{tab:parallax}, we compare the new NANOGrav parallax values with prior parallax measurements for the same objects. The previous parallaxes for PSRs~J1012+5307, J1853+1303, and J2010$-$1323 are consistent with our measurements, while spanning the gamut of measurement techniques (timing, VLBI, and optical companion parallax from {\em Gaia}).
For PSR~J0636+5128, the NANOGrav data spans $\sim3.5$~yr as opposed to only $\sim1.5$~yr available to \citet{Stovall2014}.
The parallax for PSR~B1937+21 published in the first Parkes Pulsar Timing Array (PPTA) data release \citep[DR1;][]{Reardon2016} is consistent with that presented here.

\subsection{Newly Significant Binary Parameters}
\label{subsec:newbinaryparams}

Several orbital parameters not detected in prior NANOGrav data releases have been measured in the 12.5-year data set. Of particular interest in our data set are new measurements of the secular evolution of projected orbital semimajor axis ($\dot{x}$) and of the orthometric parameters ($h_3$, and $h_4$ or $\varsigma=h_4/h_3$) that parameterize the Shapiro delay \citep{Freire2010}. We measure $\dot{x}$ for four additional pulsars relative to \elevenyr: PSRs~J0613$-$0200, B1953+29, J2145$-$0750, and J2229+2643. We now measure both $h_3$ and $h_4$ in the timing model of PSR~J1853+1303, for which only $h_3$ was measured with significance in \elevenyr.
We measure the first indication of Shapiro delay in PSR~J2145$-$0750 with a measurement of $h_3$. Additionally, for the newly-added pulsar PSR~J1946+3417, we measure $\dot{\omega}$ and Shapiro delay parameters that are consistent with those reported by \citep{Barr2017}.

Checking the literature and the Australia Telescope National Facility (ATNF) pulsar catalog\footnote{\url{https://www.atnf.csiro.au/research/pulsar/psrcat/}} \citep[][version 1.63]{Manchester2005}, we find no previously-measured values of $\dot{x}$ for PSRs~J0613$-$0200 or J2229+2643. For PSR~J2145$-$0750, our measurement of $\dot{x}=(5.8\pm1.0)\times 10^{-15}$\,lt-s\,s$^{-1}$ is consistent at the $\sim2\sigma$ level with that of \citet{Reardon2016}.
Our measurement of $h_3$ for PSR~J2145$-$0750 is the first indication of Shapiro delay for this pulsar, and aids in constraining the companion mass (Section~\ref{sec:binary}). The timing model for PSRs~J1853+1303 in the ATNF pulsar catalog contains $h_3$ and $h_4$, with values consistent with those found in this work and previously in \elevenyr\ (for $h_3$).

In addition to secular and Shapiro delay parameters, we also measure one new Laplace-Lagrange eccentricity component ($\epsilon_1$ or $\epsilon_2$) in PSRs~J0023+0923, J1738+0333, and J2214+3000 with $\gtrsim3\sigma$ significance. We also find that although $\dot{P_b}$ for PSR~J0636+5128 and $\dot{\omega}$ for J1600$-$3053 were detected at a significant level in \elevenyr, they are not measured significantly in the present data set, so are no longer included in the timing models for these pulsars.

\section{Consistency of Astrometric Parameters across Data Sets}
\label{sec:astrometry}
\begin{figure*}
\begin{center}
    \includegraphics[scale=0.65]{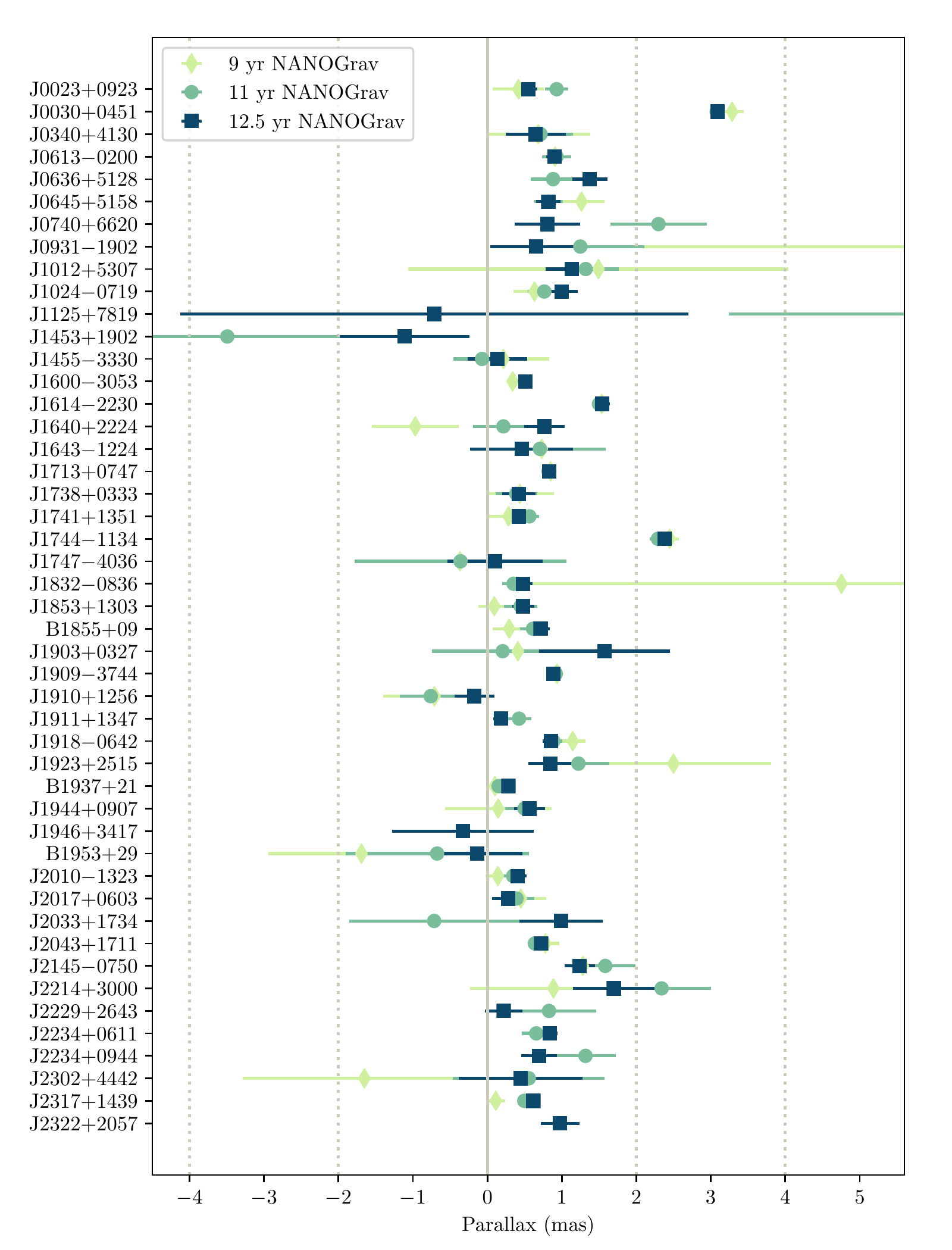}
\caption{\label{fig:astrometry:px}
  Parallax measurements and formal uncertainties for all 12.5-year pulsars from \nineyr, \elevenyr, and the current data set. While only values of parallax greater than zero are physically meaningful, all formally fit values are shown here; the preponderance of positive values serves to verify that a real physical parameter is being measured.
Two outlier values from previous data releases fall beyond the right edge of the plot, as indicated by the error bars.
   }
\end{center}
\end{figure*}
 
\begin{figure*}
\begin{center}
\includegraphics[width=0.8\textwidth]{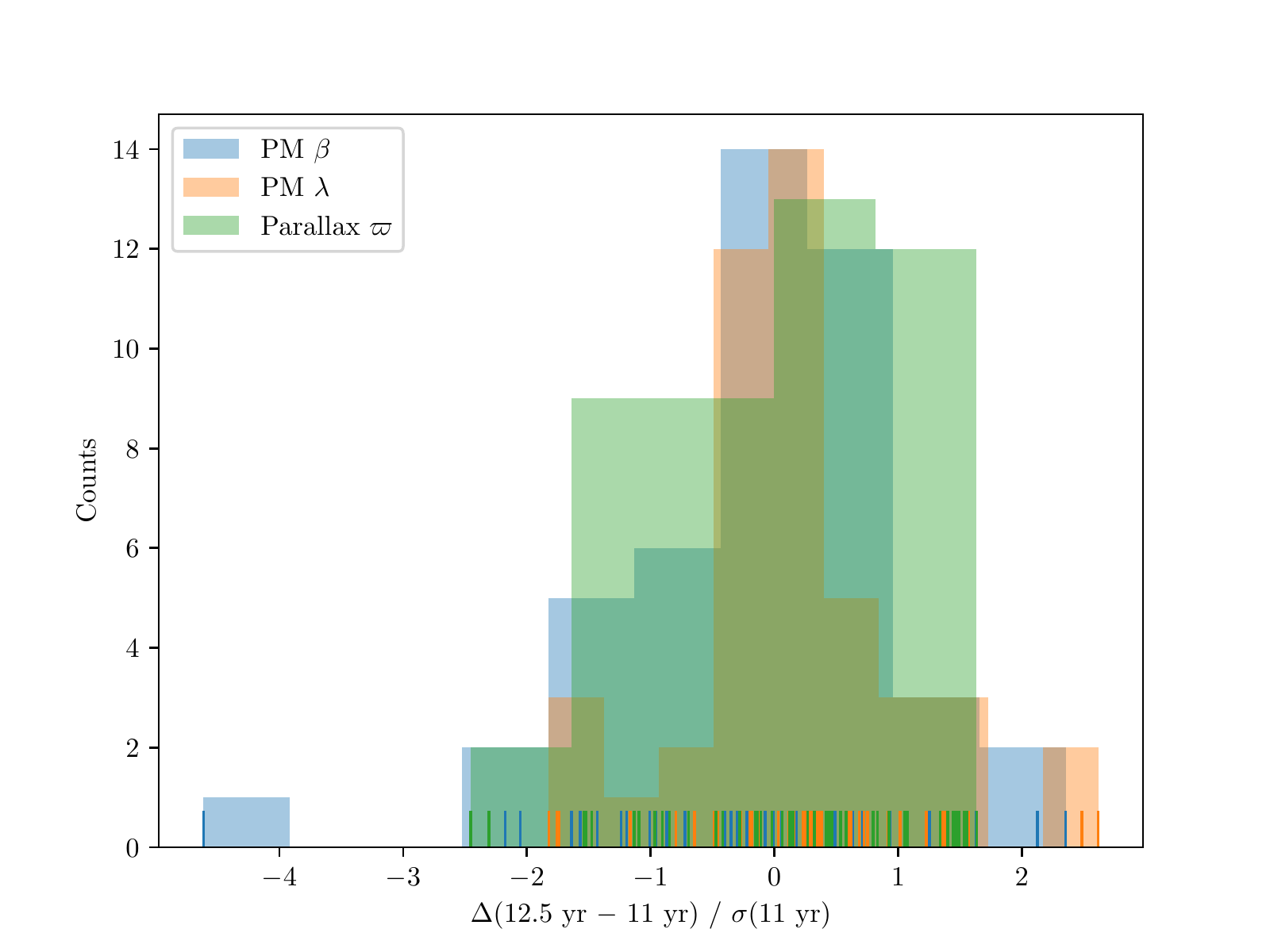}
\caption{\label{fig:astrometryhist}
Comparison of astrometric measurements across \elevenyr\ and the current 12.5-year data set. The differences in proper motion ($\mu_\beta,\; \mu_\lambda$) and parallax $\varpi$ are shown in units of the uncertainty in the 11-year measurement ($\sigma_{11}$). The figure shows binned histograms of each type of measurement, with all individual measurements superimposed as short vertical lines at the bottom of the figure.
The value of $\mu_\beta$ for PSR~J2214+3000 is an outlier at $-4.6\,\sigma$; the rest are reasonably consistent, as discussed in the text.}
\end{center}
\end{figure*}
 
As noted previously, our pulsar timing analyses always include five astrometric parameters as free parameters: two sky position parameters, two components of proper motion, and annual trigonometric parallax. Detailed analyses of the astrometry of NANOGrav pulsars, including comparisons with VLBI measurements, were presented in \citet{Matthews2016} and \elevenyr, thus we do not repeat such a detailed analysis in the present work. 
Comparisons of astrometric measurements obtained via different measurement methods (e.g., table~\ref{tab:parallax}) are potentially useful for such purposes as tying astrometric reference frames \citep{Wang2017}, using measurements made by one method as priors in analysis of other data, etc.  To make use of pulsar astrometric measurements, it is important that they be robust, accurate, and stable over time.  To test the stability of our astrometric measurements, we compare the parallax and proper motion measurements between the current and previous NANOGrav data releases (with newly-detected astrometric parameters specifically highlighted in the previous section).

For pulsar timing, the position (and hence proper motion) is naturally parameterized in terms of ecliptic coordinates. As the timing data span increases, the proper motion is expected to be measured with increasing accuracy, and the covariance between proper motion and parallax should rapidly decrease. Figure~\ref{fig:astrometry:px} shows the measured parallaxes in the 12.5-year data release, as well as previous (\nineyr\ and \elevenyr) measurements, where available. The number of measurements has increased (see Section~\ref{subsec:newastrometricparams} below) and the formal significance has generally improved.

However, a comparison of the changes in astrometric parameters between the current and previous (11-year) data releases might suggest that some caution is warranted. In Figure~\ref{fig:astrometryhist} we show histograms of the differences in proper motion ($\mu_\beta$ and $\mu_\lambda$) and parallax ($\varpi$) between the current (12.5-year) and \elevenyr\ data releases, with the differences scaled by the estimated uncertainty for the 11-year parameters (i.e., $(P_{12.5} - P_{11})/\sigma_{11}$, where $P = \{\mu_\beta, \mu_\lambda, \varpi\}$).

The single obvious outlier is the measurement of $\mu_\beta$ for PSR~J2214$+$3000, at $-4.6\sigma_{11}$. For pulsars near the ecliptic plane, the ecliptic latitude $\beta$ is poorly constrained by timing in comparison to the ecliptic longitude $\lambda$, and we expect the accuracy of $\mu_\beta$ measurements to be correspondingly worse. That alone cannot explain the discrepancy for J2214$+$3000, at $\beta = 37.7\arcdeg$.  The other notable fact about this source is that it is one of four black widow pulsars in our data set, along with J0023$+$0923, J0636$+$5128, and J2234$+$0944.  Like the other black widows in our sample, it does not exhibit eclipses \citep{Ransom11}, and, as with J2234$+$0944, it does not show orbital variability \citep{BakNielsen20,Arzoumanian2018a}.  
In \elevenyr, we noted difficulty fitting a noise model to this pulsar, possibly related to excess noise in mid-to-late 2013.  Imperfect noise modeling, combined with covariance between noise parameters and astrometric parameters, may contribute to the change in reported $\mu_\beta$ value.

Besides J2214$+$3000, the astrometric parameters appear generally consistent between the 11-year and 12.5-year measurements, with $\sim$85\% of the measurements differing by less than $\pm 1.5\,\sigma_{11}$.  Even though \elevenyr\ is a subset of the 12.5-year data release, such measurement differences are not unreasonable.  Due to the additional processing of the 12.5-year data, as described in Section~\ref{subsec:cleaningdata}, in combination with a longer baseline that further down-weights the earlier, less constraining data, such changes in the astrometric parameters can be expected.

\section{Binary Analysis of Selected Pulsars}
\label{sec:binary}

In \elevenyr, we presented a summary of modeling methods and results
for binary pulsars in the data set presented therein; we do not repeat
such detailed descriptions here. Instead, we highlight five binary
pulsars for which additional description or analysis is
warranted. PSR~J0740+6620 is an extremely high-mass MSP for which a
more up-to-date timing solution is published in \citet{cfr+20}; for
PSRs~J1909$-$3744 and J2234+0611, significant testing was required to
obtain the timing models presented in this work; and we use the
newly-measured Shapiro delay parameters of PSRs~J1853+1303 and
J2145$-$0750 to place mass and geometry constraints on these systems.

\subsection{PSR~J0740+6620}
\label{subsec:J0740}

In the course of analyzing the 12.5-yr data set, we found that the significance of the Shapiro delay in PSR~J0740+6620 had dramatically increased from its initial detection in \elevenyr. The constraints on $m_{\rm c}$, $\sin i$, and the pulsar mass ($m_{\rm p}$) from the nominal 12.5-yr data set motivated additional, targeted observations for improving the Shapiro-delay measurement. By combining 12.5-yr NANOGrav timing data with additional data obtained during specific orbital phases optimally sensitive to Shapiro delay, \cite{cfr+20} found an improved pulsar mass of $m_{\rm p} = 2.14^{+0.10}_{-0.09}\textrm{ M}_\odot$ (68.3\% credible region), representing the most massive, precisely-measured neutron star known to date.

\subsection{PSR~J1853+1303}
\label{subsec:J1853}

Both $h_3$ and $h_4$ of the orthometric parameterization of the Shapiro delay \citep{Freire2010} for PSR~J1853+1303 are significant.  Our new measurements ($h_3=0.18\pm0.04\,\mu$s, $h_4=0.17\pm0.05\,\mu$s) are consistent with those first presented in \elevenyr. We tested whether $\sin{i}$ and $m_c$ could be independently measured using \tempo, as described in Section~\ref{subsec:timing:models_and_parameters}, but this test was not successful.

The orthometric parameters $h_3$, $h_4$, and $\varsigma$ are related to the traditional post-Keplerian Shapiro delay parameters as:
\begin{align}
\varsigma & = \sqrt{\frac{1-\cos{i}}{1+\cos{i}}} \\
h_3 & = r\varsigma^3 \\
h_4 & = h_3\varsigma 
\end{align}
\noindent where $r = m_{\mathrm c}T_{\odot}$ is the ``range'' of the delay, $m_{\mathrm c}$ is the companion mass, and $T_{\odot} \simeq 4.93\,\mu$s. This parameterization constrains the orbital inclination to $i=85^{\circ}\pm14^{\circ}$ (we quote the $1\sigma$ uncertainty, derived from error propagation beginning with the $1\sigma$ parameter uncertainties from the \tempo\ timing model). The orthometric parameters are not yet sufficiently well-measured to place meaningful bounds on $r$ and, therefore, the companion mass, which we calculate to be consistent with zero (at the $1\sigma$ confidence level). More insight into the physical properties of this system may be gained by explicit gridding of the posterior distribution in future work.

\subsection{PSR~J1909$-$3744}
\label{subsec:J1909}

The value of $\dot{x}$ for PSR~J1909$-$3744 has been measured or constrained by several groups. \citet{Verbiest2009} found $\dot{x} = (5\pm4)\times 10^{-16}$\,lt-s\,s$^{-1}$, while \citet{Desvignes2016} found $\dot{x} = (0.6\pm1.7) \times 10^{-16}$\,lt-s\,s$^{-1}$. In \elevenyr, we found $\dot{x} = (-4\pm1) \times 10^{-16}$\,lt-s\,s$^{-1}$. Using the 12.5-yr data set for PSR~J1909$-$3744, we have further constrained its value to $(-2.9\pm0.8) \times 10^{-16}$\,lt-s\,s$^{-1}$.

As in \nineyr{} and \elevenyr, we detect red noise in PSR~J1909$-$3744. In this work, we find $\dot{x}$ and the red-noise terms in the timing model to be covariant.
In particular, the presence or absence of $\dot{x}$ in the timing model had a significant effect on the red noise amplitude: the amplitude was significantly lower when $\dot{x}$ was included in the model, compared to when $\dot{x}$ was not included. Additionally, if $\dot{x}$ was initially excluded from the timing model such that the red noise amplitude assumed its higher value, then adding $\dot{x}$ to the model while red noise was also included resulted in a non-measurement of $\dot{x}$ according to our $F$-test criterion. While not common, this covariant behavior is not unexpected, as $\dot{x}$ is a secular parameter that evolves slowly, as does red noise.

The value of $\dot{x}$ can be inferred from the changing geometry due to the relative motion between the pulsar system and Earth \citep{Kopeikin1995},
\begin{equation}
\dot{x}_{\mathrm{k}} = x \mu \cot{i} \sin{(\Theta_{\mu} - \Omega)}
\label{eqn:xdot_pm}
\end{equation}

\noindent where $\sin i$ is constrained from the Shapiro delay, giving $i = 86.39^{\circ}$ or $93.61^{\circ}$; the magnitude of proper motion $\mu = 36.94$\,mas\,yr$^{-1}$; the position angle of the proper motion ($\Theta_{\mu}$) is derived from timing measurements of proper motion; and $\Omega$ is the longitude of the ascending node. In the case where annual orbital parallax is detected, the three-dimensional geometry of the system can be constrained, such that the values of $i$ and $\Omega$ are measured definitively (as opposed to having two possible values of $i$ and four possible values of $\Omega$).

We attempted to directly fit for $i$ and $\Omega$ using the DDK model in \tempo, but the fit did not converge. Therefore, we instead performed the following test to determine whether we were likely measuring a physically reasonable value of $\dot{x}$ in the absence of a significant detection of annual orbital parallax. Using the T2 model in \tempotwo\footnote{We used \tempotwo\ instead of \tempo\ because the T2 model, which can be used to model the effects described by \citet{Kopeikin1995, Kopeikin1996} for low-eccentricity systems, does not exist in \tempo. PSR~J1909$-$3744 has a very low eccentricity, so using an ELL1-type model is preferable to, e.g., DD., we fixed $i$ at each of its two possible values\footnote{We chose to fix $i$ at its two possible values rather than also running a grid over $i$ values because $\sin{i}$ is well-constrained for this pulsar, at $\sin{i} = 0.99794\pm0.00007$. Thus we could expect the best-fit $\dot{x}_{\mathrm{k}}$ to also have small errors, allowing us to determine whether it is consistent with the secular $\dot{x}$ found by \tempo.}, and then ran \tempotwo\ over a grid of $\Omega$ values. From this gridding test,} we found the best-fit $(i, \Omega) = \{(86.39^{\circ}, 350^{\circ}), (93.61^{\circ}, 170^{\circ})\}$, and used them to calculate $\dot{x}_{\mathrm{k}}$. In all cases, $\dot{x}_{\mathrm{k}} \approx -3 \times 10^{-16}$\,lt-s\,s$^{-1}$, consistent with the value we measure from timing.

This result suggests that the $\dot{x}$ we measure with \tempo\ for PSR~J1909$-$3744 is robust, and should be included in the model rather than being absorbed by red noise. As noted above, it is also consistent with and an improvement upon our $\dot{x}$ measurement from NG11. We have therefore included this $\dot{x}$ value in our timing model for this pulsar.

\subsection{PSR~J2145$-$0750}
\label{subsec:J2145}

We can place loose constraints on the geometry of the PSR~J2145$-$0750 system using the proper motion and $h_3$ measurements. An upper bound on the orbital plane inclination can be calculated by inverting equation~\ref{eqn:xdot_pm} and attributing the measured $\dot{x}$ to the proper motion \citep[e.g.,][]{Fonseca2016}:
\begin{equation}
  i_{\mathrm{max}} = \arctan \frac{x \mu}{|\dot{x}_{\mathrm{obs}}|},
\end{equation}
\noindent yielding $i_{\mathrm{max}} = 74^{\circ}\pm5^{\circ}$. We can then combine equations~3 and 4 to obtain $r_{\mathrm{min}}$, yielding a lower bound on the companion mass, $m_{\mathrm{c,min}} = r_{\mathrm{min}}/T_{\odot} = 0.08\pm0.03\,M_{\odot}$.
Much more robust system constraints have previously been made with a combination of optical imaging, VLBI parallax, and radio timing: $i=21^{+7}_{-4}$\,degrees and $m_{\mathrm c} = 0.83\pm0.06\,M_{\odot}$ \citep{Deller2016}, and $i=34^{+5}_{-7}$\,degrees and $m_{\mathrm p} = 1.3^{+0.4}_{-0.5}\,M_{\odot}$ \citep{Fonseca2016}. We will place improved constraints on the geometry and mass of the PSR~J2145$-$0750 system in future studies with longer timing baselines.

\begin{comment}

However, these bounds 
are much less constraining than (though technically consistent with) previous parameter estimates: $i=21^{+7}_{-4}$\,degrees and $m_{\mathrm c} = 0.83\pm0.06\,M_{\odot}$ \citep{Deller2016}, and $i=34^{+5}_{-7}$\,degrees and $m_{\mathrm p} = 1.3^{+0.4}_{-0.5}\,M_{\odot}$ \citep{Fonseca2016}.

from \citet[][; $m_{\mathrm c} = 0.83\pm0.06\,M_{\odot}$]{Deller2016} and \citet[][; ]{Fonseca2016}

, and replacing $i$ with $i_{\mathrm{max}}$, we obtain
\begin{equation}
  r_{\mathrm{min}} = h_3 \left ( \frac{1+\cos{i_{\mathrm{max}}}}{1-\cos{i_{\mathrm{max}}}} \right )^{3/2},
\end{equation}
\noindent with which we find a lower bound on the companion mass, $m_{\mathrm{c,min}} = r_{\mathrm{min}}/T_{\odot} = 0.08\pm0.03\,M_{\odot}$.

\attn{This is much lower than the standard minimum companion mass measurement for this source, $\sim$ 0.4~M$_{\odot}$.  Have we checked other contraints?}
\attn{Use Kerr et al. omegadot to get pulsar mass?}

\end{comment}

\subsection{PSR~J2234+0611}
\label{subsec:J2234}

The eccentric orbit and high timing precision of J2234+0611 allowed \citet{Stovall2019} to measure a large number of binary-related effects from this system, including one orthometric Shapiro delay parameter ($h_3$) and annual orbital parallax. Together, these measurements allowed Stovall et al.\ to unambiguously determine the three-dimensional geometry of the system, giving $i \approx 138.7^{\circ}$ and $\Omega \approx 44^{\circ}$. These parameters correspond to $\dot{x}_\mathrm{k} = (-2.78 \pm 0.07) \times 10^{-14}$\,lt-s\,s$^{-1}$.

In this work, we find that $i$ and $\Omega$ are not constrained, and we do not obtain a significant measurement of $h_3$. There are two likely reasons that we are not able to reproduce the measurements of Stovall et al. First, their data set is a superset of that presented here, with an additional $\sim1.5$\,yr in their timing baseline. Secondly, Stovall et al.\ fix $i$ and $\varsigma=h_4/h_3$ based on the derived value of $s=\sin{i}$ from the DDGR binary model; they then constrain $\Omega$ by running \tempo{} over a grid in $i$, $\Omega$, and $m_{\mathrm{tot}} = m_{\rm p} + m_{\rm c}$, where $i$ and $\Omega$ are held constant at each step.
Based on our $F$-test criterion for including post-Keplerian parameters, we instead fit for $\dot{x}$ (which is related to annual orbital parallax and its secular variation shown in Equation~\ref{eqn:xdot_pm}); its value, $(-2.8 \pm 0.2) \times 10^{-14}$\,lt-s\,s$^{-1}$, is consistent with that found by Stovall et al.

\section{Flux Densities}
\label{sec:flux}
The algorithm used to calculate TOAs in our narrowband data (Section
\ref{subsec:toa_gen}) also yields the amplitudes of the pulsed signals
relative to the amplitudes of the template profiles used for timing.
Through suitable calibration and normalization of the template profiles,
these amplitudes can be used to estimate
the period-averaged flux
densities of the pulsed signals.  In this section, we describe our flux
density calculations.  The results are summarized in
Table~\ref{tab:flux}.

For the flux density analysis, we used only GUPPI and PUPPI data.
The narrower bands of GASP and ASP made them less suitable for
the cross-checks described below and would have yielded less robust
measurements.

{\setlength\tabcolsep{5pt}
\begin{deluxetable*}{lccccccccccccccccccc}
\tablewidth{0pt}
\tabletypesize{\scriptsize}
\tablecaption{Flux densities\label{tab:flux}}
\tablehead{\colhead{PSR} & \colhead {Obs.} &&
  \multicolumn{3}{c}{$S_{430}$} &&
  \multicolumn{3}{c}{$S_{800}$} &&
  \multicolumn{3}{c}{$S_{1400}$} &&
  \multicolumn{3}{c}{$S_{2000}$} &&
  \colhead{Spectral}
 \\ 
 & && & (mJy) & && & (mJy) & && & (mJy) & && & (mJy) & && Index\tablenotemark{\scriptsize a} \\ 
\cline{4-6}\cline{8-10}\cline{12-14}\cline{16-18}
 \rule{0pt}{10pt} & &&  16\tablenotemark{\scriptsize b} & 50\tablenotemark{\scriptsize b} & 84\tablenotemark{\scriptsize b} &&  16\tablenotemark{\scriptsize b} & 50\tablenotemark{\scriptsize b} & 84\tablenotemark{\scriptsize b} &&  16\tablenotemark{\scriptsize b} & 50\tablenotemark{\scriptsize b} & 84\tablenotemark{\scriptsize b} &&  16\tablenotemark{\scriptsize b} & 50\tablenotemark{\scriptsize b} & 84\tablenotemark{\scriptsize b}  
}
\startdata
J0023$+$0923   & AO          &&  \phn   1.93   &  \phn   2.50   &  \phn   3.95   & &      \nodata  &       \nodata  &       \nodata  & & \phn   0.21   &  \phn   0.32   &  \phn   0.54   & &      \nodata  &       \nodata  &       \nodata  & &        $-1.74$  \\ 
J0030$+$0451   & AO          &&  \phn   2.34   &  \phn   5.80   &        23.12   & &      \nodata  &       \nodata  &       \nodata  & & \phn   0.85   &  \phn   1.12   &  \phn   1.60   & & \phn   0.48   &  \phn   0.60   &  \phn   0.78   & &        $-1.39$  \\ 
J0340$+$4130   & GB          &&       \nodata  &       \nodata  &       \nodata  & & \phn   1.14   &  \phn   1.42   &  \phn   1.89   & & \phn   0.45   &  \phn   0.54   &  \phn   0.60   & &      \nodata  &       \nodata  &       \nodata  & &        $-1.73$  \\ 
J0613$-$0200   & GB          &&       \nodata  &       \nodata  &       \nodata  & & \phn   5.45   &  \phn   6.66   &  \phn   8.12   & & \phn   1.81   &  \phn   2.17   &  \phn   2.48   & &      \nodata  &       \nodata  &       \nodata  & &        $-2.00$  \\ 
J0636$+$5128   & GB          &&       \nodata  &       \nodata  &       \nodata  & & \phn   1.64   &  \phn   2.08   &  \phn   2.90   & & \phn   0.73   &  \phn   0.94   &  \phn   1.18   & &      \nodata  &       \nodata  &       \nodata  & &        $-1.42$  \\ 
J0645$+$5158   & GB          &&       \nodata  &       \nodata  &       \nodata  & & \phn   0.47   &  \phn   0.76   &  \phn   2.02   & & \phn   0.17   &  \phn   0.29   &  \phn   0.86   & &      \nodata  &       \nodata  &       \nodata  & &        $-1.72$  \\ 
J0740$+$6620   & GB          &&       \nodata  &       \nodata  &       \nodata  & & \phn   1.11   &  \phn   2.19   &  \phn   6.06   & & \phn   0.43   &  \phn   0.97   &  \phn   1.71   & &      \nodata  &       \nodata  &       \nodata  & &        $-1.46$  \\ 
J0931$-$1902   & GB          &&       \nodata  &       \nodata  &       \nodata  & & \phn   1.41   &  \phn   1.90   &  \phn   2.60   & & \phn   0.60   &  \phn   0.84   &  \phn   1.24   & &      \nodata  &       \nodata  &       \nodata  & &        $-1.46$  \\ 
J1012$+$5307   & GB          &&       \nodata  &       \nodata  &       \nodata  & & \phn   3.19   &  \phn   6.93   &        16.46   & & \phn   1.46   &  \phn   2.90   &  \phn   6.09   & &      \nodata  &       \nodata  &       \nodata  & &        $-1.56$  \\ 
J1024$-$0719   & GB          &&       \nodata  &       \nodata  &       \nodata  & & \phn   1.43   &  \phn   2.92   &  \phn   7.48   & & \phn   0.74   &  \phn   1.36   &  \phn   2.56   & &      \nodata  &       \nodata  &       \nodata  & &        $-1.37$  \\ 
J1125$+$7819   & GB          &&       \nodata  &       \nodata  &       \nodata  & & \phn   1.39   &  \phn   2.58   &  \phn   5.20   & & \phn   0.55   &  \phn   0.90   &  \phn   1.70   & &      \nodata  &       \nodata  &       \nodata  & &        $-1.88$  \\ 
J1453$+$1902   & AO          &&  \phn   1.11   &  \phn   1.64   &  \phn   2.81   & &      \nodata  &       \nodata  &       \nodata  & & \phn   0.11   &  \phn   0.17   &  \phn   0.48   & &      \nodata  &       \nodata  &       \nodata  & &        $-1.92$  \\ 
J1455$-$3330   & GB          &&       \nodata  &       \nodata  &       \nodata  & & \phn   1.42   &  \phn   2.28   &  \phn   5.03   & & \phn   0.45   &  \phn   0.73   &  \phn   1.51   & &      \nodata  &       \nodata  &       \nodata  & &        $-2.04$  \\ 
J1600$-$3053   & GB          &&       \nodata  &       \nodata  &       \nodata  & & \phn   2.66   &  \phn   3.17   &  \phn   4.02   & & \phn   1.99   &  \phn   2.37   &  \phn   2.99   & &      \nodata  &       \nodata  &       \nodata  & &        $-0.52$  \\ 
J1614$-$2230   & GB          &&       \nodata  &       \nodata  &       \nodata  & & \phn   2.02   &  \phn   2.68   &  \phn   3.51   & & \phn   0.79   &  \phn   1.11   &  \phn   1.53   & &      \nodata  &       \nodata  &       \nodata  & &        $-1.58$  \\ 
J1640$+$2224   & AO          &&  \phn   4.20   &  \phn   6.31   &        11.33   & &      \nodata  &       \nodata  &       \nodata  & & \phn   0.24   &  \phn   0.54   &  \phn   1.26   & &      \nodata  &       \nodata  &       \nodata  & &        $-2.08$  \\ 
J1643$-$1224   & GB          &&       \nodata  &       \nodata  &       \nodata  & &       11.31   &        13.02   &        14.50   & & \phn   4.06   &  \phn   4.70   &  \phn   5.38   & &      \nodata  &       \nodata  &       \nodata  & &        $-1.82$  \\ 
J1713$+$0747   & AO          &&       \nodata  &       \nodata  &       \nodata  & &      \nodata  &       \nodata  &       \nodata  & & \phn   2.47   &  \phn   4.69   &  \phn   8.70   & & \phn   2.24   &  \phn   5.01   &        10.24   & &  \phs  $ 0.19$  \\ 
J1713$+$0747   & GB          &&       \nodata  &       \nodata  &       \nodata  & & \phn   3.39   &  \phn   6.14   &        12.12   & & \phn   2.02   &  \phn   4.32   &  \phn   8.43   & &      \nodata  &       \nodata  &       \nodata  & &        $-0.63$  \\ 
J1738$+$0333   & AO          &&       \nodata  &       \nodata  &       \nodata  & &      \nodata  &       \nodata  &       \nodata  & & \phn   0.37   &  \phn   0.48   &  \phn   0.82   & & \phn   0.26   &  \phn   0.43   &  \phn   0.65   & &        $-0.31$  \\ 
J1741$+$1351   & AO          &&  \phn   1.79   &  \phn   2.60   &  \phn   4.64   & &      \nodata  &       \nodata  &       \nodata  & & \phn   0.14   &  \phn   0.21   &  \phn   0.38   & &      \nodata  &       \nodata  &       \nodata  & &        $-2.13$  \\ 
J1744$-$1134   & GB          &&       \nodata  &       \nodata  &       \nodata  & & \phn   2.94   &  \phn   5.29   &        10.48   & & \phn   0.98   &  \phn   1.96   &  \phn   4.74   & &      \nodata  &       \nodata  &       \nodata  & &        $-1.77$  \\ 
J1747$-$4036   & GB          &&       \nodata  &       \nodata  &       \nodata  & & \phn   5.63   &  \phn   6.85   &  \phn   9.05   & & \phn   1.31   &  \phn   1.55   &  \phn   2.20   & &      \nodata  &       \nodata  &       \nodata  & &        $-2.66$  \\ 
J1832$-$0836   & GB          &&       \nodata  &       \nodata  &       \nodata  & & \phn   3.14   &  \phn   3.65   &  \phn   4.23   & & \phn   0.90   &  \phn   1.11   &  \phn   1.35   & &      \nodata  &       \nodata  &       \nodata  & &        $-2.13$  \\ 
J1853$+$1303   & AO          &&  \phn   3.13   &  \phn   4.71   &  \phn   7.11   & &      \nodata  &       \nodata  &       \nodata  & & \phn   0.24   &  \phn   0.43   &  \phn   0.59   & &      \nodata  &       \nodata  &       \nodata  & &        $-2.03$  \\ 
B1855$+$09     & AO          &&        14.36   &        19.17   &        26.18   & &      \nodata  &       \nodata  &       \nodata  & & \phn   2.13   &  \phn   3.47   &  \phn   6.50   & &      \nodata  &       \nodata  &       \nodata  & &        $-1.45$  \\ 
J1903$+$0327   & AO          &&       \nodata  &       \nodata  &       \nodata  & &      \nodata  &       \nodata  &       \nodata  & & \phn   0.63   &  \phn   0.69   &  \phn   0.79   & & \phn   0.45   &  \phn   0.52   &  \phn   0.58   & &        $-0.79$  \\ 
J1909$-$3744   & GB          &&       \nodata  &       \nodata  &       \nodata  & & \phn   1.54   &  \phn   3.01   &  \phn   5.15   & & \phn   0.55   &  \phn   1.03   &  \phn   2.17   & &      \nodata  &       \nodata  &       \nodata  & &        $-1.92$  \\ 
J1910$+$1256   & AO          &&       \nodata  &       \nodata  &       \nodata  & &      \nodata  &       \nodata  &       \nodata  & & \phn   0.35   &  \phn   0.56   &  \phn   0.69   & & \phn   0.21   &  \phn   0.31   &  \phn   0.47   & &        $-1.66$  \\ 
J1911$+$1347   & AO          &&  \phn   2.10   &  \phn   2.93   &  \phn   4.16   & &      \nodata  &       \nodata  &       \nodata  & & \phn   0.51   &  \phn   0.86   &  \phn   1.26   & &      \nodata  &       \nodata  &       \nodata  & &        $-1.04$  \\ 
J1918$-$0642   & GB          &&       \nodata  &       \nodata  &       \nodata  & & \phn   3.12   &  \phn   4.13   &  \phn   5.43   & & \phn   0.89   &  \phn   1.36   &  \phn   1.99   & &      \nodata  &       \nodata  &       \nodata  & &        $-1.98$  \\ 
J1923$+$2515   & AO          &&  \phn   2.04   &  \phn   2.56   &  \phn   3.36   & &      \nodata  &       \nodata  &       \nodata  & & \phn   0.18   &  \phn   0.28   &  \phn   0.49   & &      \nodata  &       \nodata  &       \nodata  & &        $-1.87$  \\ 
B1937$+$21     & AO          &&       \nodata  &       \nodata  &       \nodata  & &      \nodata  &       \nodata  &       \nodata  & & \phn   9.14   &        12.44   &        16.56   & & \phn   3.24   &  \phn   4.72   &  \phn   6.25   & &        $-2.72$  \\ 
B1937$+$21     & GB          &&       \nodata  &       \nodata  &       \nodata  & &       41.93   &        56.37   &        74.68   & &       11.33   &        14.83   &        18.65   & &      \nodata  &       \nodata  &       \nodata  & &        $-2.39$  \\ 
J1944$+$0907   & AO          &&        14.57   &        18.82   &        28.53   & &      \nodata  &       \nodata  &       \nodata  & & \phn   1.08   &  \phn   1.94   &  \phn   3.41   & &      \nodata  &       \nodata  &       \nodata  & &        $-1.92$  \\ 
J1946$+$3417   & AO          &&       \nodata  &       \nodata  &       \nodata  & &      \nodata  &       \nodata  &       \nodata  & & \phn   0.84   &  \phn   0.92   &  \phn   0.96   & & \phn   0.56   &  \phn   0.61   &  \phn   0.66   & &        $-1.15$  \\ 
B1953$+$29     & AO          &&        10.77   &        12.00   &        13.28   & &      \nodata  &       \nodata  &       \nodata  & & \phn   0.69   &  \phn   0.94   &  \phn   1.21   & &      \nodata  &       \nodata  &       \nodata  & &        $-2.16$  \\ 
J2010$-$1323   & GB          &&       \nodata  &       \nodata  &       \nodata  & & \phn   1.20   &  \phn   1.63   &  \phn   2.12   & & \phn   0.50   &  \phn   0.71   &  \phn   0.92   & &      \nodata  &       \nodata  &       \nodata  & &        $-1.49$  \\ 
J2017$+$0603   & AO          &&       \nodata  &       \nodata  &       \nodata  & &      \nodata  &       \nodata  &       \nodata  & & \phn   0.22   &  \phn   0.32   &  \phn   0.46   & & \phn   0.21   &  \phn   0.31   &  \phn   0.42   & &        $-0.09$  \\ 
J2033$+$1734   & AO          &&  \phn   2.98   &  \phn   3.62   &  \phn   4.23   & &      \nodata  &       \nodata  &       \nodata  & & \phn   0.26   &  \phn   0.30   &  \phn   0.38   & &      \nodata  &       \nodata  &       \nodata  & &        $-2.11$  \\ 
J2043$+$1711   & AO          &&  \phn   1.51   &  \phn   1.99   &  \phn   3.01   & &      \nodata  &       \nodata  &       \nodata  & & \phn   0.15   &  \phn   0.21   &  \phn   0.40   & &      \nodata  &       \nodata  &       \nodata  & &        $-1.91$  \\ 
J2145$-$0750   & GB          &&       \nodata  &       \nodata  &       \nodata  & & \phn   5.10   &        14.25   &        38.84   & & \phn   1.69   &  \phn   4.86   &        12.15   & &      \nodata  &       \nodata  &       \nodata  & &        $-1.92$  \\ 
J2214$+$3000   & AO          &&       \nodata  &       \nodata  &       \nodata  & &      \nodata  &       \nodata  &       \nodata  & & \phn   0.28   &  \phn   0.38   &  \phn   0.70   & & \phn   0.20   &  \phn   0.35   &  \phn   0.51   & &        $-0.23$  \\ 
J2229$+$2643   & AO          &&  \phn   2.19   &  \phn   3.61   &  \phn   7.96   & &      \nodata  &       \nodata  &       \nodata  & & \phn   0.22   &  \phn   0.41   &  \phn   0.70   & &      \nodata  &       \nodata  &       \nodata  & &        $-1.84$  \\ 
J2234$+$0611   & AO          &&  \phn   0.72   &  \phn   0.90   &  \phn   1.82   & &      \nodata  &       \nodata  &       \nodata  & & \phn   0.12   &  \phn   0.33   &  \phn   0.89   & &      \nodata  &       \nodata  &       \nodata  & &        $-0.85$  \\ 
J2234$+$0944   & AO          &&       \nodata  &       \nodata  &       \nodata  & &      \nodata  &       \nodata  &       \nodata  & & \phn   0.81   &  \phn   1.15   &  \phn   2.83   & & \phn   0.40   &  \phn   0.76   &  \phn   1.14   & &        $-1.16$  \\ 
J2302$+$4442   & GB          &&       \nodata  &       \nodata  &       \nodata  & & \phn   2.59   &  \phn   3.29   &  \phn   4.31   & & \phn   1.06   &  \phn   1.41   &  \phn   1.84   & &      \nodata  &       \nodata  &       \nodata  & &        $-1.51$  \\ 
J2317$+$1439   & AO          &&  \phn   4.81   &  \phn   7.20   &  \phn   9.96   & &      \nodata  &       \nodata  &       \nodata  & & \phn   0.26   &  \phn   0.45   &  \phn   0.80   & &      \nodata  &       \nodata  &       \nodata  & &        $-2.35$  \\ 
J2322$+$2057   & AO          &&  \phn   1.37   &  \phn   1.94   &  \phn   3.54   & &      \nodata  &       \nodata  &       \nodata  & & \phn   0.14   &  \phn   0.26   &  \phn   0.72   & & \phn   0.31   &  \phn   0.32   &  \phn   1.06   & &        $-1.70$  \\ 
\enddata
\tablenotetext{a}{Calculated  from $S_{800}$ and $S_{1400}$ for GB pulsars,  $S_{430}$ and $S_{1400}$ for AO pulsars with 430~MHz data, and $S_{1400}$ and $S_{2000}$ for other pulsars.}
\tablenotetext{b}{In flux density columns,  16, 50, and 84 refer to 16th, 50th, and 84th percentile  of epoch-averaged flux density values.}
\end{deluxetable*}
} 
\begin{figure*}
\begin{center}
\includegraphics[width=6.0in]{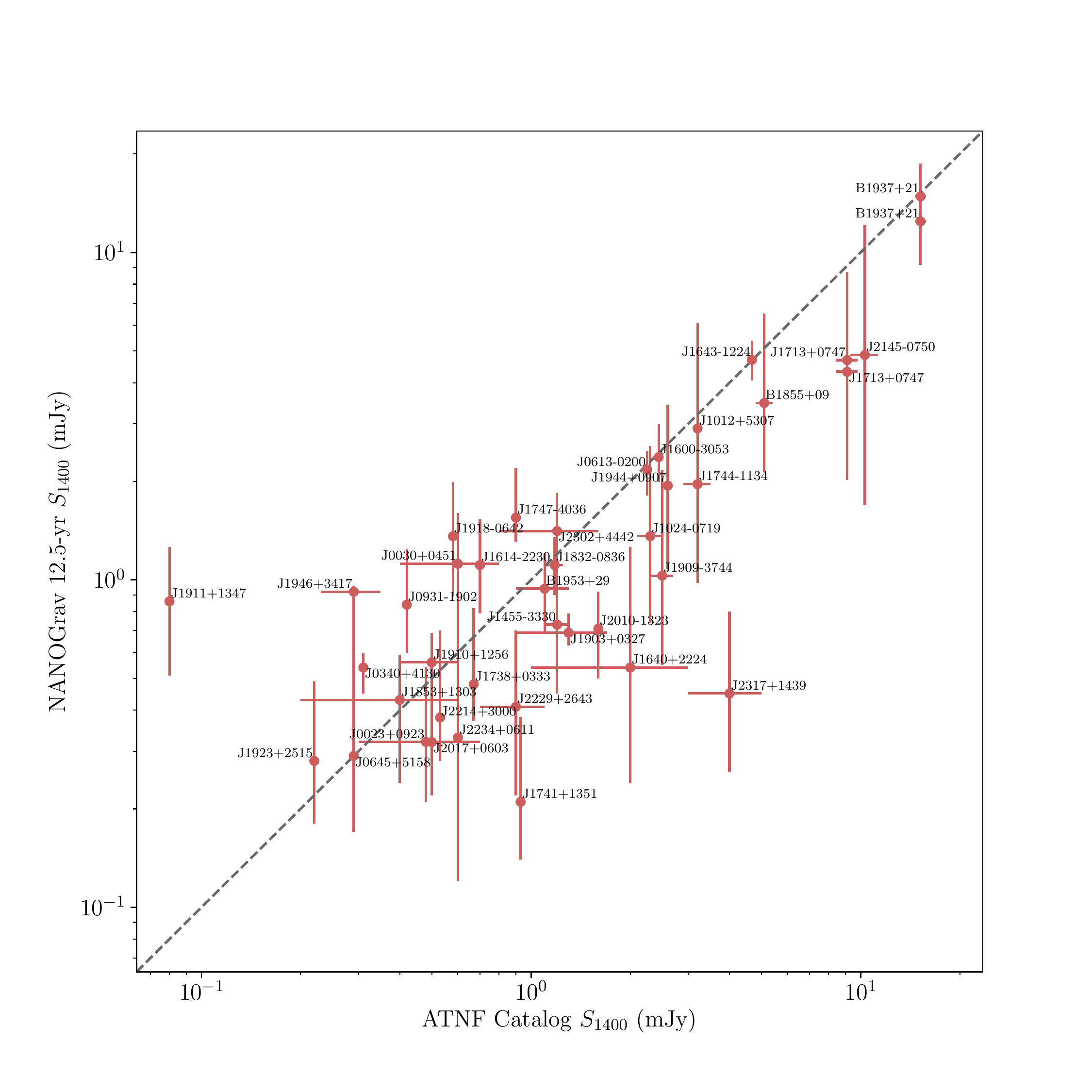}
\caption{\label{fig:flux:compare}
Comparison of 1400~MHz flux density measurements ($S_{\rm 1400}$) from 
the present paper (vertical axis) with values in the ATNF pulsar catalog \citep[horizontal axis;][version 1.63]{Manchester2005}.  
Vertical error bars indicate the central 68\% of measured flux density values (Table~\ref{tab:flux}).
Horizontal error bars, where present, indicate the uncertainties reported in the ATNF pulsar catalog.  Two pulsars are shown
twice (J1713+0747 and B9137+21), with similar values, because we made
separate analyses of our Arecibo and GBT measurements.
}
\end{center}
\end{figure*}
 
\begin{deluxetable*}{lcccccc}
\tablewidth{0pt}
\tablecaption{Comparison of Spectral Indexes\label{tab:spectidx}}
\tablehead{\colhead{PSR} & \colhead {Obs.} &&
  \multicolumn{1}{c}{NG12.5} &&
  \multicolumn{2}{c}{\cite{Frail2016}} 
 \\ 
\cline{4-4}\cline{6-7} & && $\alpha$ && $\alpha$ & $\sigma_\alpha$ 
}
\startdata
J0030$+$0451   &  AO &&  -1.39   &&  -1.93 &  0.15 \\ 
J1643$-$1224   &  GB &&  -1.82   &&  -1.64 &  0.06 \\ 
J1747$-$4036   &  GB &&  -2.66   &&  -2.81 &  0.22 \\ 
B1937$+$21     &  AO &&  -2.72   &&  -2.51 &  0.17 \\ 
B1937$+$21     &  GB &&  -2.39   &&  -2.51 &  0.17 \\ 
J1944$+$0907   &  AO &&  -1.92   &&  -3.03 &  0.06 \\ 
B1953$+$29     &  AO &&  -2.16   &&  -1.77 &  0.09 \\ 
J2145$-$0750   &  GB &&  -1.92   &&  -1.47 &  0.61 \\ 
J2317$+$1439   &  AO &&  -2.35   &&  -0.86 &  0.12 \\ 
\enddata
\end{deluxetable*}

\subsection{Absolute Calibration}

As described in Section \ref{sec:obs}, each individual observation
was preceded by a pulsed-cal measurement using an artificial noise
signal.  The pulsed-cal measurements were used to calibrate the
amplitude scale of the pulsar observations.  The noise signals themselves were
calibrated by measuring them in on- and off-source observations of a standard
continuum calibrator radio source.  The noise signals were checked
against the continuum calibrator approximately once per month at
each observatory.  Thus our flux density measurements depend directly
on our assumptions about the continuum calibrator flux density.

For continuum calibration, we used the compact radio source
J1445+0958 (B1442+101) for all GUPPI observations and all
PUPPI observations after MJD 56619 (2013 November 29).  We
only used data calibrated with this source in the flux density 
analysis.
The VLA calibrator
manual\footnote{\url{https://science.nrao.edu/facilities/vla/observing/callist}}
gives the
flux density of J1445+0958 as 2.60, 1.20, 0.73 0.40, and 0.10~Jy at
wavelengths of 20.0, 6.0, 3.7, 2.0, and 0.7~cm, respectively.  We
modeled this using the analytic expression
$\log S=\sum_{i=0}^{3} a_i\left(\log \nu_{\rm GHz}\right)^i$, where
$\nu_{\rm GHz}$ is the radio frequency in gigahertz and the
coefficients for J1445+0958 are $a_0=0.389314$, $a_1=-0.0280647$,
$a_2=-0.600809$, and $a_3=0.0262127$.  

\subsection{Data Checks and Cleaning}

Flux density measurements are particularly susceptible to calibration
errors, so we undertook further checks of the data used for
this purpose.

The signal-to-noise ratio of a pulsar observation is expected to be
${\rm S/N} = S G/T_{\rm sys}\sqrt{2 B t}\sqrt{(P-w)/w}$, where $S$, $G$, $T_{\rm sys}$,
$B$, $t$, $P$, and $w$  are the pulsar flux, telescope gain, system
temperature, bandwidth, observing time, pulse period, and pulse width,
respectively, and where $\sqrt{(P-w)/w}$ is a proxy for
a general pulse-shape-dependent factor.  For
a given telescope, receiver, and pulsar, at a given radio
frequency, the ratio $(S/[{\rm S/N}])\sqrt{2Bt}$ should
be constant (after considering small variations due to radiometer noise
and variations in $G$ and $T_{\rm sys}$ due to telescope elevation).
Radio-frequency interference, calibration errors, and other
defects in the data can cause this ratio to vary.  We used this
to flag potentially discrepant flux density values.  We removed
any data point for which this ratio was less than 0.75 or more than 1.75 times
the median value for that frequency and nearby frequencies at low
S/N, with a gradually increasing allowed upper value for high
S/N (analyzed
over ranges of 100 MHz above 1 GHz, 50 MHz for the GBT 820~MHz receiver,
and 10~MHz for the Arecibo 327 and 430 MHz receivers).  At $\nu>2300$~MHz, we used
a lower limit on the ratio of 0.50 rather than 0.75.   Further, we
entirely eliminated any observation for which five or more individual
flux values were flagged as potentially discrepant.
These specific choices were based on
empirical analysis of the data.  They eliminated obvious outliers
while allowing us to retain most of the flux density measurements.

\subsection{Flux Density Measurement Results}

For each epoch, we fit the observed set of narrowband flux values
to a power law, $S_{\rm obs}(\nu)=S_{\nu_0}(\nu/\nu_0)^{\alpha}$.
Here $S_{\nu_0}$ is the fit flux at fiducial frequency $\nu_0$ and
$\alpha$ is the spectral index.
We used fiducial frequencies of 430, 800, 1400, and 2000~MHz, chosen for their locations near the centers of the observing bands and, in some cases, because they are standard frequencies used in pulsar catalogs.
We calculated separate values of
$S_{\nu_0}$ and $\alpha$ for each receiver at each observing epoch
(defined as observations within a 3-day span).

For most pulsars, the values of $\alpha$ within these single-receiver,
single-epoch fits varied widely due to
diffractive scintillation within the receiver bands; we do not
use those values further.

Table~\ref{tab:flux} reports the median observed flux density
for each pulsar and each receiver, along with 16th and 84th percentile
values.  The table also includes spectral indexes calculated using
the median flux density values in two bands (as specified in the table).
For the two pulsars observed at both Arecibo and GBT (PSRs J1713+0747
and B1937+21), we analyzed the measurements from the observatories
separately as a check against systematic errors.  Their $S_{1400}$
measurements show good agreement between the two observatories.

PSR~J1713+0747 shows a significant difference between the spectral
index calculated from Arecibo values of $S_{1400}$ and $S_{2000}$ and
the spectral index calculated from GBT values of $S_{800}$ and
$S_{1400}$.  This suggests that a single power law
is not sufficient to model the flux density across a wide
range of frequencies.  PSR~B1937+21 also shows a difference
in spectral indexes between the two observatories, albeit
somewhat smaller (and in the opposite direction) than
that of J1713+0747.

Because our flux density analysis only includes measurements from
observations that yielded good narrowband TOA values, it excludes
observations in which the pulsar had a very low S/N or was not visible
at all.  This could potentially bias our measurements high (since low
flux density values due to, e.g., extreme scintillation are excluded). Furthermore, we
used a constant template profile for the narrowband TOA measurements, which
assumes that the true profile shape does not evolve with frequency across the
band; the amplitude, and thus the estimated flux density, will depend on the
degree to which this assumption holds.
We believe both of these factors only have a small effect in our reported
measurements, but a detailed analysis of these biases is beyond the scope of
the present work.

The spectral indexes of sources observed at Arecibo at 1400 and 2000~MHz tend
to have smaller magnitudes than the spectral indexes of other pulsars.
This is almost certainly a selection effect.  We preferentially
observe Arecibo pulsars with this combination of receivers, but only
when the pulsar is strong enough to be consistently detected with
high signal-to-noise at 2000~MHz.  This implies that such sources
have relatively high flux densities at 2000~MHz, and therefore
relatively shallow spectra.

\subsection{Comparison with Previous Work}

A comparison between our $S_{1400}$ measurements and previously
reported values for 41 pulsars is given in Figure~\ref{fig:flux:compare}.
There is some scatter in the values, likely attributable to
scintillation-induced variations in flux density measurements, 
particularly in previous measurements which might be based
on a small number of samples.  The median ratio of $S_{1400}$
from our measurements to the $S_{1400}$ of previous measurements was
0.82, and the average value of this ratio was 1.19.  The latter is
evidently dominated by a few sources with the most significant discrepancies.
The most significant outlier values are for PSR~J1911+1347, for
which we report median $S_{1400}=0.86$~mJy, while the previous reported 
value was 0.08~mJy, more than an order of magnitude lower
\citep{Lorimer2006}; and for PSR~J2317+1439, for which we
report median $S_{1400}=0.45$~mJy, while the previous reported
value was $4\pm1$~mJy, more than an order of magnitude
greater \citep{Kramer1998}.  The reason for the relatively 
large discrepancies in reported values for these two pulsars
is not known.

We made similar comparisons between our flux density measurements and
cataloged values at 430~MHz (11 pulsars) and 2000~MHz (1 pulsar).
The results were comparable to those at 1400~MHz: general agreement
with modest scatter between new and previous measurements.

\citet[][herein S2017]{Shaifullah2017} reports measurements
of spectral indexes
of a dozen MSPs at Arecibo using the 327~MHz and 1400~MHz receivers.
Of those, five MSPs were observed for a large number of epochs
(thus mitigating scintillation issues).  Within that group,
two overlap with the present work, and both show
good agreement with our measurements:  For PSR~J1453+1902, 
we find $\alpha=-1.92$, and S2017
reports $\alpha=-1.7\pm 0.4$; for J2322+2057, 
we find $\alpha=-1.70$, and S2017 reports
$\alpha=-1.7\pm 0.1$.  (Other pulsars in S2017 with fewer 
observations show larger discrepancies.)

\cite{Frail2016} reported spectral indexes of many pulsars
calculated by combining Giant Metrewave Radio Telescope
(GMRT) 150~MHz observations with
measurements in the literature at other frequencies.
A summary is presented in Table~\ref{tab:spectidx},
which lists the spectral indexes from the 
present work (NG12.5), along with the spectral
indexes and their uncertainties, $\sigma_\alpha$
reported by \cite{Frail2016}.
Many measurements are in agreement, but a few are not.
It is not clear
whether this is because of differences in observing
frequency (and the possible inadequacy of a single power-law
to describe flux density) or whether it is caused by
something else.

\section{Summary and Conclusions}
\label{sec:conclusion}
In this paper, we have introduced the NANOGrav 12.5-year data set, which contains TOAs and timing models for 47 MSPs
with baselines between $\sim2$ and $\sim13$~years (Section~\ref{sec:obs}).
In particular, the present work follows in the footsteps of NANOGrav's preceding three data releases (\fiveyr, \nineyr, and \elevenyr).  We have given the TOAs discussed in this paper the designation ``narrowband'' to distinguish them from the ``wideband'' data set that uses a number of new developments to process the same pulse profile data \citep[see the parallel paper,][]{Alam_2020_wideband}.

These data introduced two new pulsars into our PTA (PSRs~J1946$+$3417 and J2322$+$2057) and extended our baseline by 1.5~years.  A number of new procedural changes and quality control measures were introduced over \elevenyr.
In addition to the wideband processing, for this data set we:
\begin{enumerate}
    \item{Removed low-amplitude artifact images from the profile data that were introduced by the interleaved samplers of the ADCs (Section~\ref{subsec:data_reduction}).}
  \item{Automated and systematized the timing analysis procedure with Jupyter notebooks (Section~\ref{sec:timing}).} 
  \item{Excised whole epochs of data based on their disproportionately large influence on timing fit $\chi^2$, as indicated by an $F$-test (Section~\ref{subsec:cleaningdata}).}
  \item{Identified and excised specious TOAs by examining calibration scans and flux densities (Section~\ref{subsec:cleaningdata}).}
  \item{Introduced \texttt{-cut} flags that document why a TOA has been removed from the data set (Section~\ref{subsec:cleaningdata}).}
    \item{Transitioned to the \enterprise\ PTA analysis software for noise modeling (Section~\ref{subsec:noise}).}
  \item{Cross-checked timing results with the new, \tempo- and \tempo2-independent pulsar timing software \pint\ (Section~\ref{sec:pint}).}
  \item{Included per-TOA flux density measurements (Section~\ref{sec:flux}).}
\end{enumerate}

Some of these changes led to improvements in the white noise model parameters, as indicated by a re-analysis of the data in \elevenyr\ compared to the 11-year ``slice'' of the present data set (Section~\ref{subsec:noise_improvements}).
The red noise detected in eleven sources in \elevenyr\ continues to be present (Section~\ref{subsec:noise}). Red noise is also detected in an additional three pulsars---PSRs~J1744$-$1134, J1853$+$1303, and J2317$+$1439---the first and third of which are bright, precisely timed pulsars with some of the longest baselines in the data set.
A number of other pulsars have sub-threshold hints of red noise, which may become significant in future data sets.

Two of the main astrophysical results from this data set---the observation of a second chromatic ISM event in PSR~J1713$+$0747 \citep{Lam2018}, and the discovery that PSR~J0740$+$6620 is the most massive, precisely-measured neutron star known to date \citep{cfr+20}---were published prior to the present work.
Additionally, the entire data set was analyzed for pulse phase jitter in \citet{Lam19b}; \citet{Deneva19} used data from this release to compare the radio timing stability of PSR~B1937$+$21 with that seen by the NICER X-ray instrument; and \citet{Stovall2019} used a superset of these observations to solve the 3-D orbit of PSR~J2234$+$0611.
In Section~\ref{sec:newsigpars} we highlighted a number of other new NANOGrav measurements, which include the first published measurements of trigonometric parallax for PSRs~J1832$-$0836 and J2322$+$2057, of $\dot{x}$ for J0613$-$0200 and J2229+2643, and of $h_3$ indicating a marginally-detected Shapiro delay in J2145$-$0750.

NANOGrav is committed to continued public data releases, both for individual studies of high-precision pulsar timing and for the sake of gravitational wave detection.\footnote{Data from this paper are available at \url{http://data.nanograv.org} and preserved to Zenodo at \url{doi:10.5281/zenodo.4312297}.}
Analyses of these data to model a variety of gravitational wave signals will be presented in forthcoming publications, with our latest results from searching for a stochastic background presented in \citet{Arzoumanian20}.  Furthermore, advanced noise modeling techniques in which bespoke models are applied to each pulsar are anticipated to further increase our sensitivity, and will also be presented elsewhere \citep[][in preparation]{Simon2020}. Increasing the number of pulsars in the array has the largest impact on determining the prospects for detection of the stochastic background.
To this end, we are already synthesizing our next data set, which will come with the single largest increase in the number of pulsars ($\sim$50\%) since we doubled the size of the array between \fiveyr\ and \nineyr.
The concomitant increase in the sensitivity and complexity of our PTA analyses promises to deliver an exciting
upcoming few years of nanohertz gravitational wave astrophysics.

\acknowledgments

{\it Author contributions.}
The alphabetical-order author list reflects the broad variety of
contributions of authors to the NANOGrav project.  Some specific
contributions to this paper, particularly incremental work
between \elevenyr\ and the present work, are as follows.

ZA,
HB,
PTB,
HTC,
MED,
PBD,
TD,
RDF,
ECF,
EF,
PAG,
MLJ,
MAL,
DRL,
RSL,
MAM,
CN,
DJN,
TTP,
SMR,
KS,
IHS,
JKS,
RS,
SJV,
and
WZ
each ran at least 10 sessions and/or 20 hours of observations for this project.
MFA,
KEB,
KC,
RSC,
RLC,
WF,
YG,
DH,
CJ,
KM,
BMXN,
JR,
and
MT
were undergraduate observing-team leaders, supervised by FC, TD, DLK, JKS, and XS.

PTB, PBD, MED, MTL, JL, MAM, TTP, and KS
developed and refined procedures and computational tools for the timing pipeline.
HB,
PRB,
HTC,
MED,
PBD,
EF,
DCG,
MLJ,
MTL,
MAM,
DJN,
NSP,
TTP,
SMR,
IHS,
and
KS
generated and checked timing solutions for individual pulsars.
MED, PBD, PAG, MTL, DJN, TTP, and KS performed the various data quality checks described in Section~\ref{subsec:cleaningdata}.

MTL developed the systematic pipeline for developing timing models.
MV assisted in implementing the outlier analysis code into the 12.5-year data set analysis.
PBD wrote observing proposals, performed calibration and TOA
generation, coordinated the data flow, developed and implemented the
methodology used to mitigate the artifact images in the data, and
produced the finalized versions of the data files for public release.
PBD and KS coordinated GBT observations.
MED coordinated development of the data set and the writing of this
paper, wrote much of the text, and coordinated observations at the
Arecibo Observatory.
DJN and TTP also aided in coordination of data analyses and paper writing.
JS
and
SRT
assisted in interpreting the red noise modeling results.
JSH performed the noise parameter comparison between the 11-year data set and 11-year ``slice'' analysis of the 12.5-year data set.
JL undertook the comparison between \pint\ and \tempo.
SC performed the comparison of astrometric parameters between data sets.
JAG and DJN produced the flux density analysis.
MED and EF undertook analysis of binary systems.
TTP and SMR also contributed text, especially for the introduction and conclusion.
HTC and TTP produced the timing residual plots in Appendix~\ref{sec:resid}.
NGD oversaw and maintained much of the computational infrastructure related and essential to this work, including installing and setting up the outlier analysis software, regularly updating \tempo, and maintaining the server with the Jupyter notebooks used in our systematic analysis pipeline.
 
The NANOGrav project receives support from National Science Foundation (NSF) Physics Frontiers Center award number 1430284.
The Arecibo Observatory is a facility of the NSF operated under cooperative agreement (\#AST-1744119) by the University of Central Florida (UCF) in alliance with Universidad Ana G. M\'{e}ndez (UAGM) and Yang Enterprises (YEI), Inc.
The Green Bank Observatory is a facility of the NSF operated under cooperative agreement by Associated Universities, Inc.
The National Radio Astronomy Observatory is a facility of the NSF operated under cooperative agreement by Associated Universities, Inc.
Part of this research was carried out at the Jet Propulsion Laboratory, California Institute of Technology, under a contract with the National Aeronautics and Space Administration.
Pulsar research at UBC is supported by an NSERC Discovery Grant and by the Canadian Institute for Advanced Research.
TTP acknowledges support from the MTA-ELTE Extragalactic Astrophysics Research Group, funded by the Hungarian Academy of Sciences (Magyar Tudom\'{a}nyos Akad\'{e}mia), that was used during the development of this research.
WWZ is supported by the CAS Pioneer Hundred Talents Program and the Strategic Priority Research Program of the Chinese Academy of Sciences Grant No. XDB23000000.

We thank the telescope operators and all the staff at the Arecibo Observatory and the Green Bank Observatory for the essential role they played in collecting the data for this data release. We also thank the anonymous referee for their very useful comments that helped to improve the quality of this paper.

The NANOGrav Collaboration dedicates this work to the Arecibo Observatory, its employees and staff, and the many students, teachers, and others who have drawn inspiration from it.

\facilities{Arecibo Observatory, Green Bank Observatory}

\software{\texttt{ENTERPRISE} \citep{enterprise}, \texttt{libstempo} \citep{libstempo}, \texttt{matplotlib} \citep{matplotlib}, \texttt{nanopipe} \citep{nanopipe}, \texttt{piccard} \citep{piccard}, \texttt{PINT} \citep{pint}, \texttt{PSRCHIVE} \citep{Hotan04, vanStraten11psrchive}, \texttt{PTMCMC} \citep{ptmcmc}, \texttt{PyPulse} \citep{pypulse}, \texttt{Tempo} \citep{tempo}, \texttt{Tempo2} \citep{tempo2}, \texttt{tempo\_utils}\footnote{\href{https://github.com/demorest/tempo\_utils}{https://github.com/demorest/tempo\_utils}, git commit 51e0d9c on 2018-07-17}}

\clearpage

\appendix

\section{Averaged Residuals}

\label{sec:resid}
\begin{figure*}[p]
\centering
\includegraphics[scale=0.76]{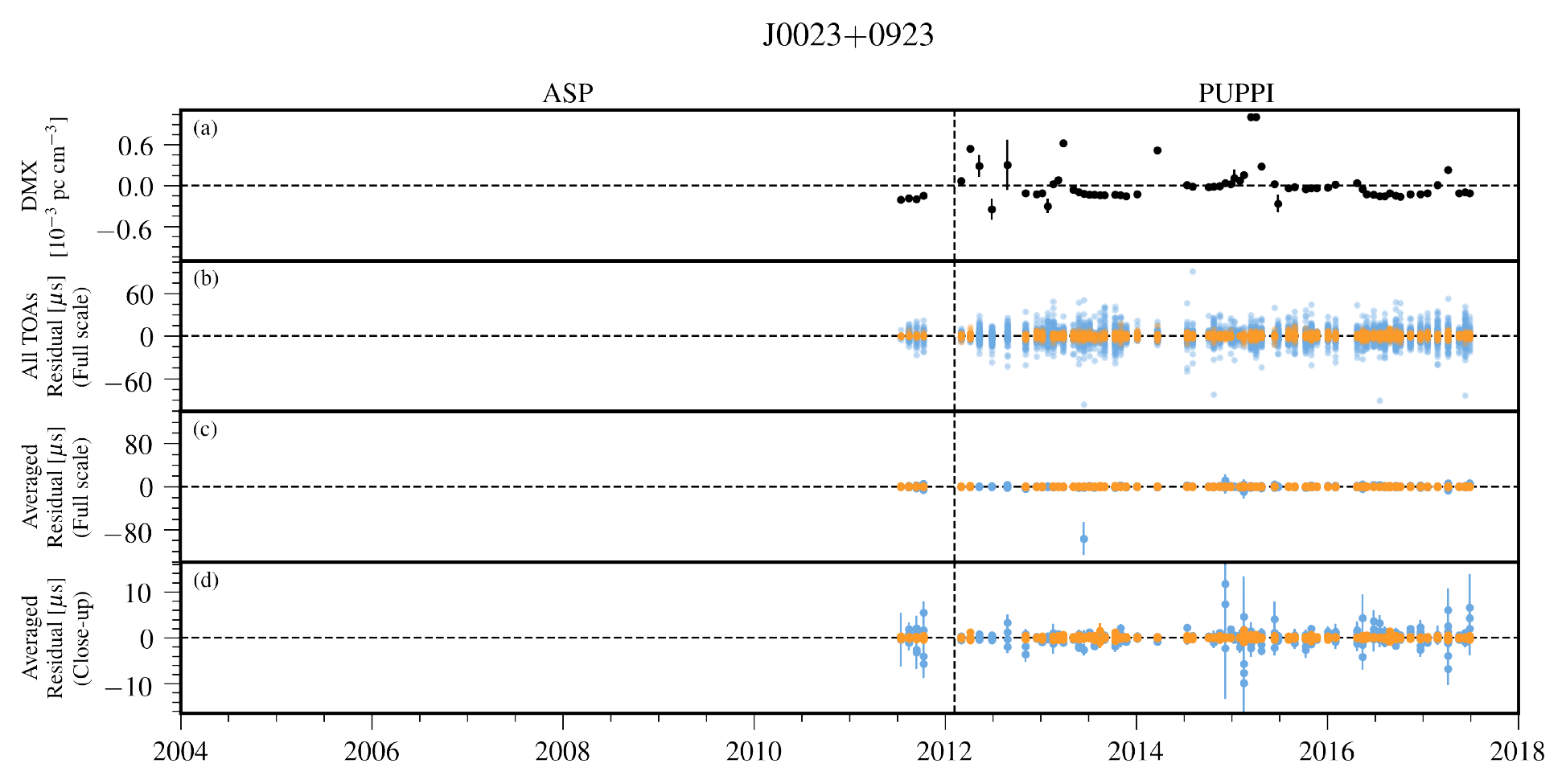}
\caption{Timing residuals and DM variations for PSR J0023$+$0923. See appendix~\ref{sec:resid} text for details.  In residual plots, colored points indicate the receiver of each observation: 430~MHz (Orange) and 1.4~GHz (Light blue).  (a) Variations in DMX.  (b) Residual arrival times for all TOAs.  Points are semi-transparent; dark regions arise from the overlap of many points.  (c,d) Average residual arrival times shown full scale (panel c) and close-up of low residuals (panel d). }
\label{fig:summary-J0023+0923}
\end{figure*}

\begin{figure*}[p]
\centering
\includegraphics[scale=0.76]{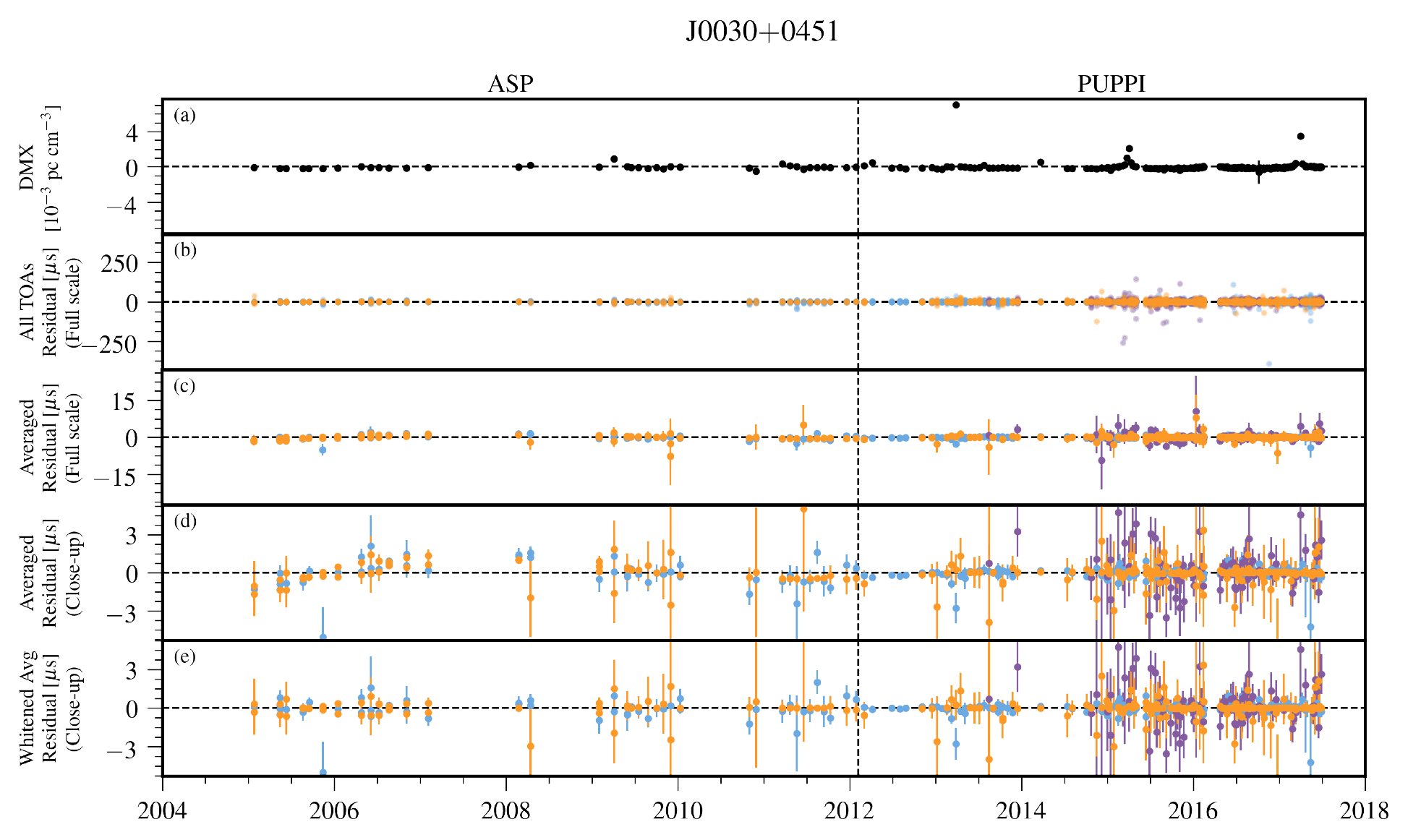}
\caption{Timing residuals and DM variations for PSR J0030$+$0451. See appendix~\ref{sec:resid} text for details.  In residual plots, colored points indicate the receiver of each observation: 430~MHz (Orange), 1.4~GHz (Light blue), and 2.1~GHz (Purple).  (a) Variations in DMX.  (b) Residual arrival times for all TOAs.  Points are semi-transparent; dark regions arise from the overlap of many points.  (c,d) Average residual arrival times shown full scale (panel c) and close-up of low residuals (panel d).  (e) Whitened average residual arrival times after removing the red noise model (close-up of low residuals).}
\label{fig:summary-J0030+0451}
\end{figure*}

\begin{figure*}[p]
\centering
\includegraphics[scale=0.76]{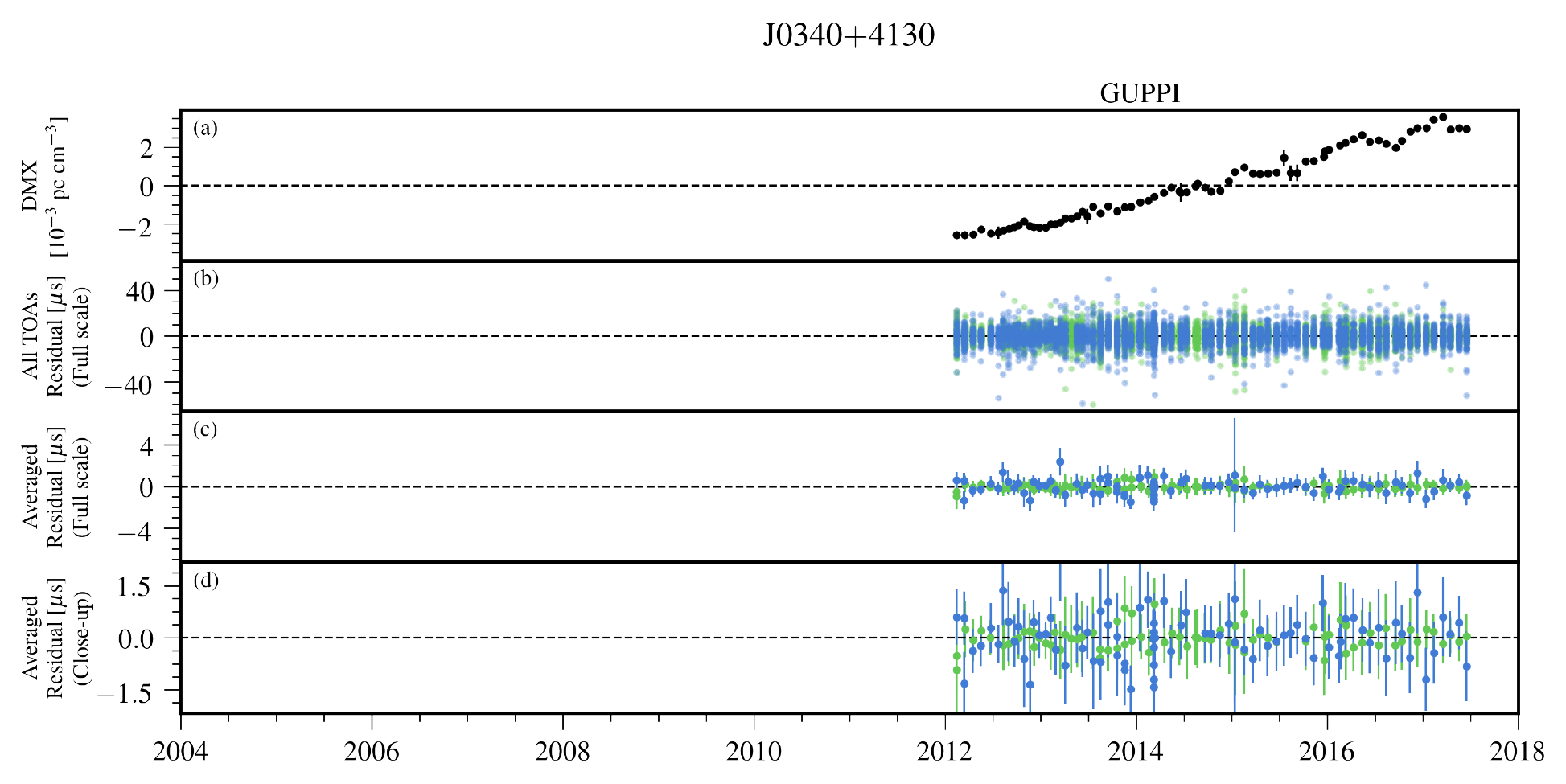}
\caption{Timing residuals and DM variations for PSR J0340$+$4130. See appendix~\ref{sec:resid} text for details.  In residual plots, colored points indicate the receiver of each observation: 820~MHz (Green) and 1.4~GHz (Dark blue).  (a) Variations in DMX.  (b) Residual arrival times for all TOAs.  Points are semi-transparent; dark regions arise from the overlap of many points.  (c,d) Average residual arrival times shown full scale (panel c) and close-up of low residuals (panel d). }
\label{fig:summary-J0340+4130}
\end{figure*}

\begin{figure*}[p]
\centering
\includegraphics[scale=0.76]{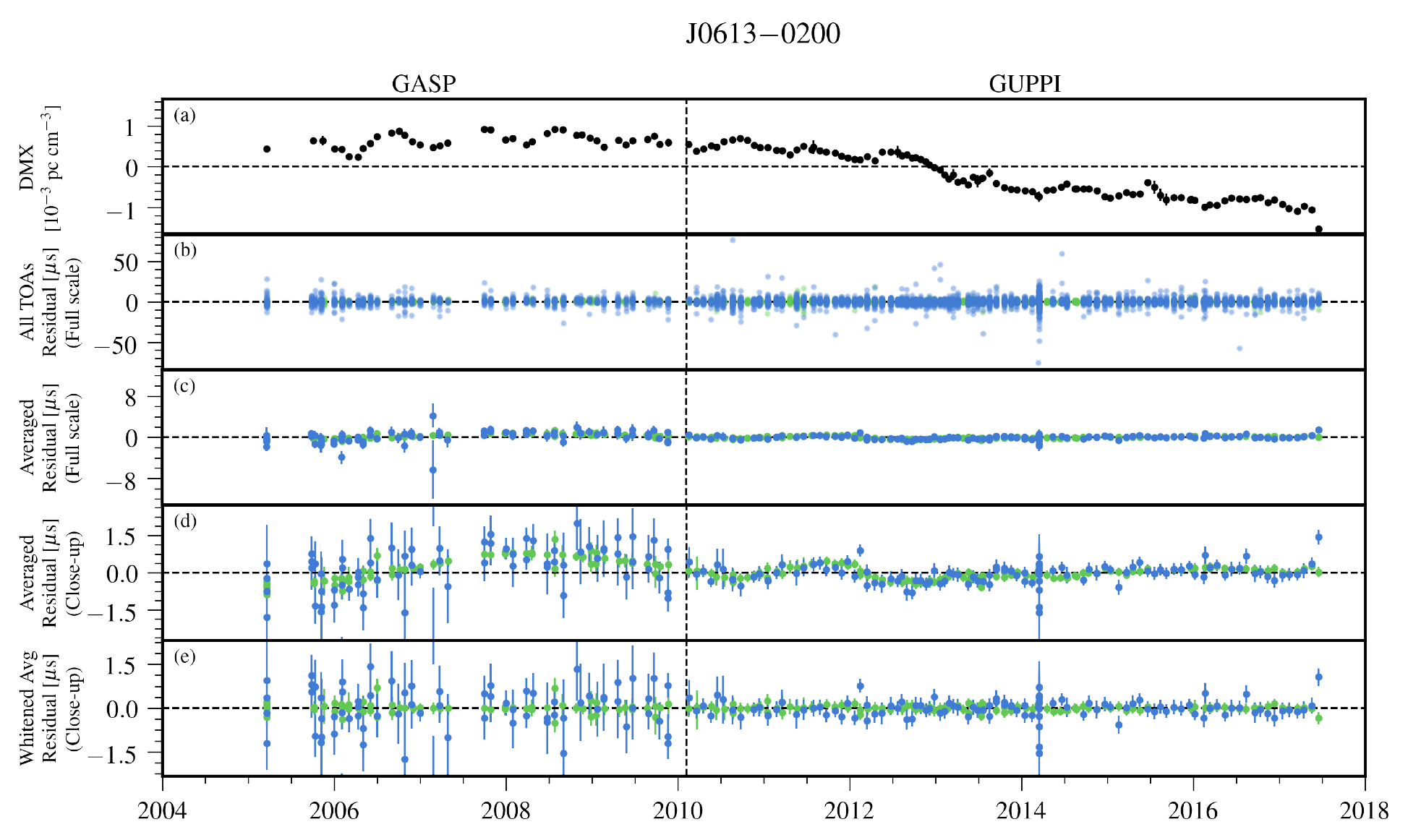}
\caption{Timing residuals and DM variations for PSR J0613$-$0200. See appendix~\ref{sec:resid} text for details.  In residual plots, colored points indicate the receiver of each observation: 820~MHz (Green) and 1.4~GHz (Dark blue).  (a) Variations in DMX.  (b) Residual arrival times for all TOAs.  Points are semi-transparent; dark regions arise from the overlap of many points.  (c,d) Average residual arrival times shown full scale (panel c) and close-up of low residuals (panel d).  (e) Whitened average residual arrival times after removing the red noise model (close-up of low residuals).}
\label{fig:summary-J0613-0200}
\end{figure*}

\begin{figure*}[p]
\centering
\includegraphics[scale=0.76]{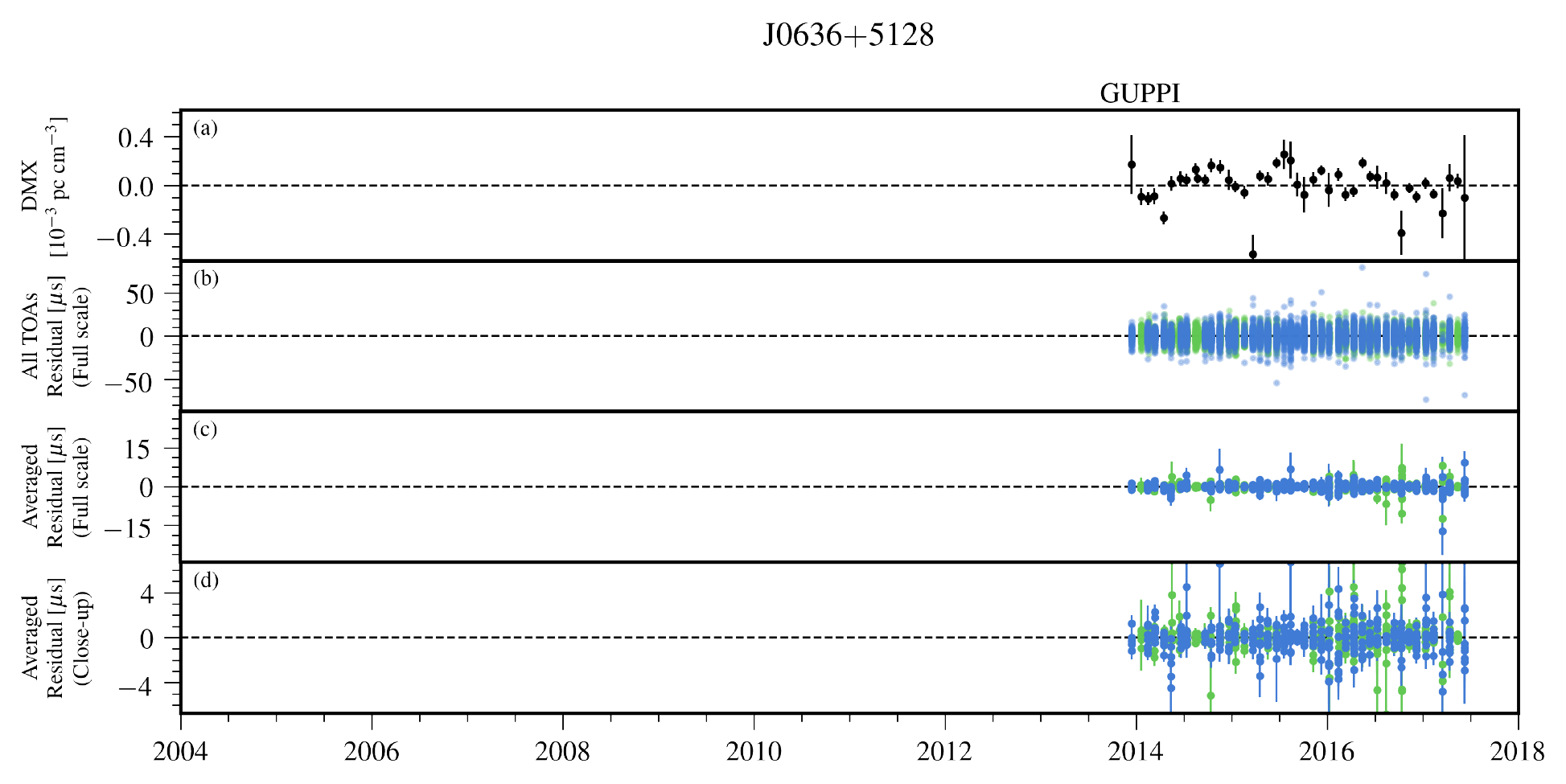}
\caption{Timing residuals and DM variations for PSR J0636$+$5128. See appendix~\ref{sec:resid} text for details.  In residual plots, colored points indicate the receiver of each observation: 820~MHz (Green) and 1.4~GHz (Dark blue).  (a) Variations in DMX.  (b) Residual arrival times for all TOAs.  Points are semi-transparent; dark regions arise from the overlap of many points.  (c,d) Average residual arrival times shown full scale (panel c) and close-up of low residuals (panel d). }
\label{fig:summary-J0636+5128}
\end{figure*}

\begin{figure*}[p]
\centering
\includegraphics[scale=0.76]{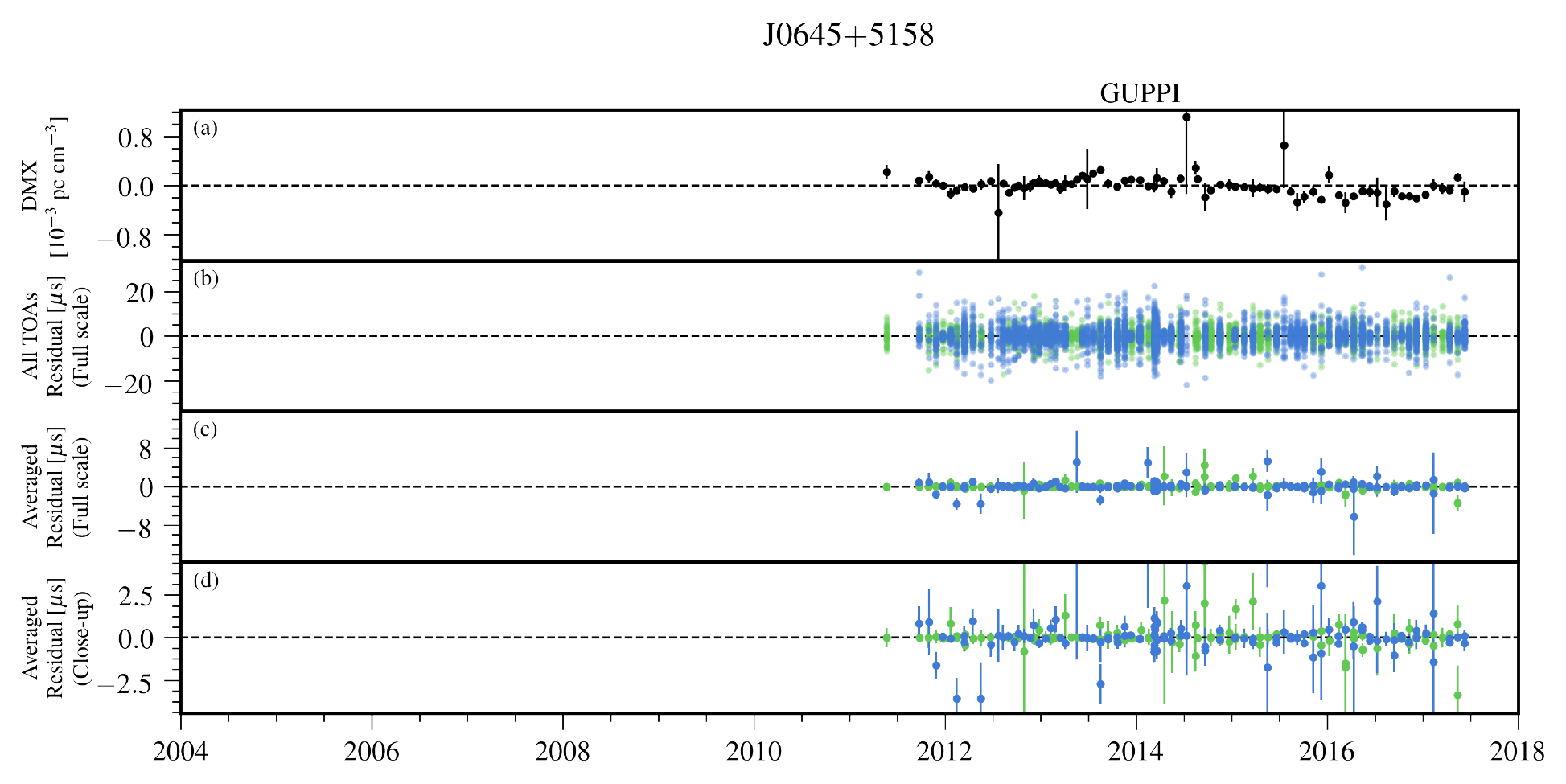}
\caption{Timing residuals and DM variations for PSR J0645$+$5158. See appendix~\ref{sec:resid} text for details.  In residual plots, colored points indicate the receiver of each observation: 820~MHz (Green) and 1.4~GHz (Dark blue).  (a) Variations in DMX.  (b) Residual arrival times for all TOAs.  Points are semi-transparent; dark regions arise from the overlap of many points.  (c,d) Average residual arrival times shown full scale (panel c) and close-up of low residuals (panel d). }
\label{fig:summary-J0645+5158}
\end{figure*}

\begin{figure*}[p]
\centering
\includegraphics[scale=0.76]{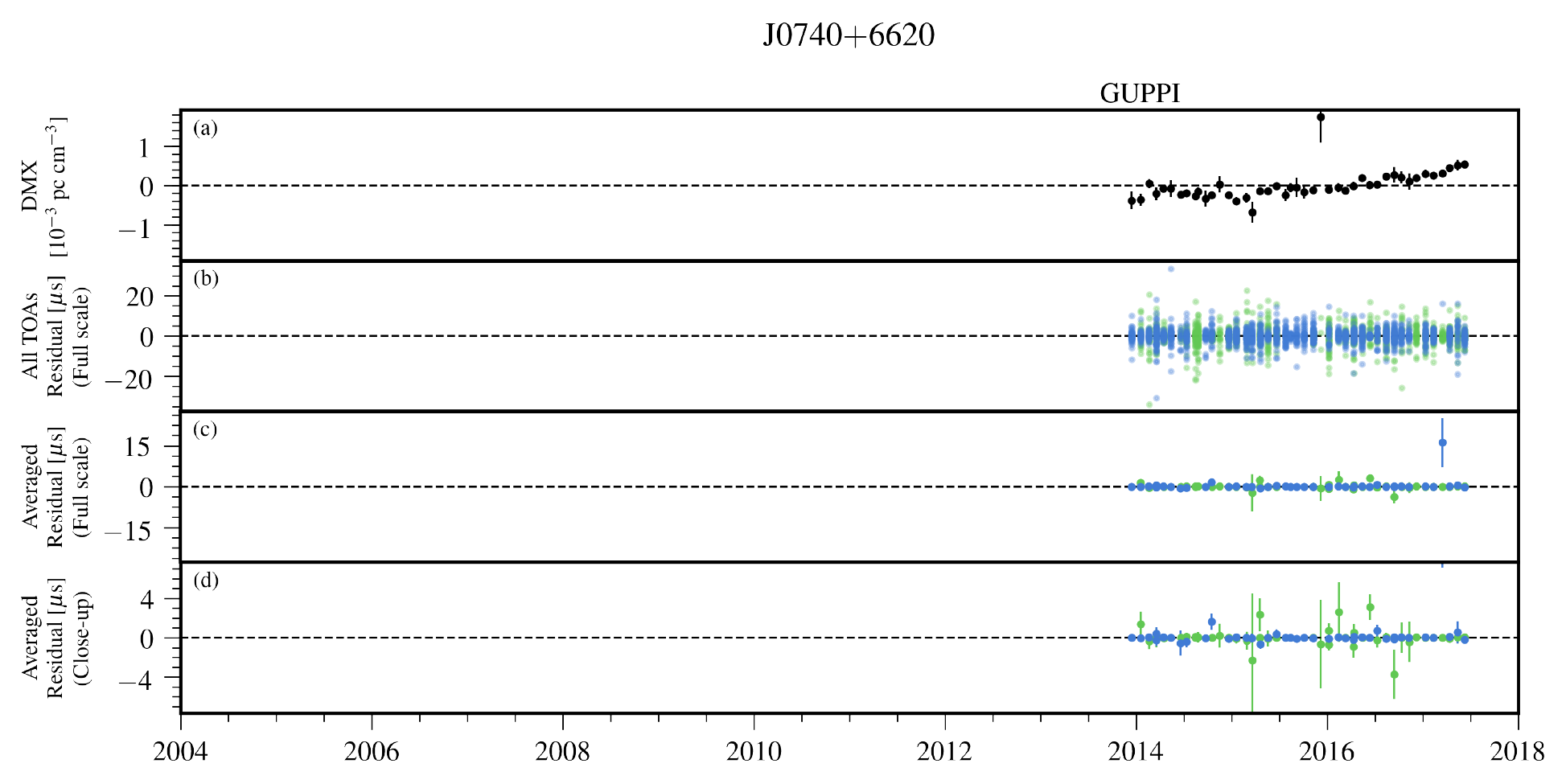}
\caption{Timing residuals and DM variations for PSR J0740$+$6620. See appendix~\ref{sec:resid} text for details.  In residual plots, colored points indicate the receiver of each observation: 820~MHz (Green) and 1.4~GHz (Dark blue).  (a) Variations in DMX.  (b) Residual arrival times for all TOAs.  Points are semi-transparent; dark regions arise from the overlap of many points.  (c,d) Average residual arrival times shown full scale (panel c) and close-up of low residuals (panel d). }
\label{fig:summary-J0740+6620}
\end{figure*}

\begin{figure*}[p]
\centering
\includegraphics[scale=0.76]{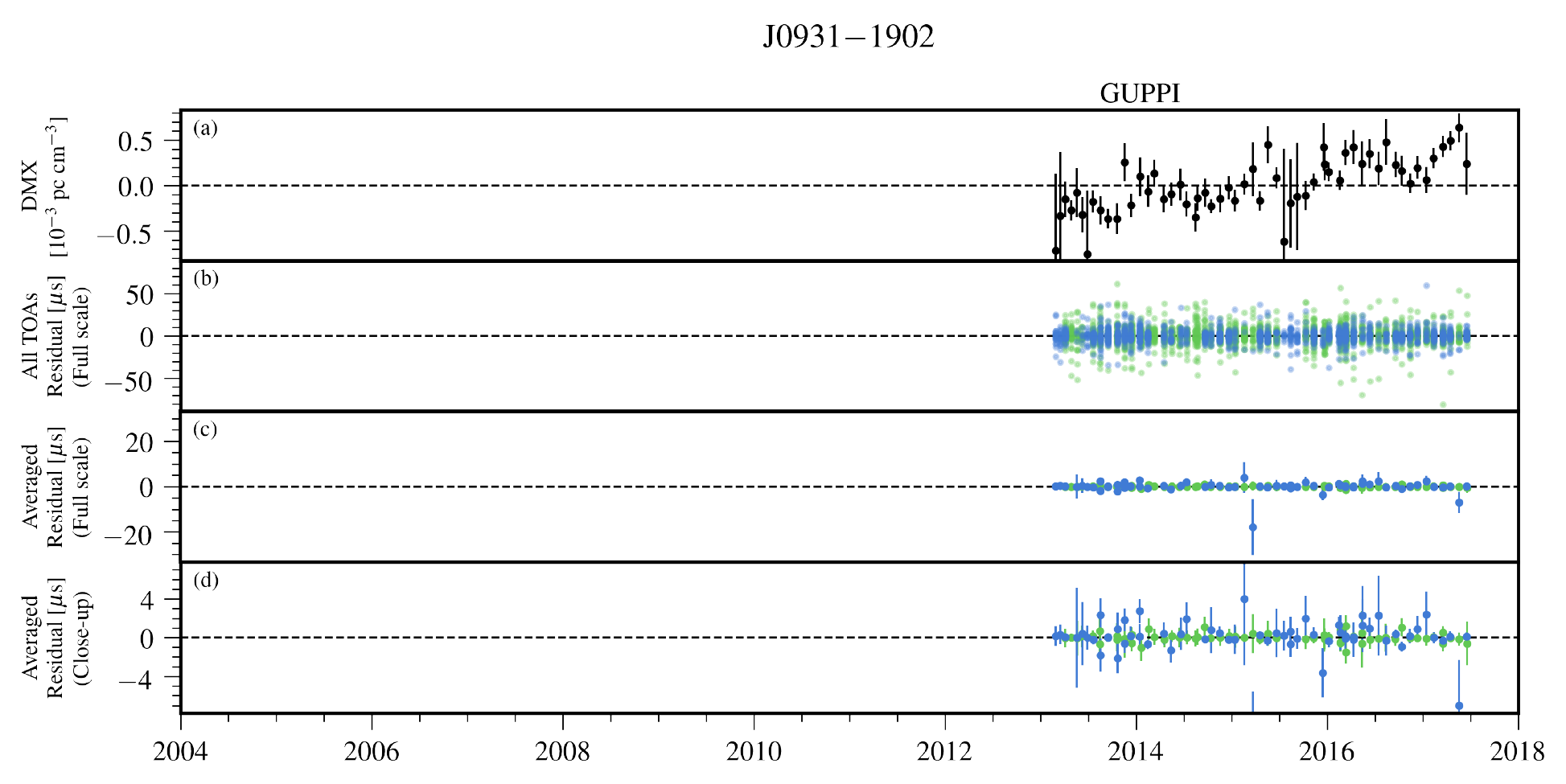}
\caption{Timing residuals and DM variations for PSR J0931$-$1902. See appendix~\ref{sec:resid} text for details.  In residual plots, colored points indicate the receiver of each observation: 820~MHz (Green) and 1.4~GHz (Dark blue).  (a) Variations in DMX.  (b) Residual arrival times for all TOAs.  Points are semi-transparent; dark regions arise from the overlap of many points.  (c,d) Average residual arrival times shown full scale (panel c) and close-up of low residuals (panel d). }
\label{fig:summary-J0931-1902}
\end{figure*}

\begin{figure*}[p]
\centering
\includegraphics[scale=0.76]{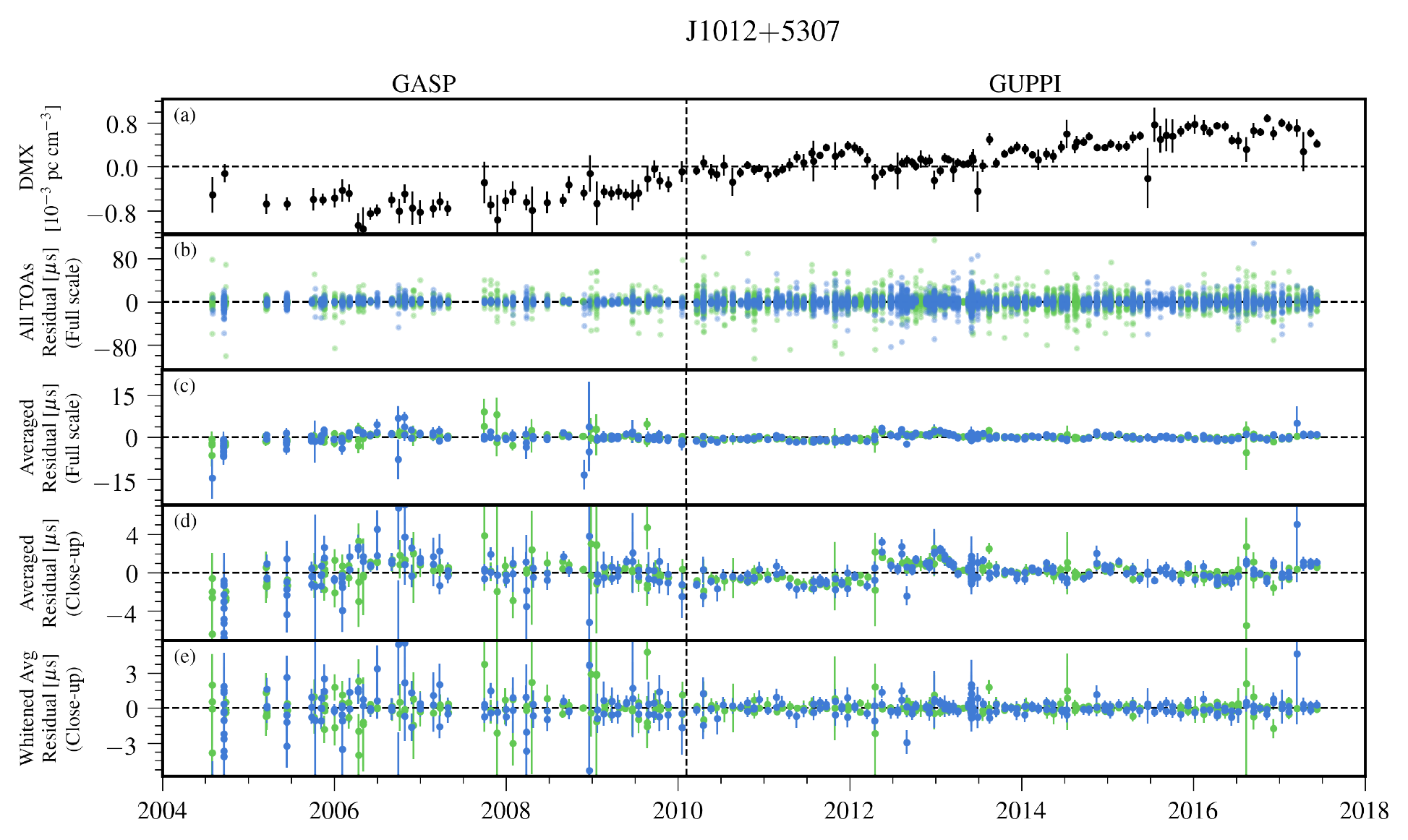}
\caption{Timing residuals and DM variations for PSR J1012$+$5307. See appendix~\ref{sec:resid} text for details.  In residual plots, colored points indicate the receiver of each observation: 820~MHz (Green) and 1.4~GHz (Dark blue).  (a) Variations in DMX.  (b) Residual arrival times for all TOAs.  Points are semi-transparent; dark regions arise from the overlap of many points.  (c,d) Average residual arrival times shown full scale (panel c) and close-up of low residuals (panel d).  (e) Whitened average residual arrival times after removing the red noise model (close-up of low residuals).}
\label{fig:summary-J1012+5307}
\end{figure*}

\begin{figure*}[p]
\centering
\includegraphics[scale=0.76]{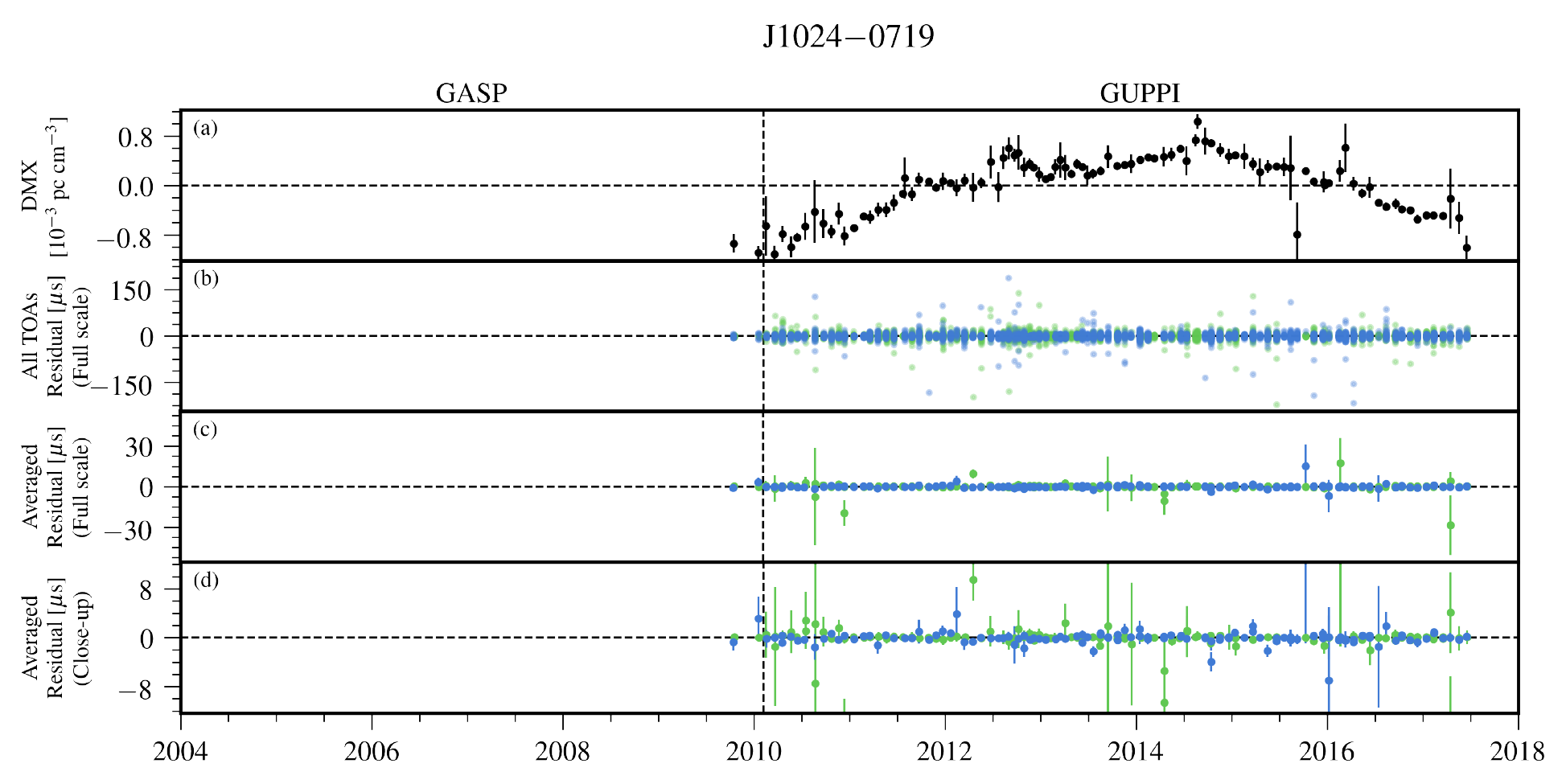}
\caption{Timing residuals and DM variations for PSR J1024$-$0719. See appendix~\ref{sec:resid} text for details.  In residual plots, colored points indicate the receiver of each observation: 820~MHz (Green) and 1.4~GHz (Dark blue).  (a) Variations in DMX.  (b) Residual arrival times for all TOAs.  Points are semi-transparent; dark regions arise from the overlap of many points.  (c,d) Average residual arrival times shown full scale (panel c) and close-up of low residuals (panel d). }
\label{fig:summary-J1024-0719}
\end{figure*}

\begin{figure*}[p]
\centering
\includegraphics[scale=0.76]{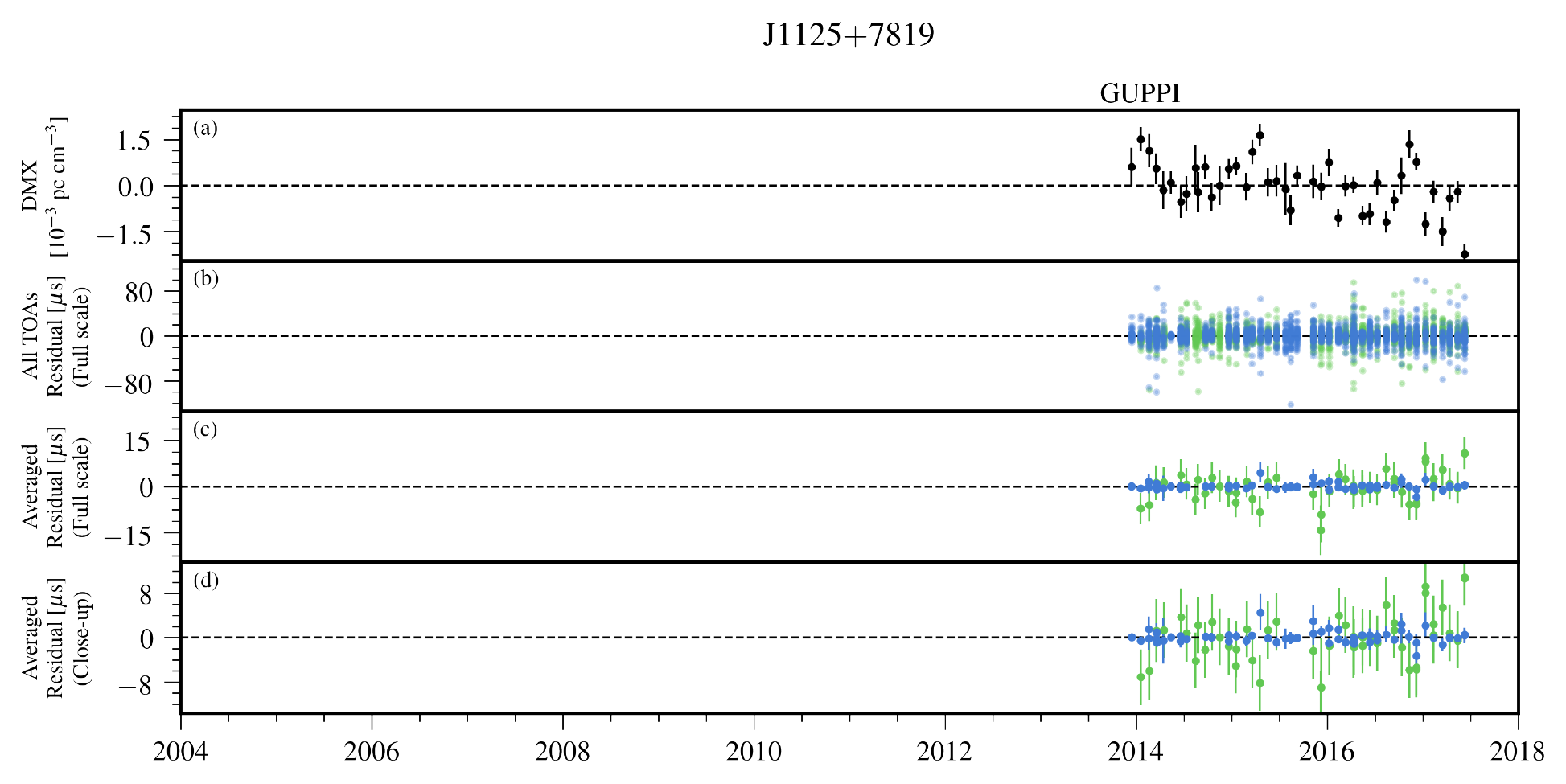}
\caption{Timing residuals and DM variations for PSR J1125$+$7819. See appendix~\ref{sec:resid} text for details.  In residual plots, colored points indicate the receiver of each observation: 820~MHz (Green) and 1.4~GHz (Dark blue).  (a) Variations in DMX.  (b) Residual arrival times for all TOAs.  Points are semi-transparent; dark regions arise from the overlap of many points.  (c,d) Average residual arrival times shown full scale (panel c) and close-up of low residuals (panel d). }
\label{fig:summary-J1125+7819}
\end{figure*}

\begin{figure*}[p]
\centering
\includegraphics[scale=0.76]{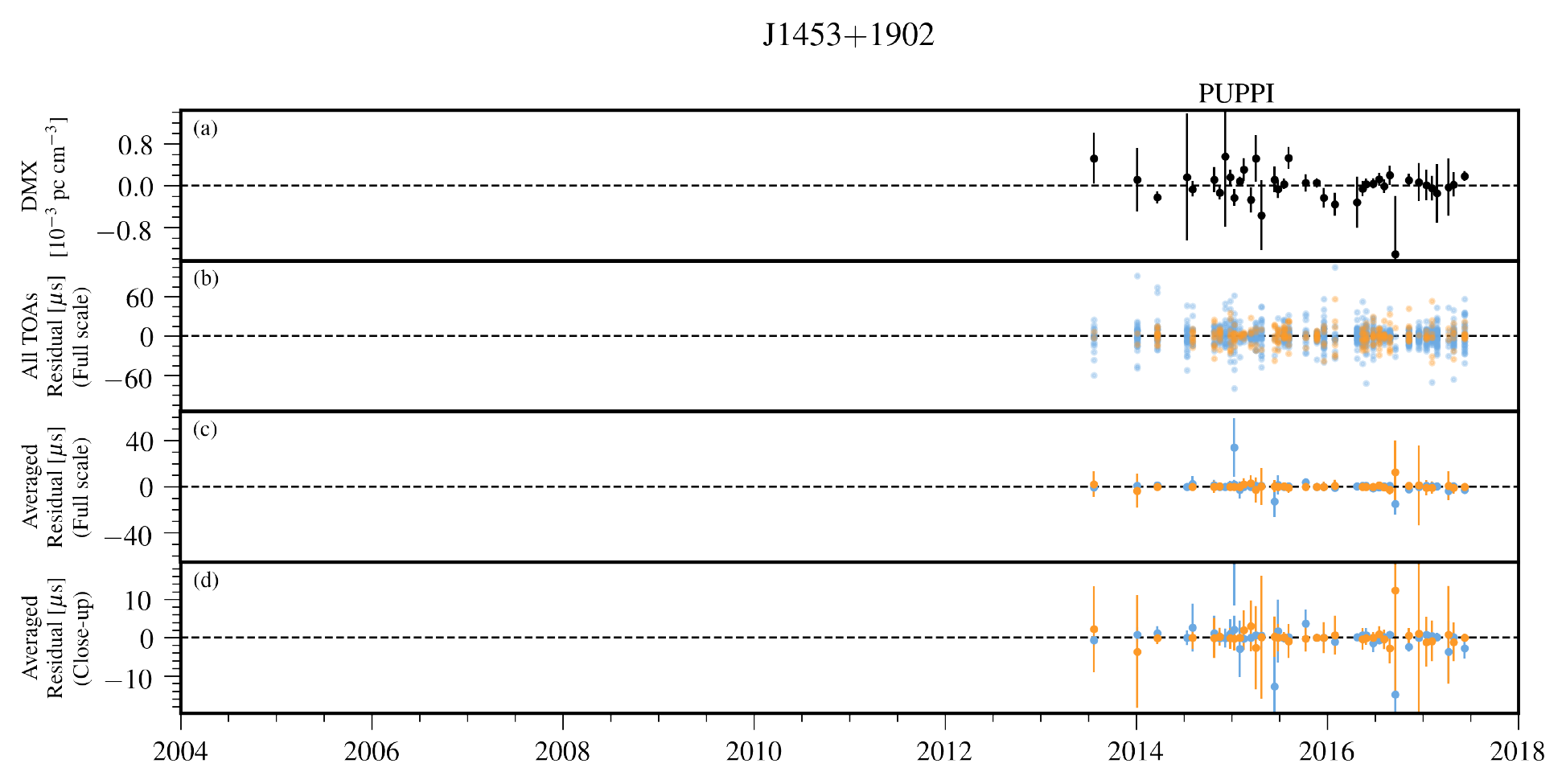}
\caption{Timing residuals and DM variations for PSR J1453$+$1902. See appendix~\ref{sec:resid} text for details.  In residual plots, colored points indicate the receiver of each observation: 430~MHz (Orange) and 1.4~GHz (Light blue).  (a) Variations in DMX.  (b) Residual arrival times for all TOAs.  Points are semi-transparent; dark regions arise from the overlap of many points.  (c,d) Average residual arrival times shown full scale (panel c) and close-up of low residuals (panel d). }
\label{fig:summary-J1453+1902}
\end{figure*}

\begin{figure*}[p]
\centering
\includegraphics[scale=0.76]{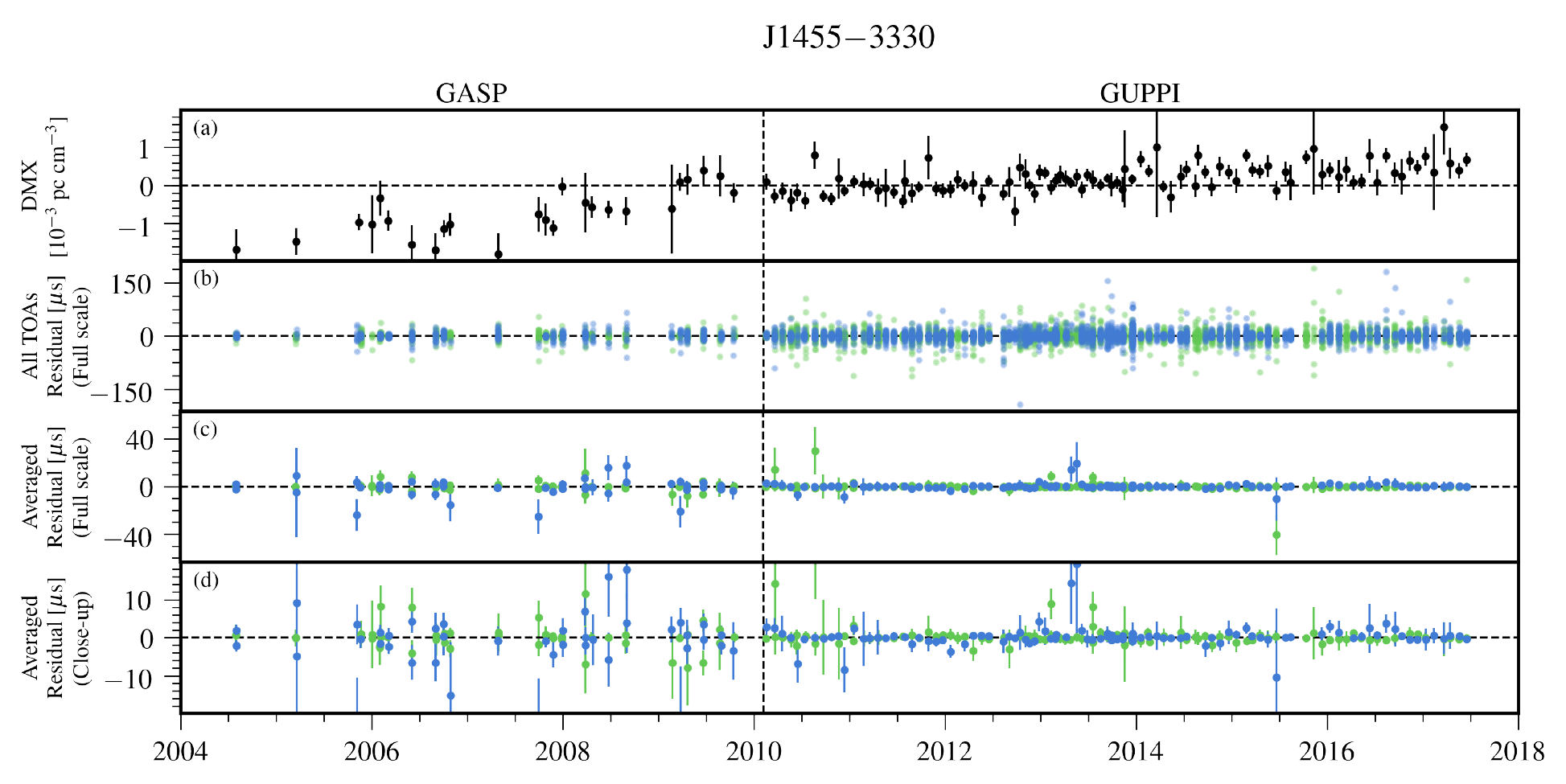}
\caption{Timing residuals and DM variations for PSR J1455$-$3330. See appendix~\ref{sec:resid} text for details.  In residual plots, colored points indicate the receiver of each observation: 820~MHz (Green) and 1.4~GHz (Dark blue).  (a) Variations in DMX.  (b) Residual arrival times for all TOAs.  Points are semi-transparent; dark regions arise from the overlap of many points.  (c,d) Average residual arrival times shown full scale (panel c) and close-up of low residuals (panel d). }
\label{fig:summary-J1455-3330}
\end{figure*}

\begin{figure*}[p]
\centering
\includegraphics[scale=0.76]{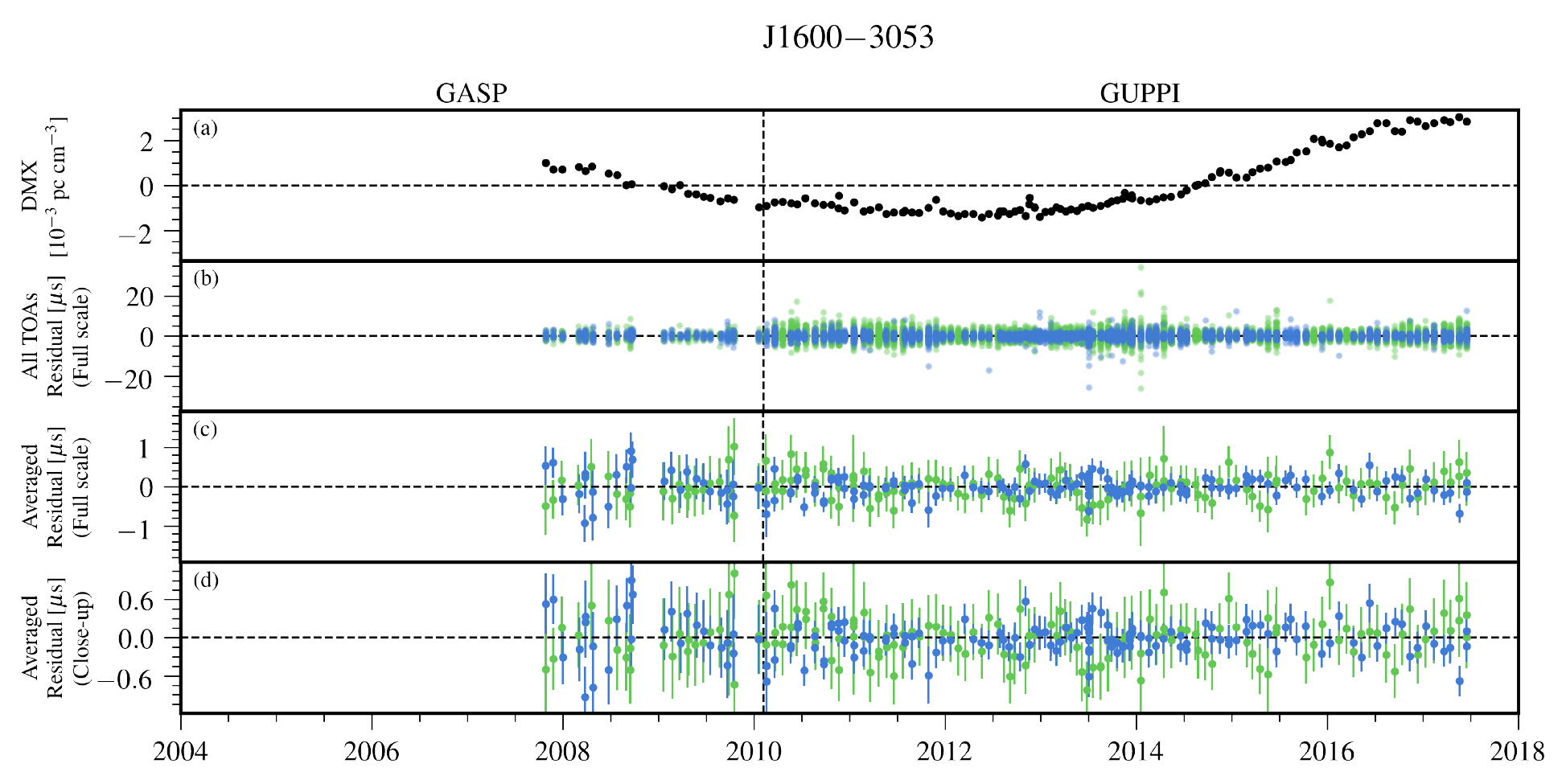}
\caption{Timing residuals and DM variations for PSR J1600$-$3053. See appendix~\ref{sec:resid} text for details.  In residual plots, colored points indicate the receiver of each observation: 820~MHz (Green) and 1.4~GHz (Dark blue).  (a) Variations in DMX.  (b) Residual arrival times for all TOAs.  Points are semi-transparent; dark regions arise from the overlap of many points.  (c,d) Average residual arrival times shown full scale (panel c) and close-up of low residuals (panel d). }
\label{fig:summary-J1600-3053}
\end{figure*}

\begin{figure*}[p]
\centering
\includegraphics[scale=0.76]{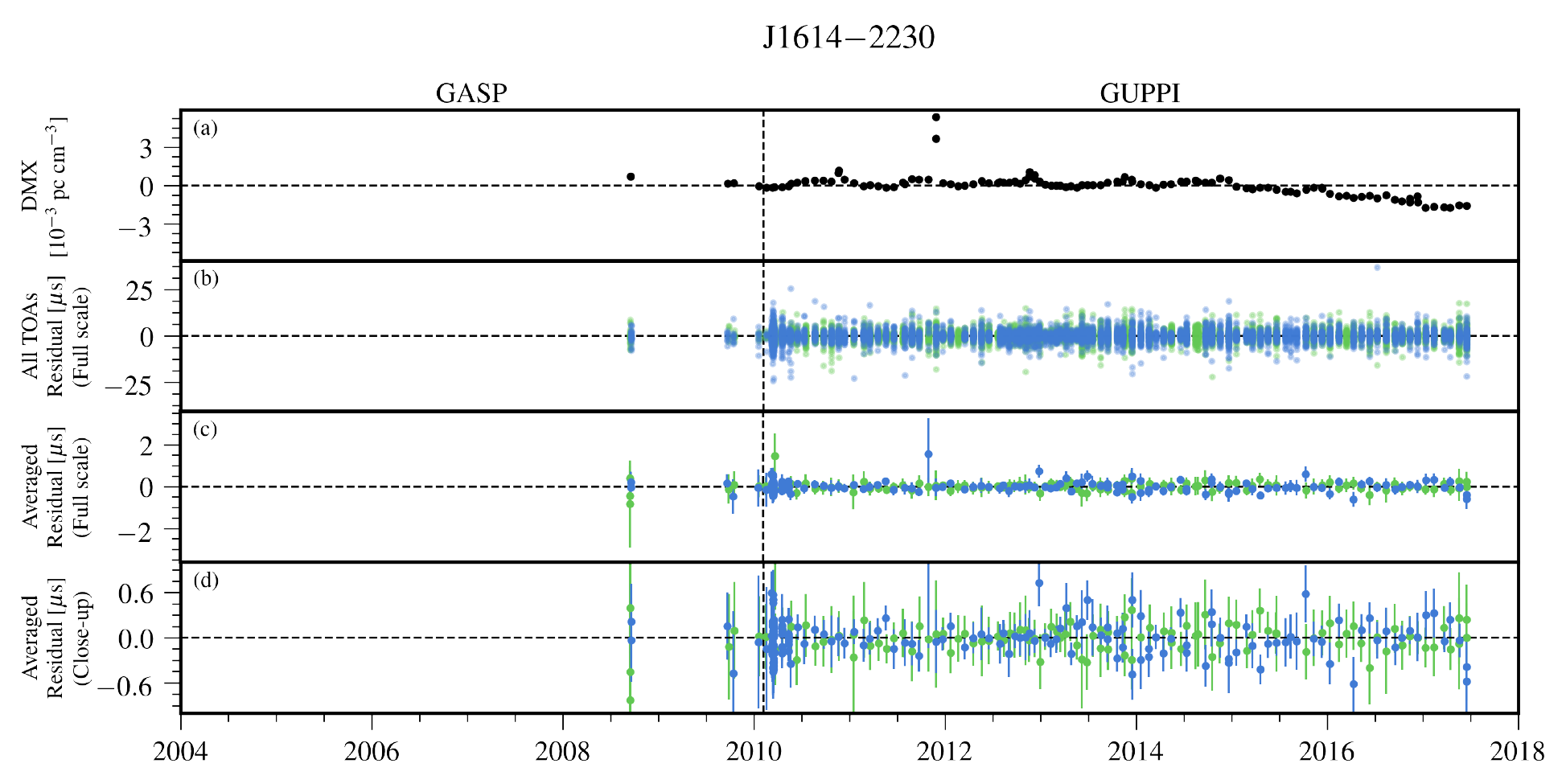}
\caption{Timing residuals and DM variations for PSR J1614$-$2230. See appendix~\ref{sec:resid} text for details.  In residual plots, colored points indicate the receiver of each observation: 820~MHz (Green) and 1.4~GHz (Dark blue).  (a) Variations in DMX.  (b) Residual arrival times for all TOAs.  Points are semi-transparent; dark regions arise from the overlap of many points.  (c,d) Average residual arrival times shown full scale (panel c) and close-up of low residuals (panel d). }
\label{fig:summary-J1614-2230}
\end{figure*}

\begin{figure*}[p]
\centering
\includegraphics[scale=0.76]{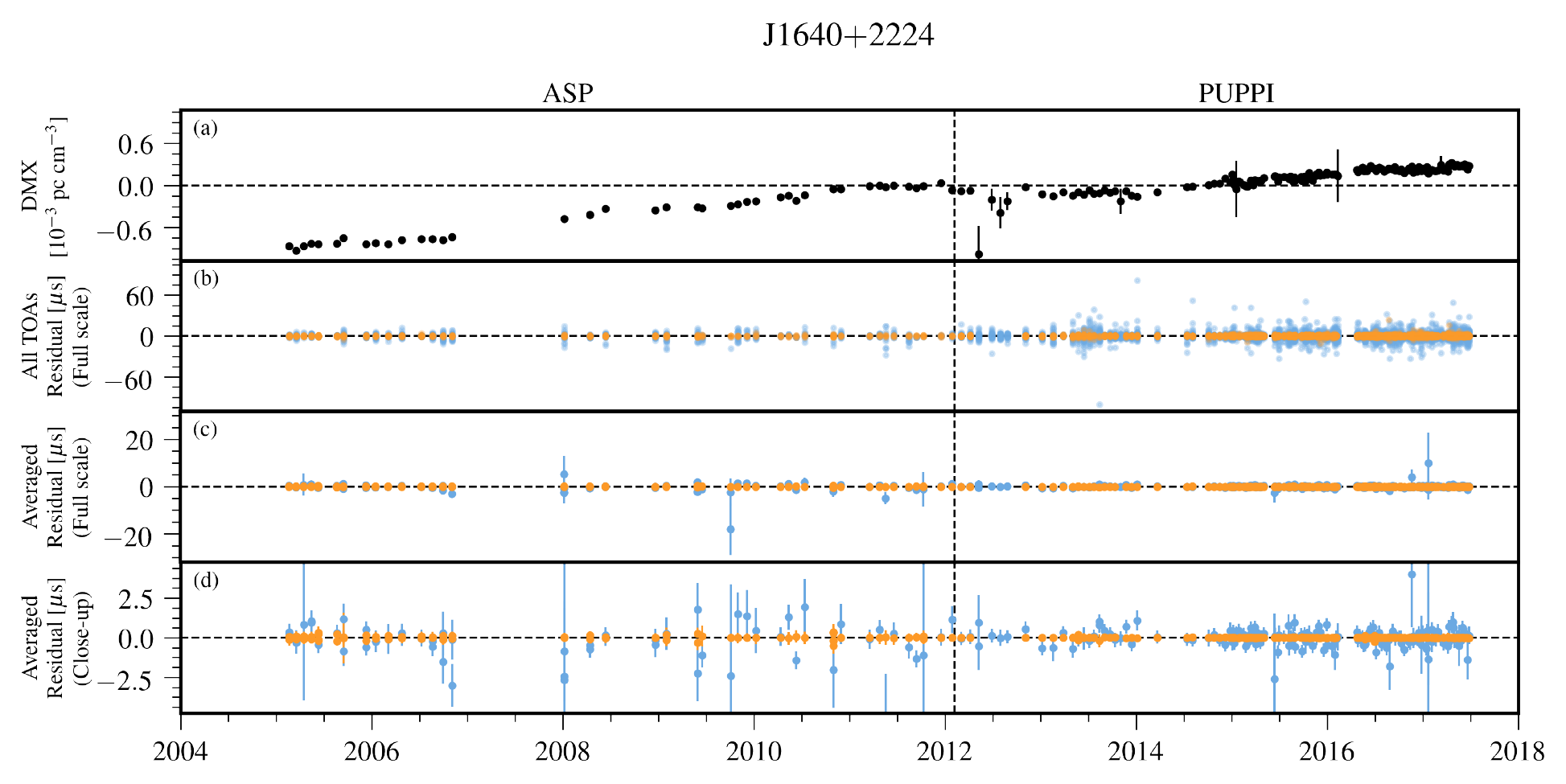}
\caption{Timing residuals and DM variations for PSR J1640$+$2224. See appendix~\ref{sec:resid} text for details.  In residual plots, colored points indicate the receiver of each observation: 430~MHz (Orange) and 1.4~GHz (Light blue).  (a) Variations in DMX.  (b) Residual arrival times for all TOAs.  Points are semi-transparent; dark regions arise from the overlap of many points.  (c,d) Average residual arrival times shown full scale (panel c) and close-up of low residuals (panel d). }
\label{fig:summary-J1640+2224}
\end{figure*}

\begin{figure*}[p]
\centering
\includegraphics[scale=0.76]{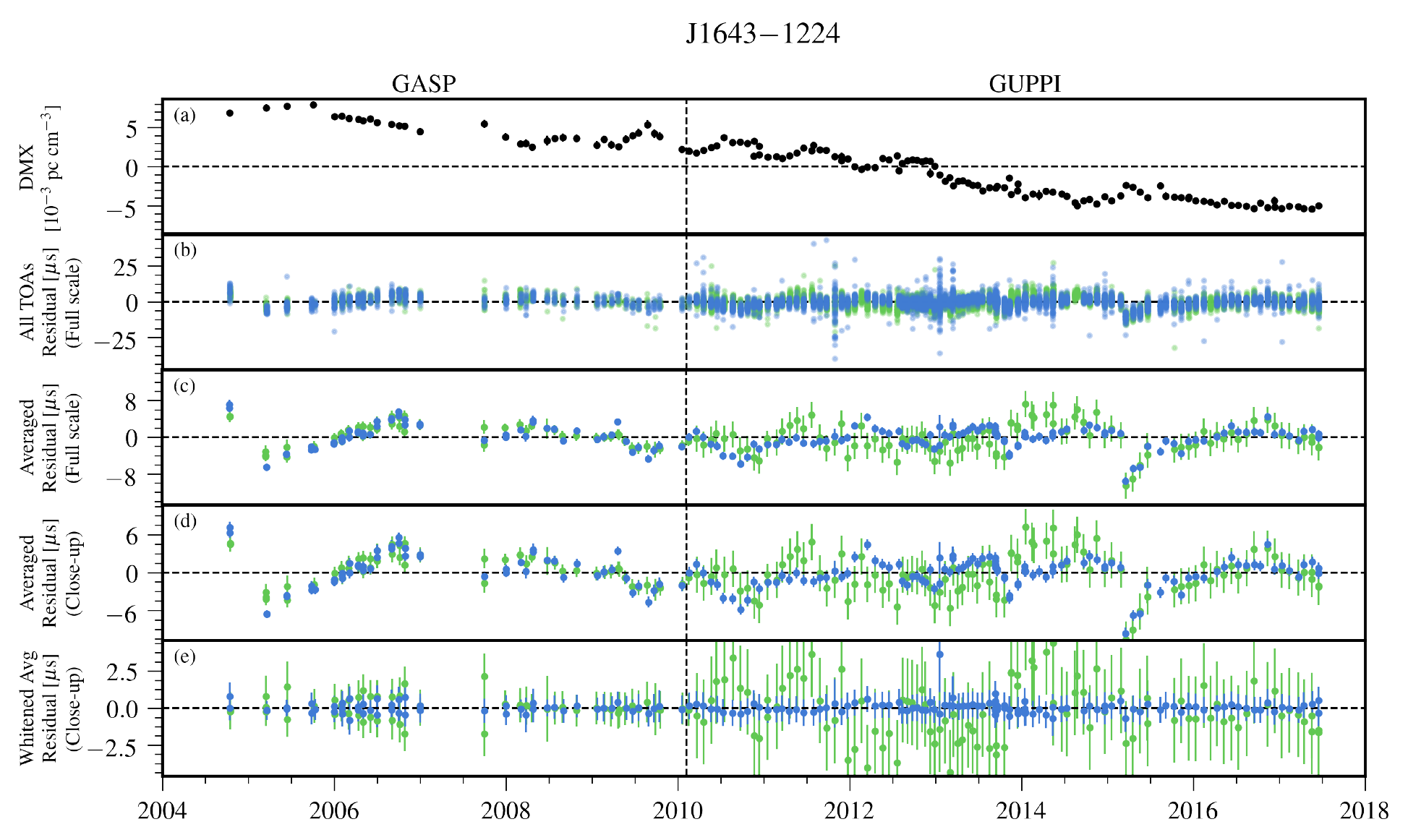}
\caption{Timing residuals and DM variations for PSR J1643$-$1224. See appendix~\ref{sec:resid} text for details.  In residual plots, colored points indicate the receiver of each observation: 820~MHz (Green) and 1.4~GHz (Dark blue).  (a) Variations in DMX.  (b) Residual arrival times for all TOAs.  Points are semi-transparent; dark regions arise from the overlap of many points.  (c,d) Average residual arrival times shown full scale (panel c) and close-up of low residuals (panel d).  (e) Whitened average residual arrival times after removing the red noise model (close-up of low residuals).}
\label{fig:summary-J1643-1224}
\end{figure*}

\begin{figure*}[p]
\centering
\includegraphics[scale=0.76]{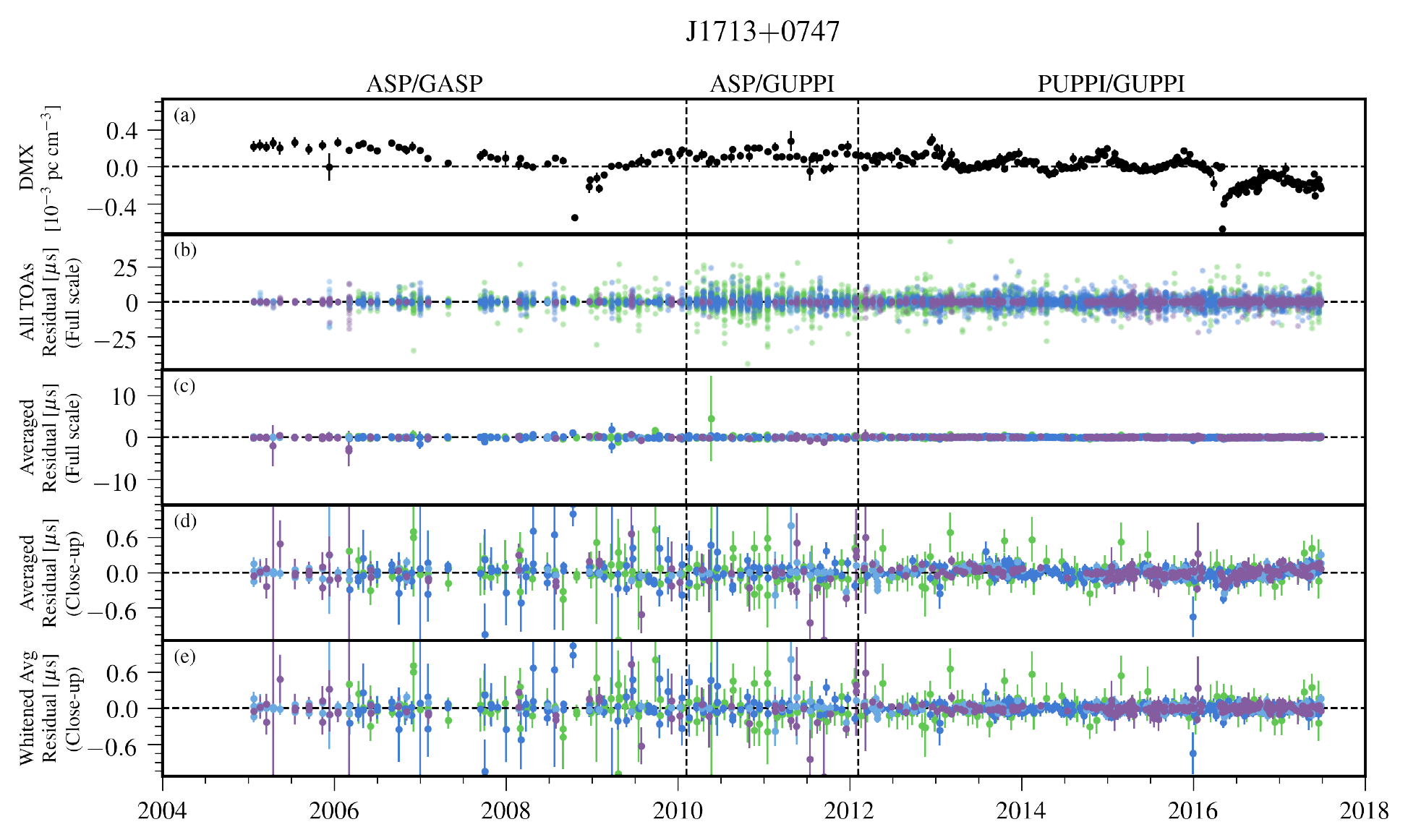}
\caption{Timing residuals and DM variations for PSR J1713$+$0747. See appendix~\ref{sec:resid} text for details.  In residual plots, colored points indicate the receiver of each observation: 820~MHz (Green), 1.4~GHz (Dark blue), 1.4~GHz (Light blue), and 2.1~GHz (Purple).  (a) Variations in DMX.  (b) Residual arrival times for all TOAs.  Points are semi-transparent; dark regions arise from the overlap of many points.  (c,d) Average residual arrival times shown full scale (panel c) and close-up of low residuals (panel d).  (e) Whitened average residual arrival times after removing the red noise model (close-up of low residuals).}
\label{fig:summary-J1713+0747}
\end{figure*}

\begin{figure*}[p]
\centering
\includegraphics[scale=0.76]{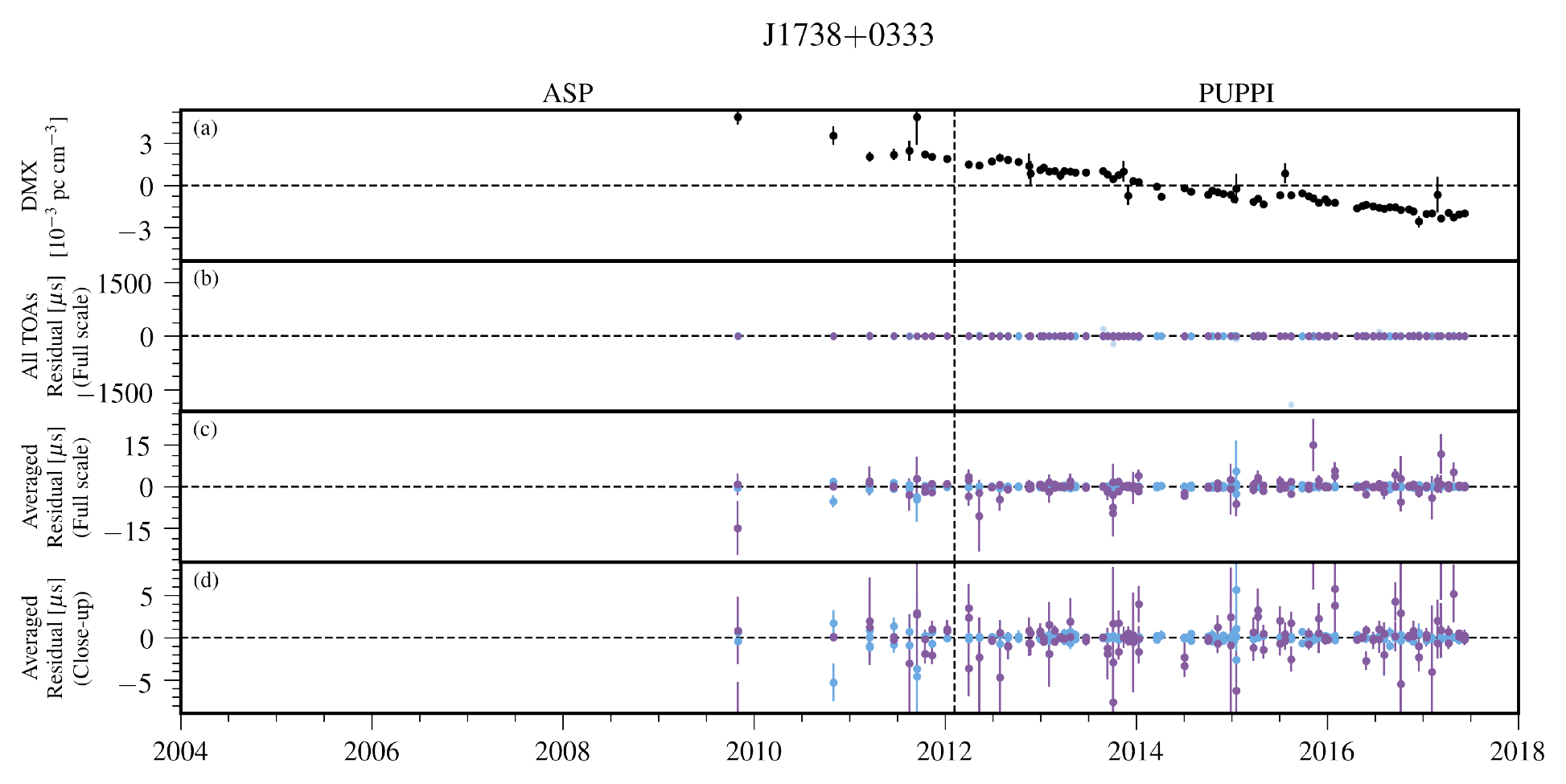}
\caption{Timing residuals and DM variations for PSR J1738$+$0333. See appendix~\ref{sec:resid} text for details.  In residual plots, colored points indicate the receiver of each observation: 1.4~GHz (Light blue) and 2.1~GHz (Purple).  (a) Variations in DMX.  (b) Residual arrival times for all TOAs.  Points are semi-transparent; dark regions arise from the overlap of many points.  (c,d) Average residual arrival times shown full scale (panel c) and close-up of low residuals (panel d). }
\label{fig:summary-J1738+0333}
\end{figure*}

\begin{figure*}[p]
\centering
\includegraphics[scale=0.76]{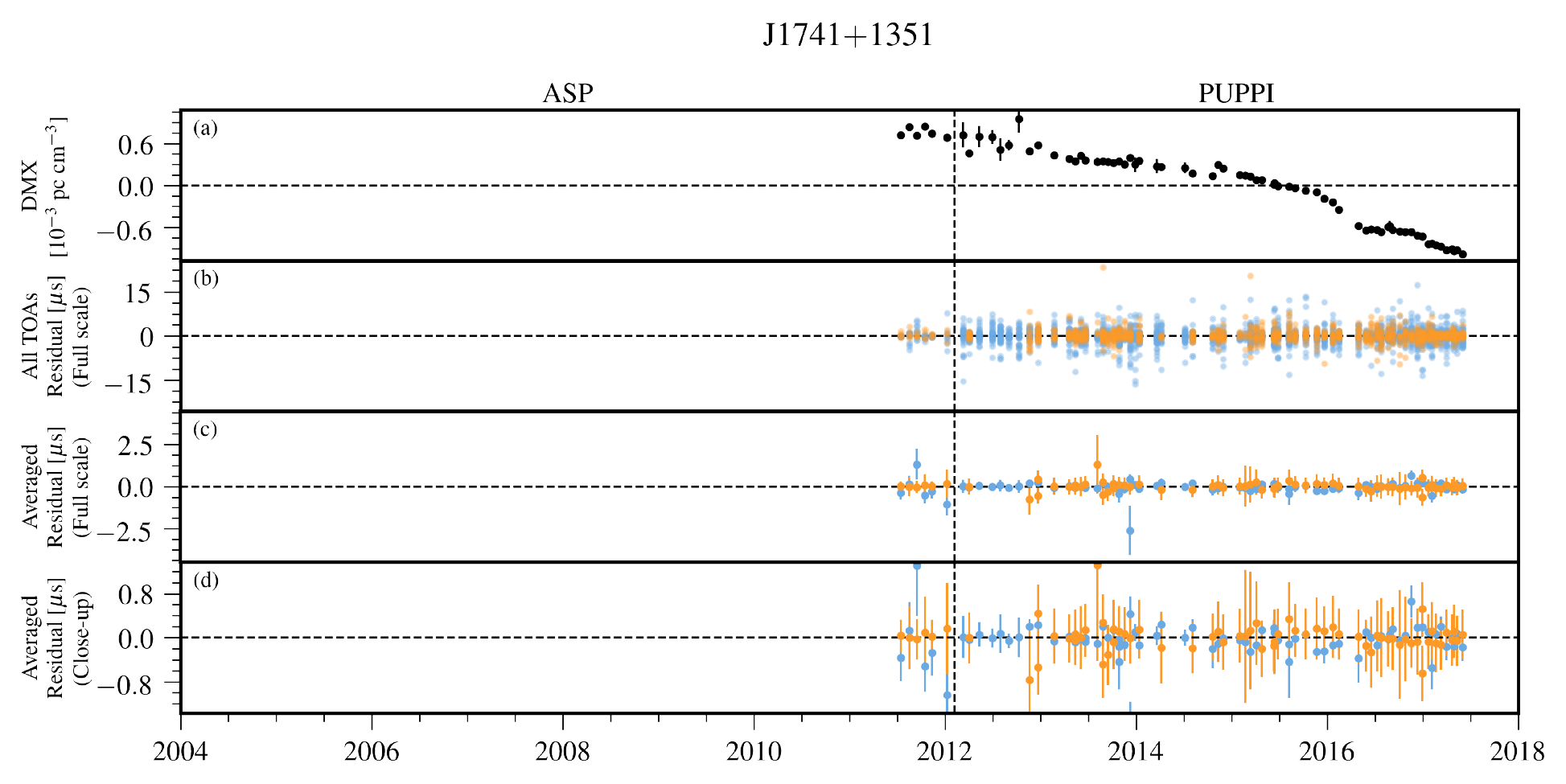}
\caption{Timing residuals and DM variations for PSR J1741$+$1351. See appendix~\ref{sec:resid} text for details.  In residual plots, colored points indicate the receiver of each observation: 430~MHz (Orange) and 1.4~GHz (Light blue).  (a) Variations in DMX.  (b) Residual arrival times for all TOAs.  Points are semi-transparent; dark regions arise from the overlap of many points.  (c,d) Average residual arrival times shown full scale (panel c) and close-up of low residuals (panel d). }
\label{fig:summary-J1741+1351}
\end{figure*}

\begin{figure*}[p]
\centering
\includegraphics[scale=0.76]{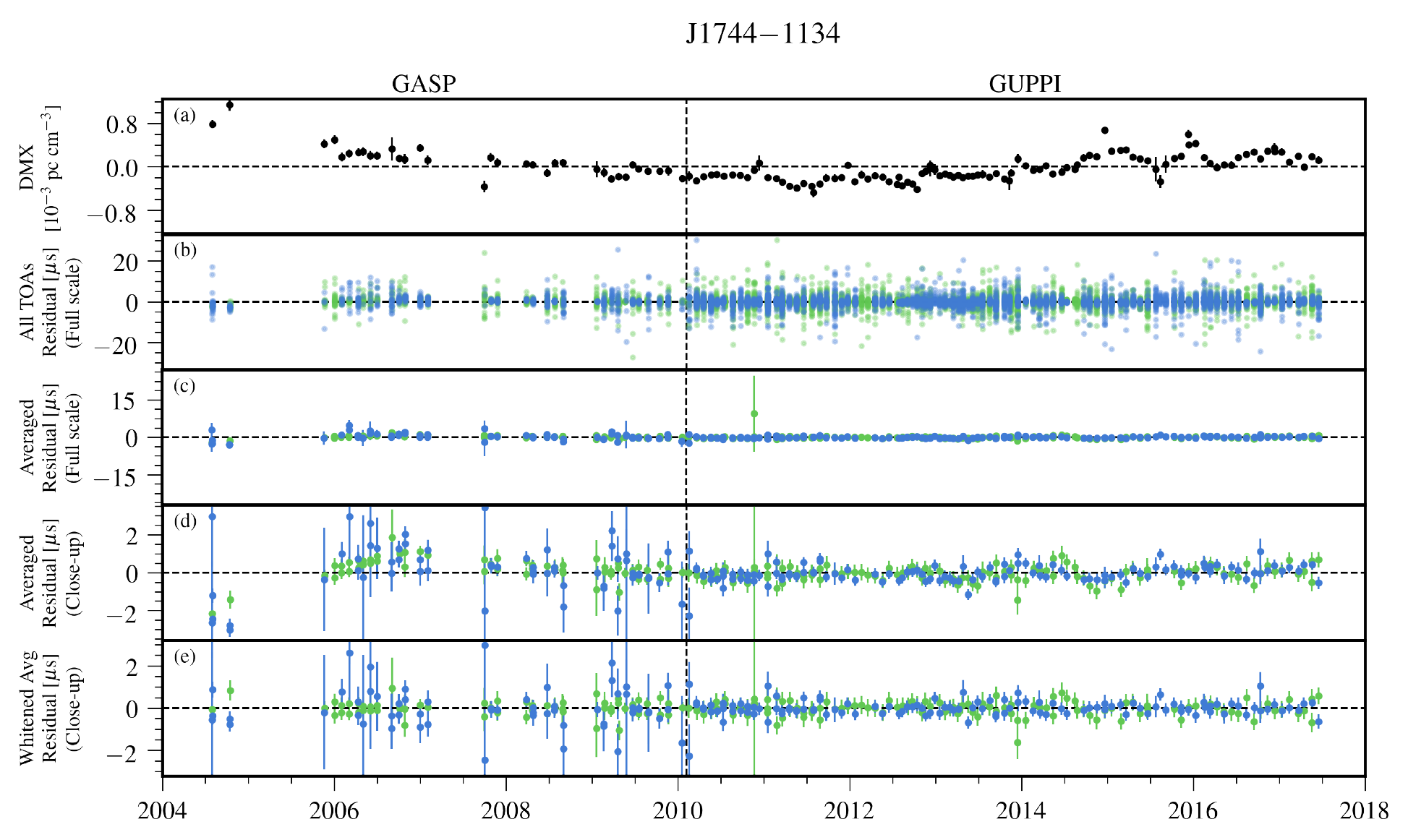}
\caption{Timing residuals and DM variations for PSR J1744$-$1134. See appendix~\ref{sec:resid} text for details.  In residual plots, colored points indicate the receiver of each observation: 820~MHz (Green) and 1.4~GHz (Dark blue).  (a) Variations in DMX.  (b) Residual arrival times for all TOAs.  Points are semi-transparent; dark regions arise from the overlap of many points.  (c,d) Average residual arrival times shown full scale (panel c) and close-up of low residuals (panel d).  (e) Whitened average residual arrival times after removing the red noise model (close-up of low residuals).}
\label{fig:summary-J1744-1134}
\end{figure*}

\begin{figure*}[p]
\centering
\includegraphics[scale=0.76]{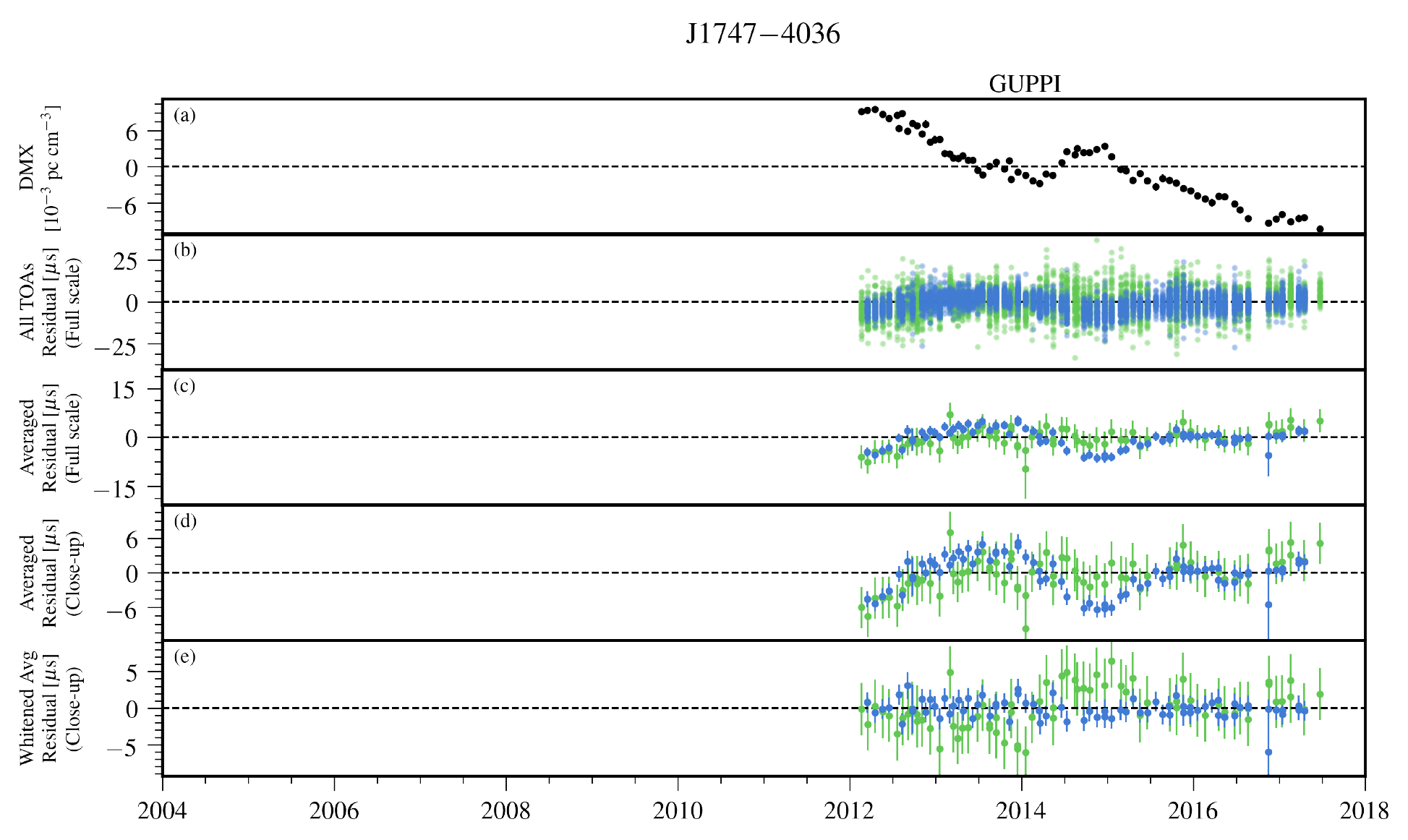}
\caption{Timing residuals and DM variations for PSR J1747$-$4036. See appendix~\ref{sec:resid} text for details.  In residual plots, colored points indicate the receiver of each observation: 820~MHz (Green) and 1.4~GHz (Dark blue).  (a) Variations in DMX.  (b) Residual arrival times for all TOAs.  Points are semi-transparent; dark regions arise from the overlap of many points.  (c,d) Average residual arrival times shown full scale (panel c) and close-up of low residuals (panel d).  (e) Whitened average residual arrival times after removing the red noise model (close-up of low residuals).}
\label{fig:summary-J1747-4036}
\end{figure*}

\begin{figure*}[p]
\centering
\includegraphics[scale=0.76]{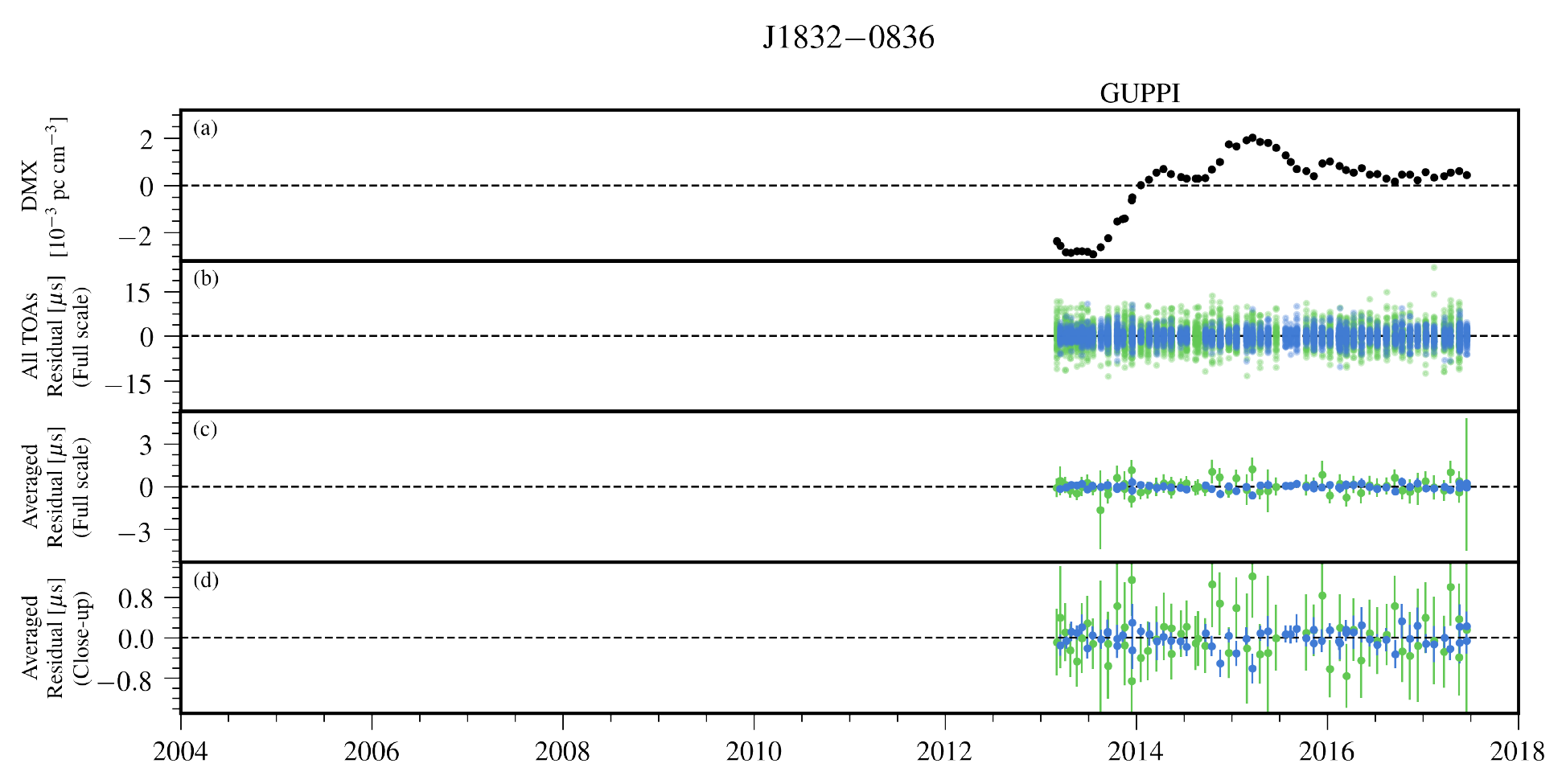}
\caption{Timing residuals and DM variations for PSR J1832$-$0836. See appendix~\ref{sec:resid} text for details.  In residual plots, colored points indicate the receiver of each observation: 820~MHz (Green) and 1.4~GHz (Dark blue).  (a) Variations in DMX.  (b) Residual arrival times for all TOAs.  Points are semi-transparent; dark regions arise from the overlap of many points.  (c,d) Average residual arrival times shown full scale (panel c) and close-up of low residuals (panel d). }
\label{fig:summary-J1832-0836}
\end{figure*}

\begin{figure*}[p]
\centering
\includegraphics[scale=0.76]{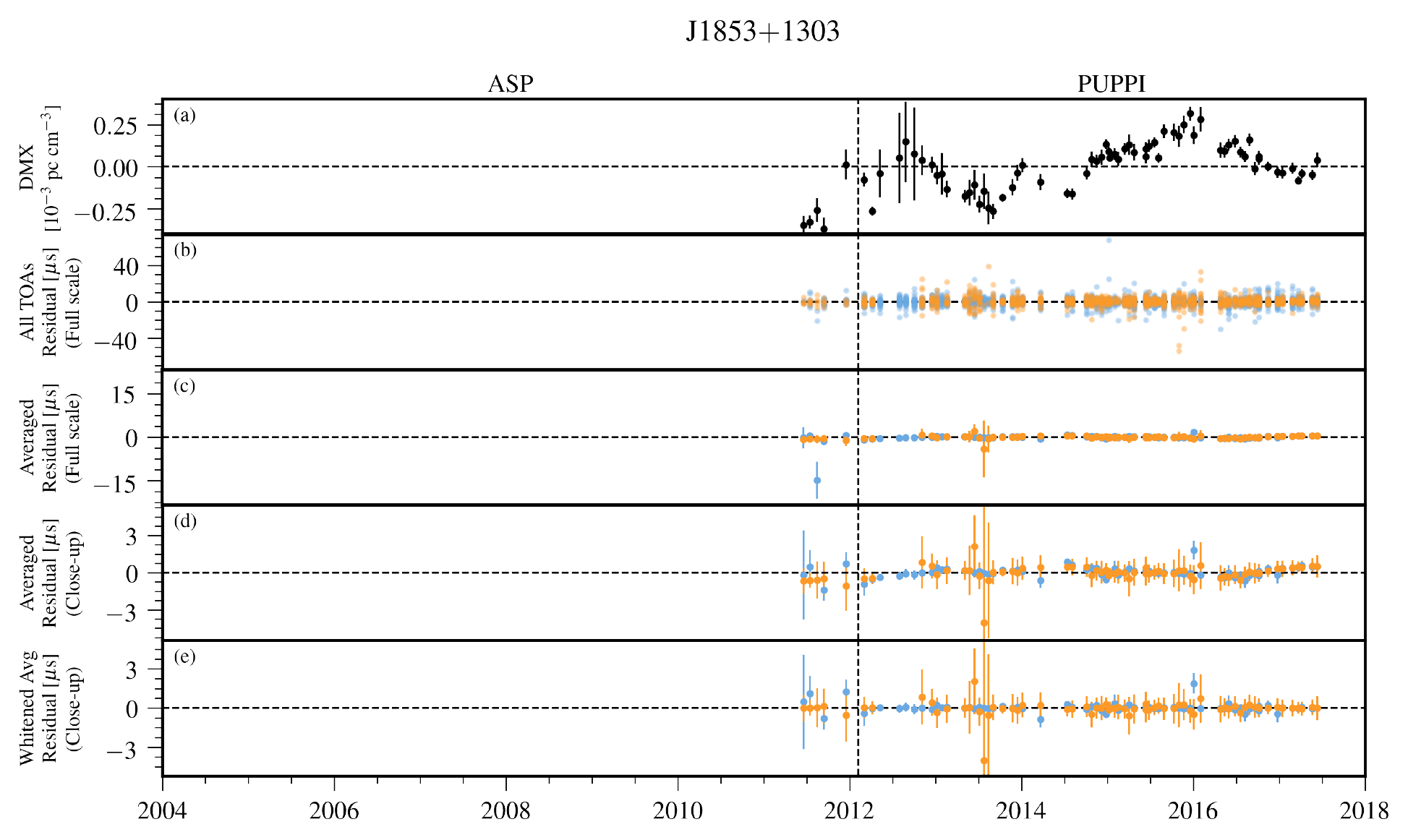}
\caption{Timing residuals and DM variations for PSR J1853$+$1303. See appendix~\ref{sec:resid} text for details.  In residual plots, colored points indicate the receiver of each observation: 430~MHz (Orange) and 1.4~GHz (Light blue).  (a) Variations in DMX.  (b) Residual arrival times for all TOAs.  Points are semi-transparent; dark regions arise from the overlap of many points.  (c,d) Average residual arrival times shown full scale (panel c) and close-up of low residuals (panel d).  (e) Whitened average residual arrival times after removing the red noise model (close-up of low residuals).}
\label{fig:summary-J1853+1303}
\end{figure*}

\begin{figure*}[p]
\centering
\includegraphics[scale=0.76]{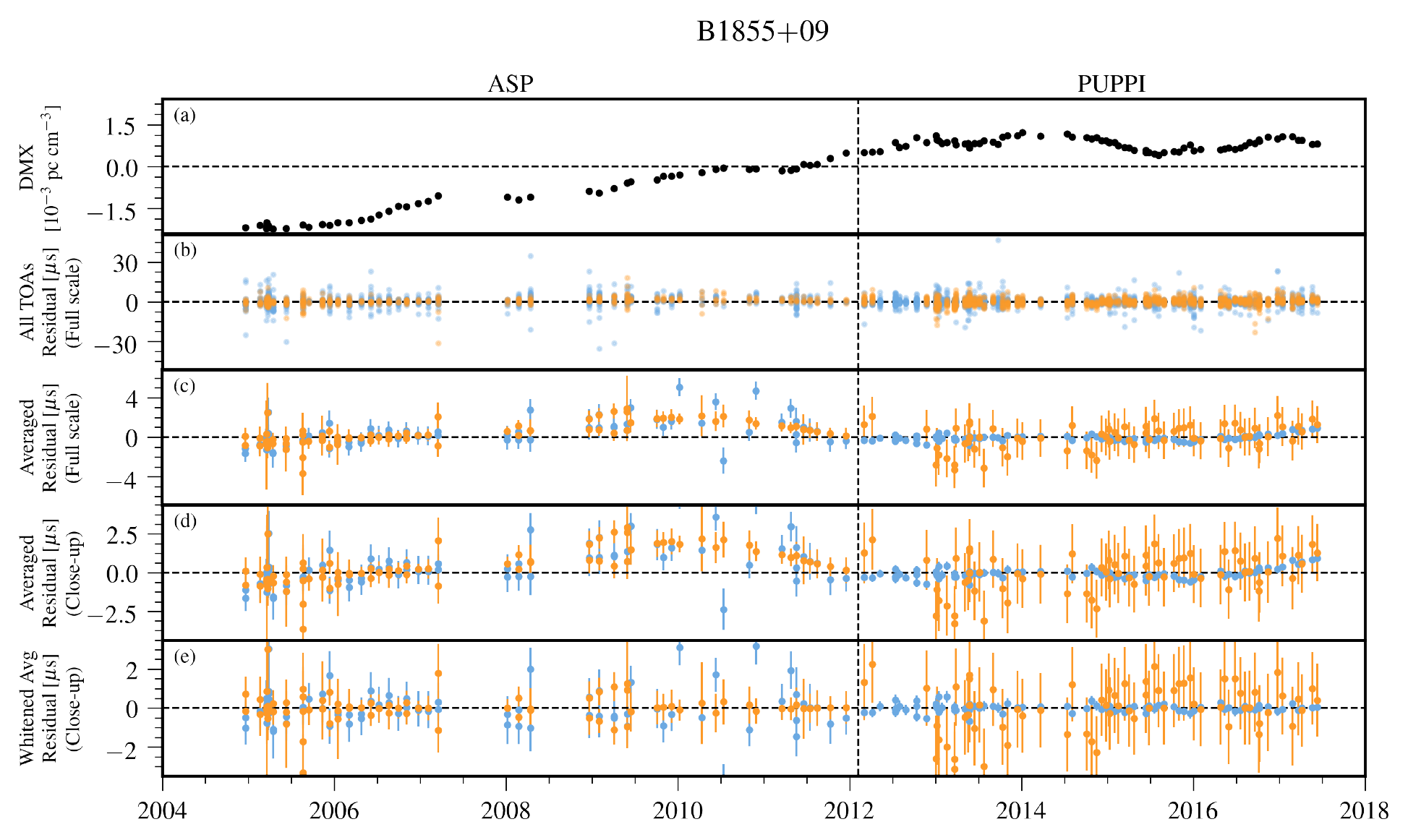}
\caption{Timing residuals and DM variations for PSR B1855$+$09. See appendix~\ref{sec:resid} text for details.  In residual plots, colored points indicate the receiver of each observation: 430~MHz (Orange) and 1.4~GHz (Light blue).  (a) Variations in DMX.  (b) Residual arrival times for all TOAs.  Points are semi-transparent; dark regions arise from the overlap of many points.  (c,d) Average residual arrival times shown full scale (panel c) and close-up of low residuals (panel d).  (e) Whitened average residual arrival times after removing the red noise model (close-up of low residuals).}
\label{fig:summary-B1855+09}
\end{figure*}

\begin{figure*}[p]
\centering
\includegraphics[scale=0.76]{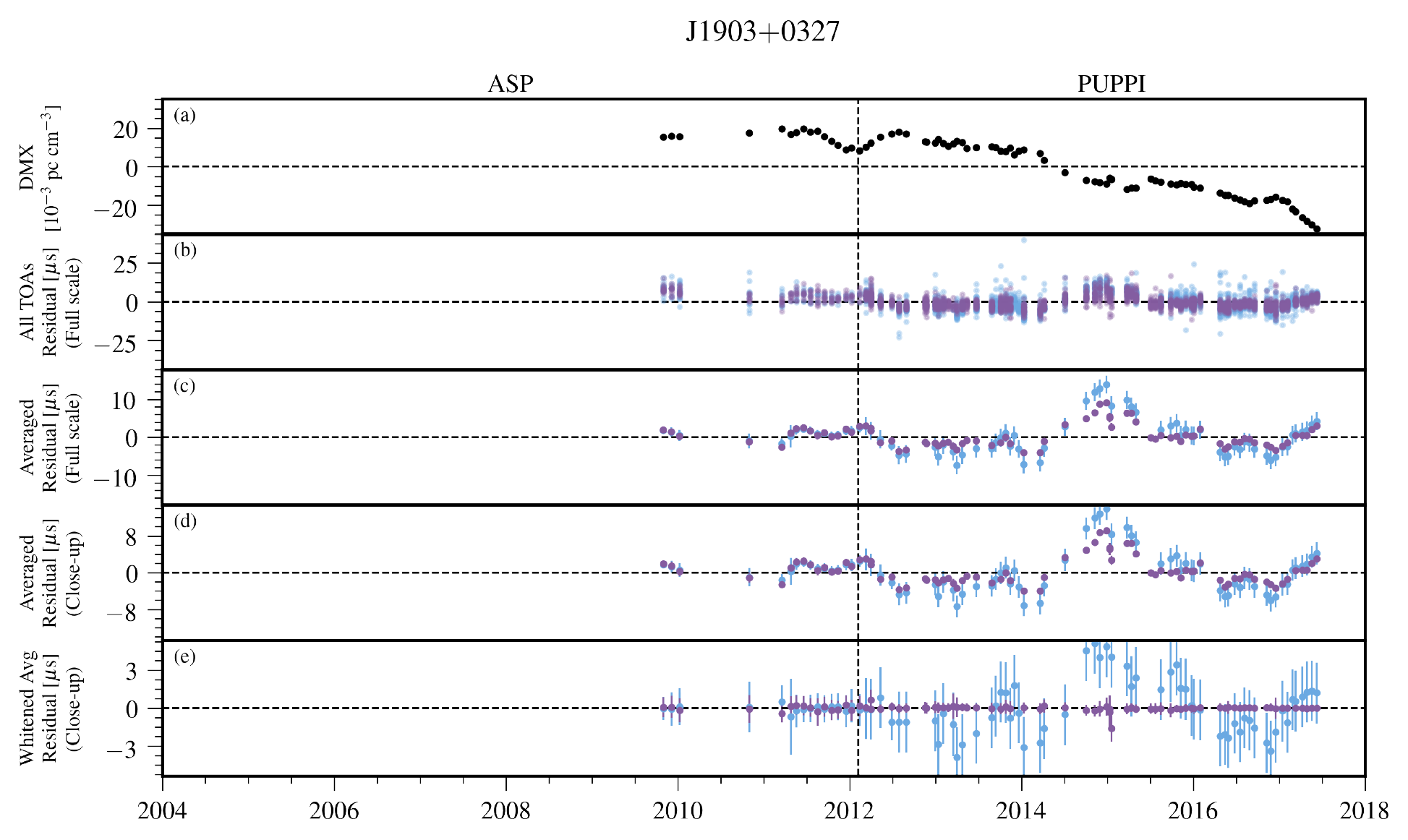}
\caption{Timing residuals and DM variations for PSR J1903$+$0327. See appendix~\ref{sec:resid} text for details.  In residual plots, colored points indicate the receiver of each observation: 1.4~GHz (Light blue) and 2.1~GHz (Purple).  (a) Variations in DMX.  (b) Residual arrival times for all TOAs.  Points are semi-transparent; dark regions arise from the overlap of many points.  (c,d) Average residual arrival times shown full scale (panel c) and close-up of low residuals (panel d).  (e) Whitened average residual arrival times after removing the red noise model (close-up of low residuals).}
\label{fig:summary-J1903+0327}
\end{figure*}

\begin{figure*}[p]
\centering
\includegraphics[scale=0.76]{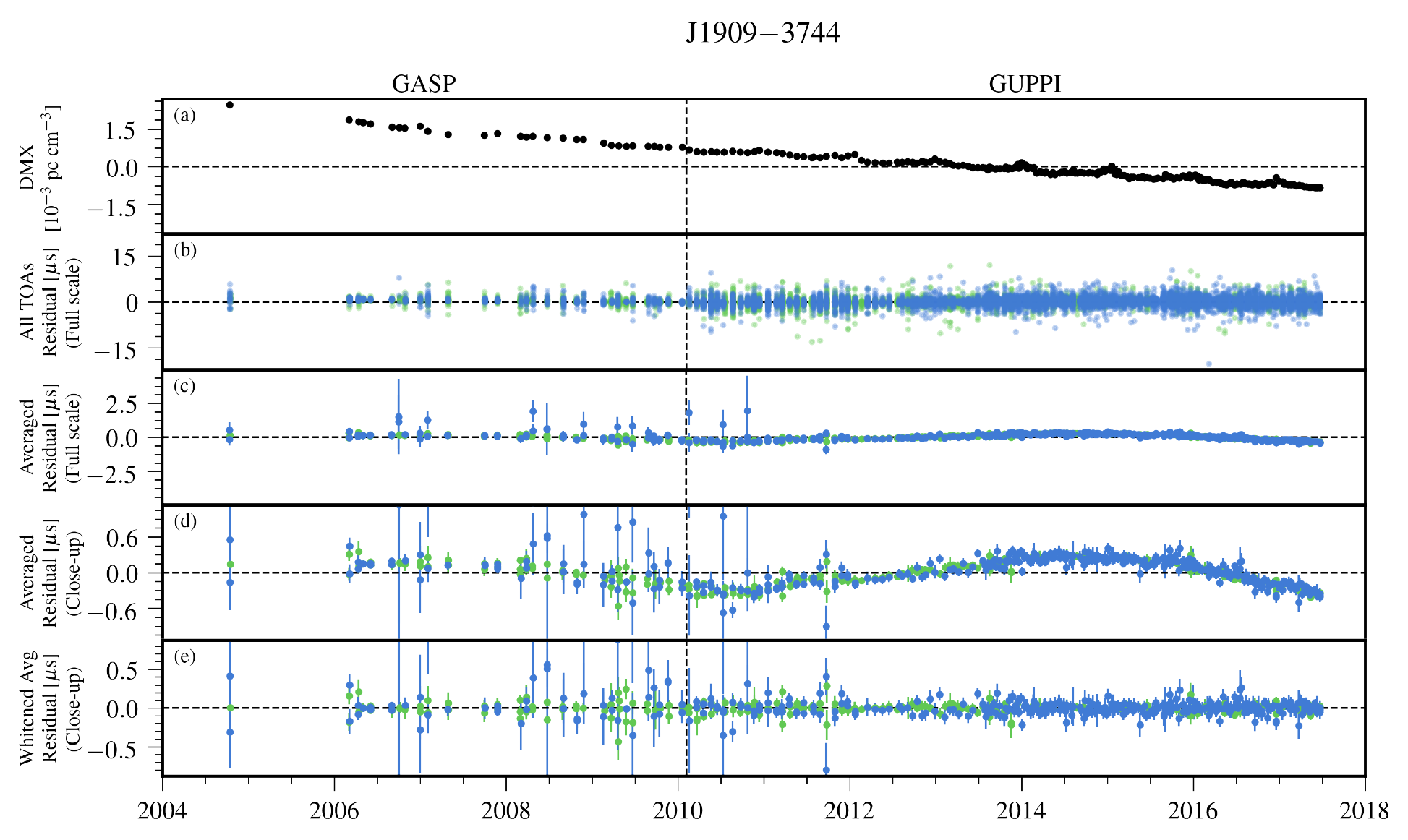}
\caption{Timing residuals and DM variations for PSR J1909$-$3744. See appendix~\ref{sec:resid} text for details.  In residual plots, colored points indicate the receiver of each observation: 820~MHz (Green) and 1.4~GHz (Dark blue).  (a) Variations in DMX.  (b) Residual arrival times for all TOAs.  Points are semi-transparent; dark regions arise from the overlap of many points.  (c,d) Average residual arrival times shown full scale (panel c) and close-up of low residuals (panel d).  (e) Whitened average residual arrival times after removing the red noise model (close-up of low residuals).}
\label{fig:summary-J1909-3744}
\end{figure*}

\begin{figure*}[p]
\centering
\includegraphics[scale=0.76]{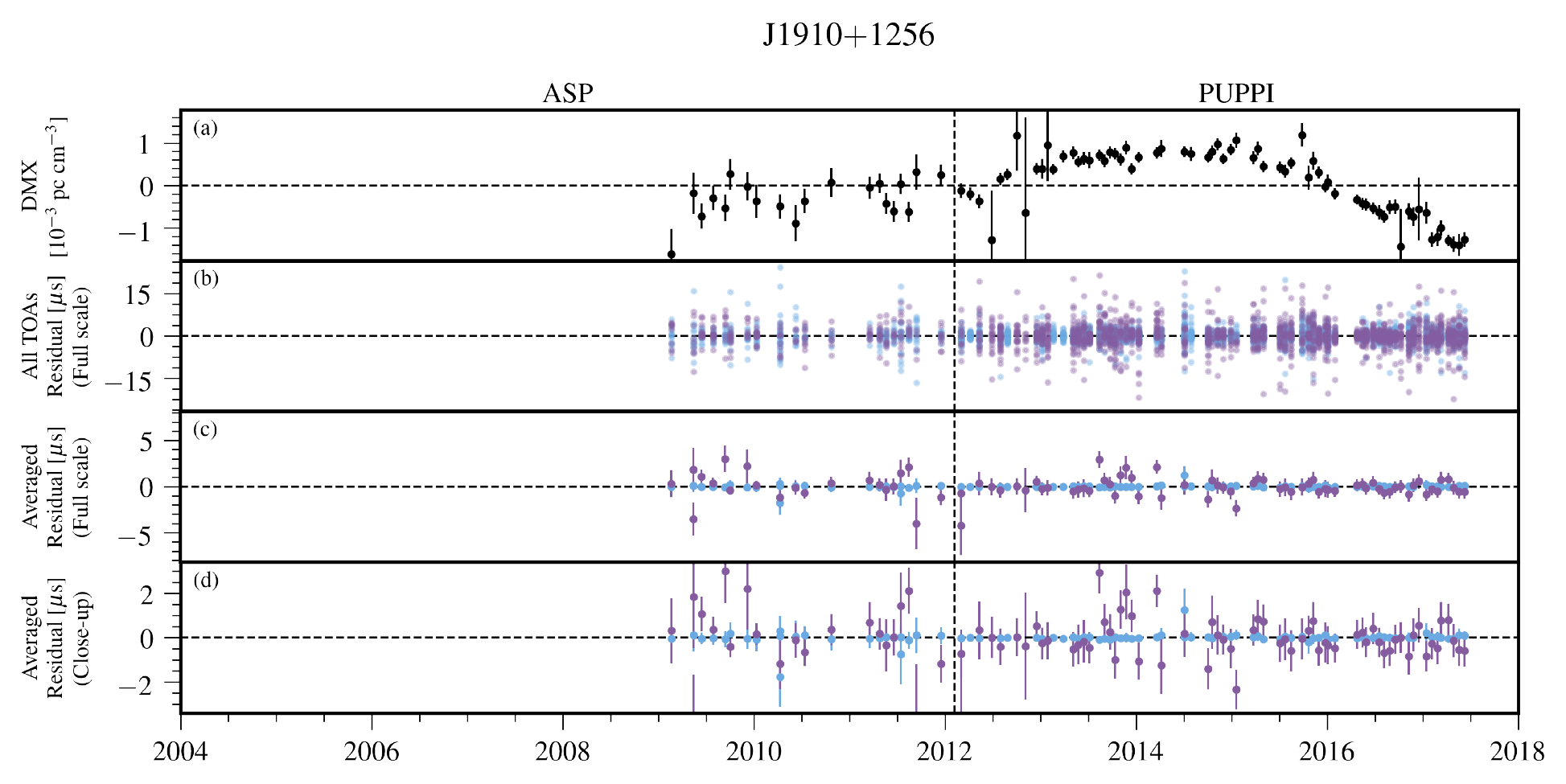}
\caption{Timing residuals and DM variations for PSR J1910$+$1256. See appendix~\ref{sec:resid} text for details.  In residual plots, colored points indicate the receiver of each observation: 1.4~GHz (Light blue) and 2.1~GHz (Purple).  (a) Variations in DMX.  (b) Residual arrival times for all TOAs.  Points are semi-transparent; dark regions arise from the overlap of many points.  (c,d) Average residual arrival times shown full scale (panel c) and close-up of low residuals (panel d). }
\label{fig:summary-J1910+1256}
\end{figure*}

\begin{figure*}[p]
\centering
\includegraphics[scale=0.76]{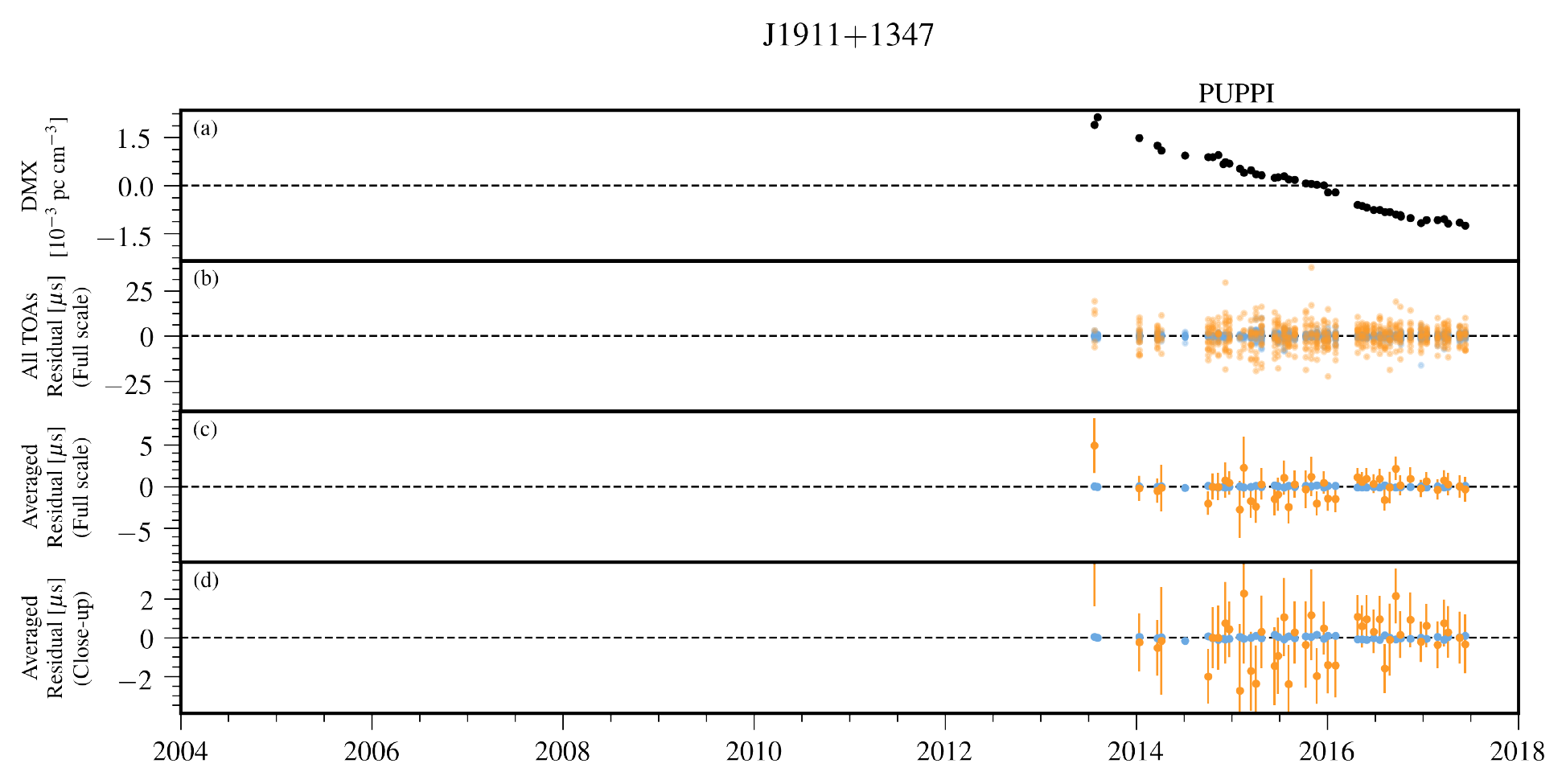}
\caption{Timing residuals and DM variations for PSR J1911$+$1347. See appendix~\ref{sec:resid} text for details.  In residual plots, colored points indicate the receiver of each observation: 430~MHz (Orange) and 1.4~GHz (Light blue).  (a) Variations in DMX.  (b) Residual arrival times for all TOAs.  Points are semi-transparent; dark regions arise from the overlap of many points.  (c,d) Average residual arrival times shown full scale (panel c) and close-up of low residuals (panel d). }
\label{fig:summary-J1911+1347}
\end{figure*}

\begin{figure*}[p]
\centering
\includegraphics[scale=0.76]{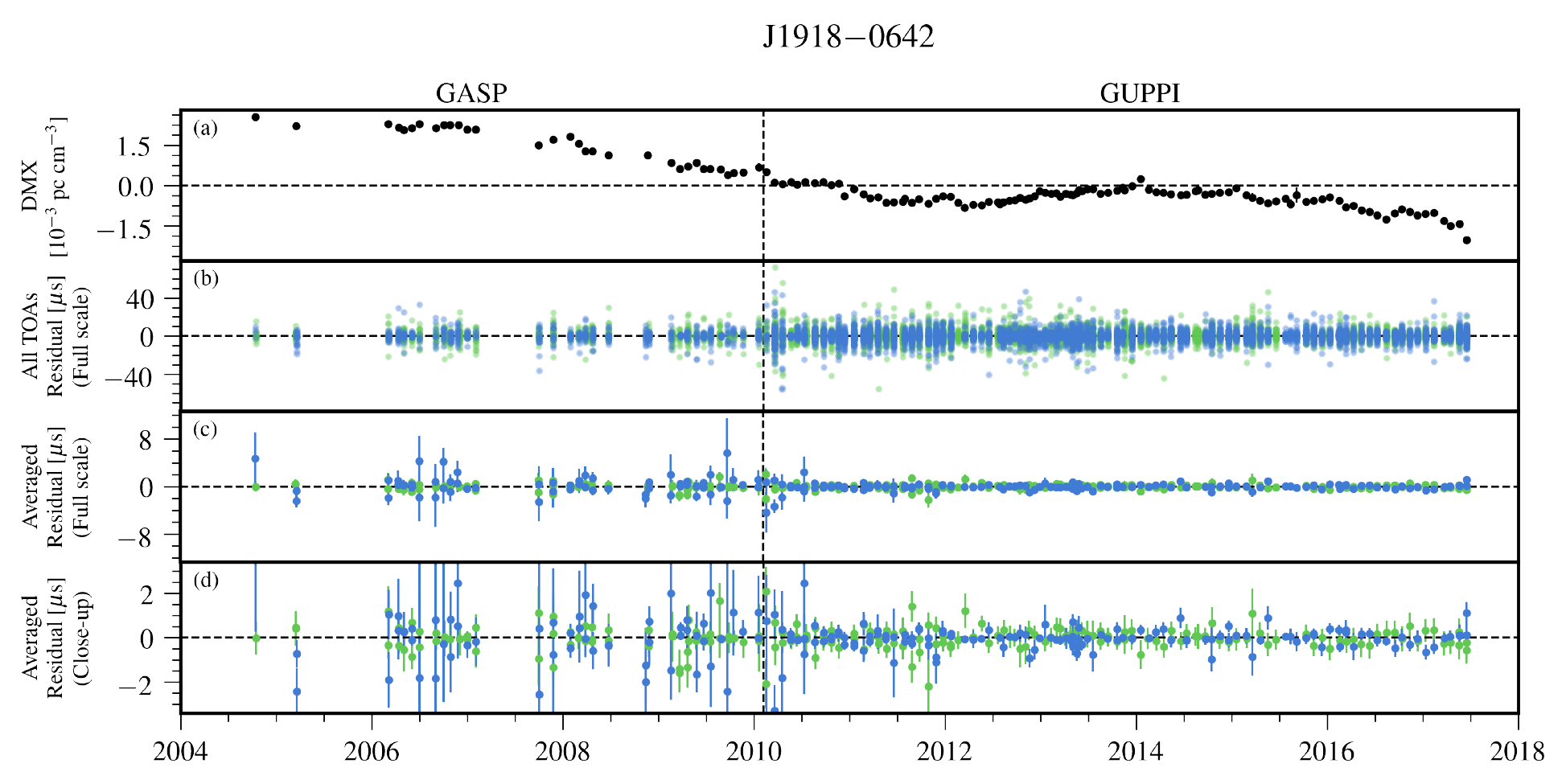}
\caption{Timing residuals and DM variations for PSR J1918$-$0642. See appendix~\ref{sec:resid} text for details.  In residual plots, colored points indicate the receiver of each observation: 820~MHz (Green) and 1.4~GHz (Dark blue).  (a) Variations in DMX.  (b) Residual arrival times for all TOAs.  Points are semi-transparent; dark regions arise from the overlap of many points.  (c,d) Average residual arrival times shown full scale (panel c) and close-up of low residuals (panel d). }
\label{fig:summary-J1918-0642}
\end{figure*}

\begin{figure*}[p]
\centering
\includegraphics[scale=0.76]{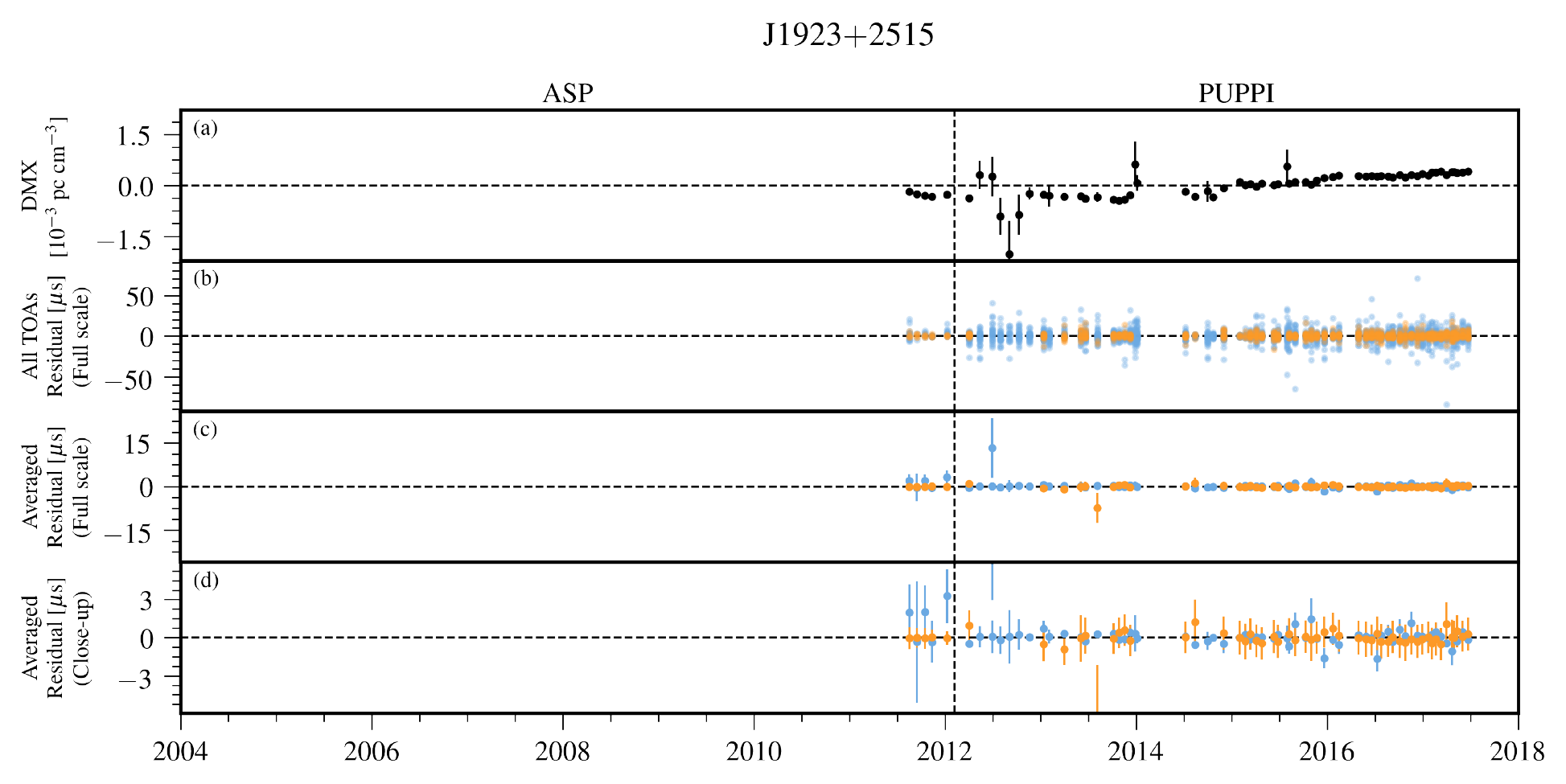}
\caption{Timing residuals and DM variations for PSR J1923$+$2515. See appendix~\ref{sec:resid} text for details.  In residual plots, colored points indicate the receiver of each observation: 430~MHz (Orange) and 1.4~GHz (Light blue).  (a) Variations in DMX.  (b) Residual arrival times for all TOAs.  Points are semi-transparent; dark regions arise from the overlap of many points.  (c,d) Average residual arrival times shown full scale (panel c) and close-up of low residuals (panel d). }
\label{fig:summary-J1923+2515}
\end{figure*}

\begin{figure*}[p]
\centering
\includegraphics[scale=0.76]{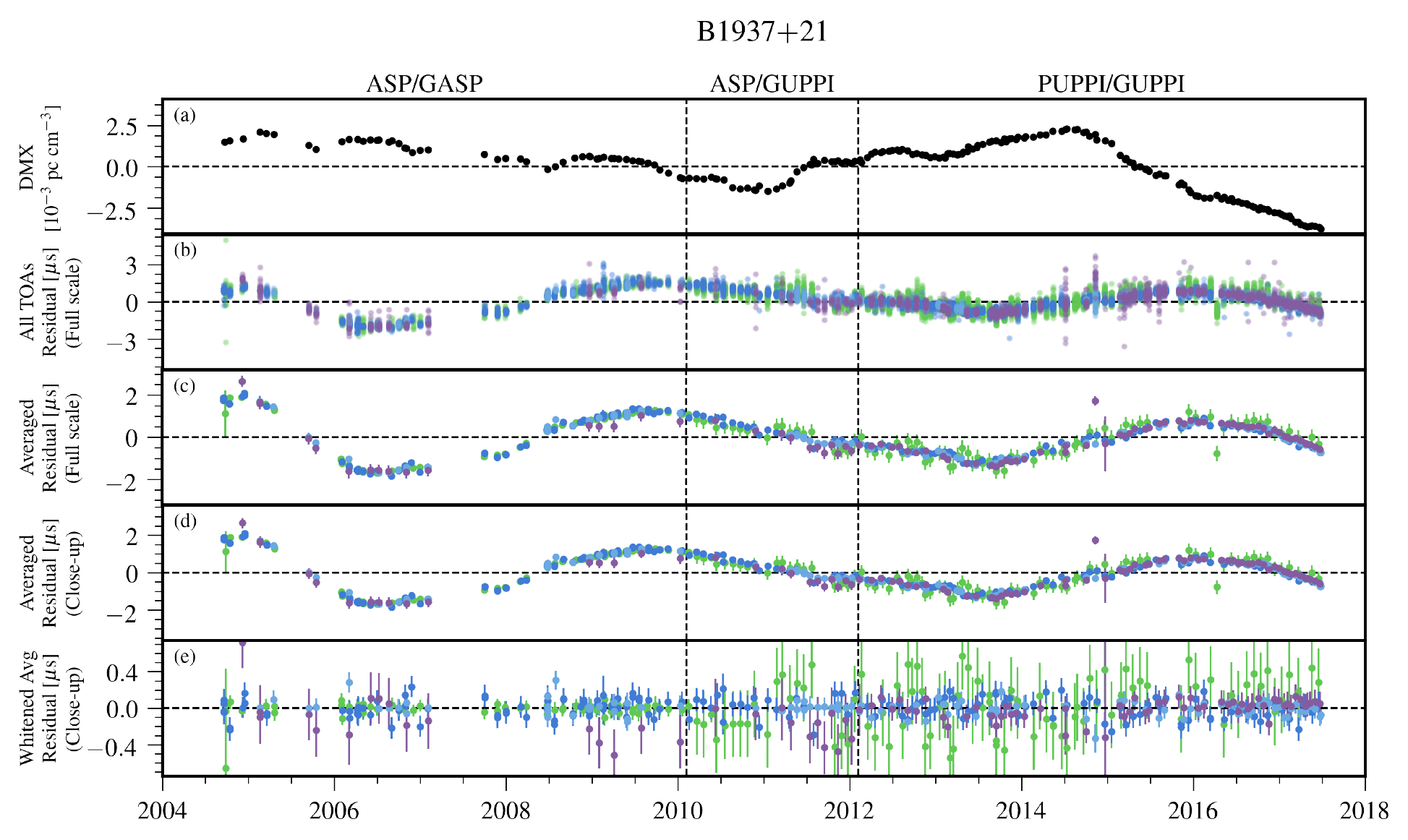}
\caption{Timing residuals and DM variations for PSR B1937$+$21. See appendix~\ref{sec:resid} text for details.  In residual plots, colored points indicate the receiver of each observation: 820~MHz (Green), 1.4~GHz (Dark blue), 1.4~GHz (Light blue), and 2.1~GHz (Purple).  (a) Variations in DMX.  (b) Residual arrival times for all TOAs.  Points are semi-transparent; dark regions arise from the overlap of many points.  (c,d) Average residual arrival times shown full scale (panel c) and close-up of low residuals (panel d).  (e) Whitened average residual arrival times after removing the red noise model (close-up of low residuals).}
\label{fig:summary-B1937+21}
\end{figure*}

\begin{figure*}[p]
\centering
\includegraphics[scale=0.76]{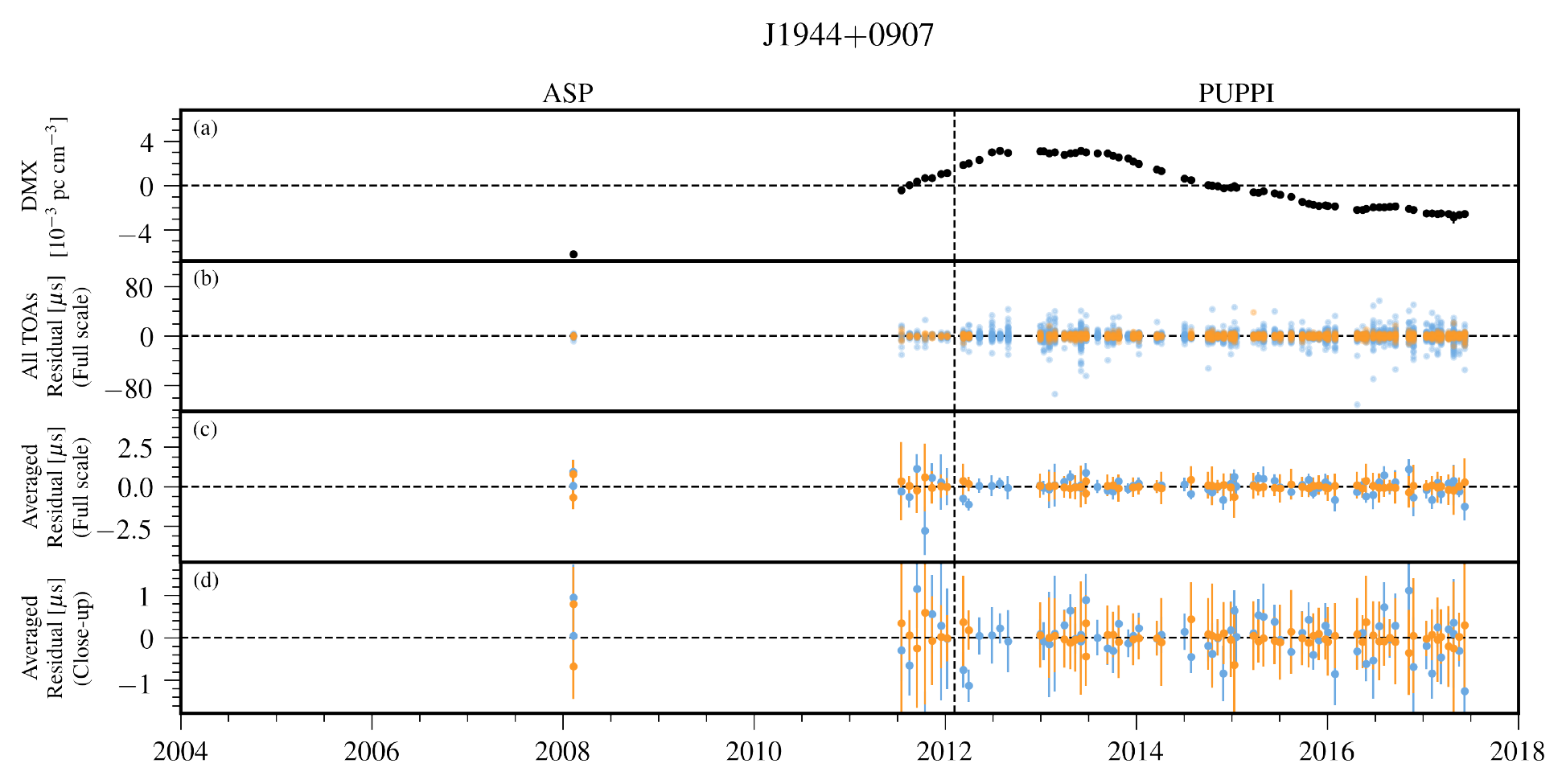}
\caption{Timing residuals and DM variations for PSR J1944$+$0907. See appendix~\ref{sec:resid} text for details.  In residual plots, colored points indicate the receiver of each observation: 430~MHz (Orange) and 1.4~GHz (Light blue).  (a) Variations in DMX.  (b) Residual arrival times for all TOAs.  Points are semi-transparent; dark regions arise from the overlap of many points.  (c,d) Average residual arrival times shown full scale (panel c) and close-up of low residuals (panel d). }
\label{fig:summary-J1944+0907}
\end{figure*}

\begin{figure*}[p]
\centering
\includegraphics[scale=0.76]{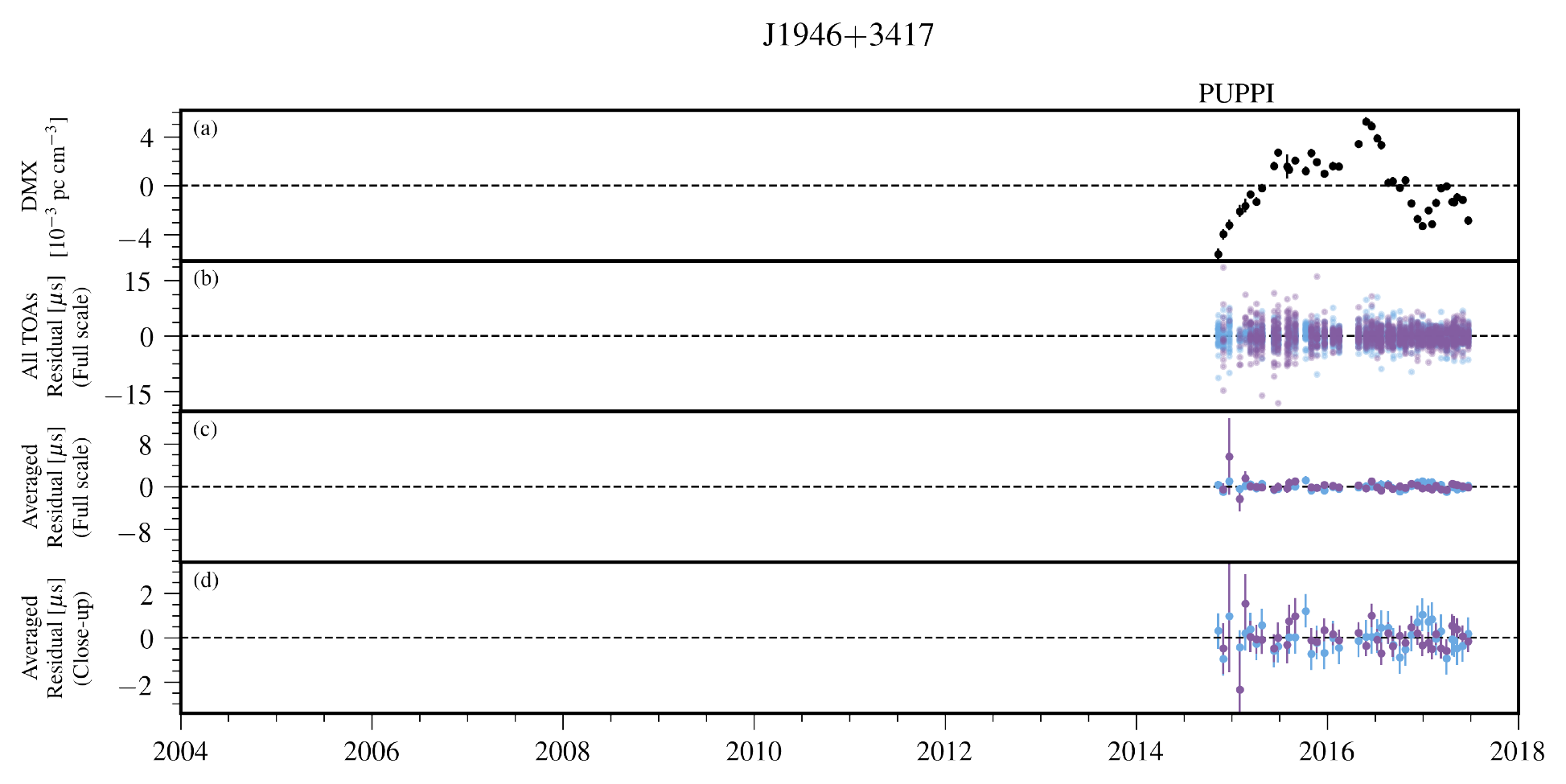}
\caption{Timing residuals and DM variations for PSR J1946$+$3417. See appendix~\ref{sec:resid} text for details.  In residual plots, colored points indicate the receiver of each observation: 1.4~GHz (Light blue) and 2.1~GHz (Purple).  (a) Variations in DMX.  (b) Residual arrival times for all TOAs.  Points are semi-transparent; dark regions arise from the overlap of many points.  (c,d) Average residual arrival times shown full scale (panel c) and close-up of low residuals (panel d). }
\label{fig:summary-J1946+3417}
\end{figure*}

\begin{figure*}[p]
\centering
\includegraphics[scale=0.76]{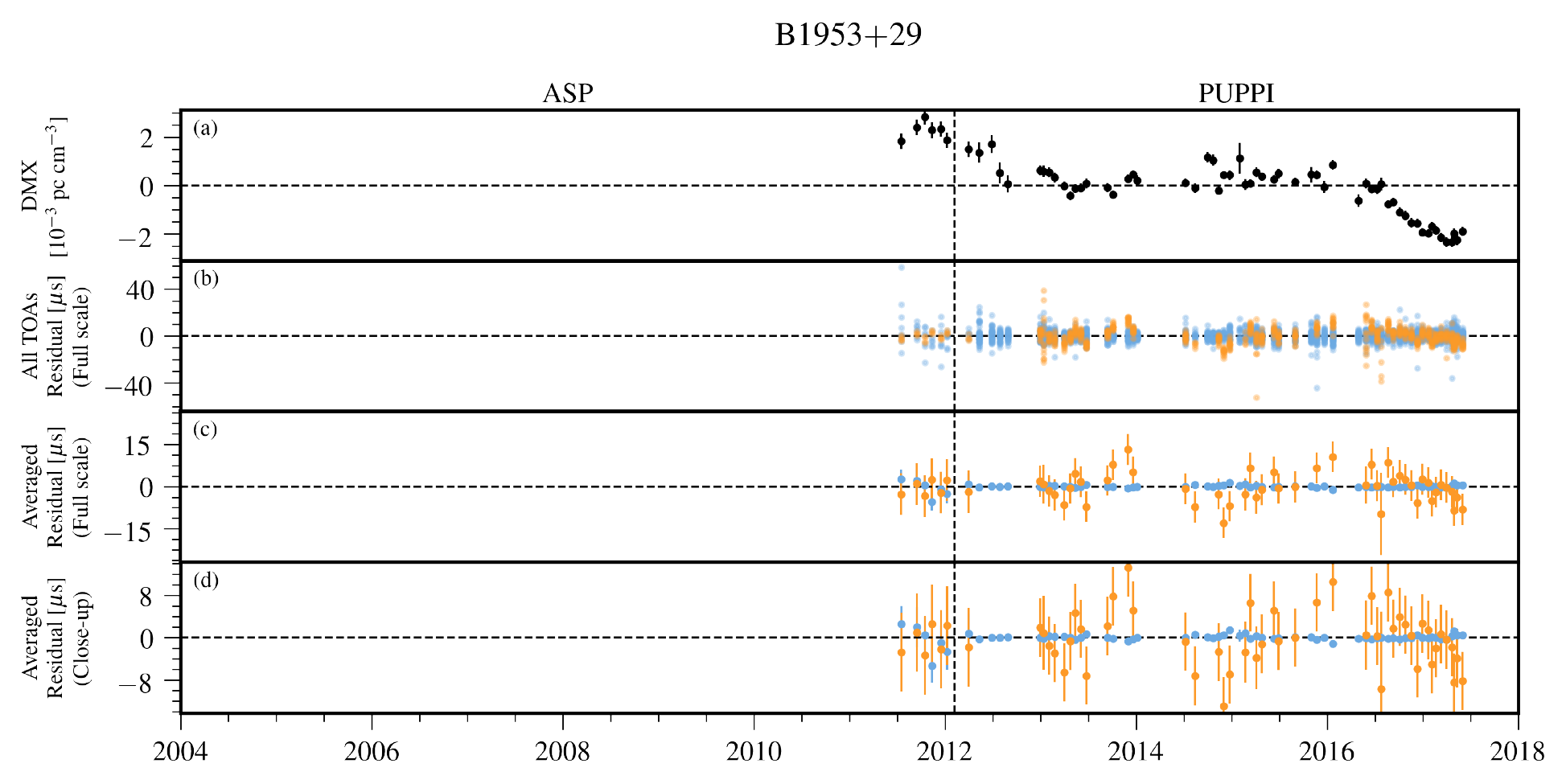}
\caption{Timing residuals and DM variations for PSR B1953$+$29. See appendix~\ref{sec:resid} text for details.  In residual plots, colored points indicate the receiver of each observation: 430~MHz (Orange) and 1.4~GHz (Light blue).  (a) Variations in DMX.  (b) Residual arrival times for all TOAs.  Points are semi-transparent; dark regions arise from the overlap of many points.  (c,d) Average residual arrival times shown full scale (panel c) and close-up of low residuals (panel d). }
\label{fig:summary-B1953+29}
\end{figure*}

\begin{figure*}[p]
\centering
\includegraphics[scale=0.76]{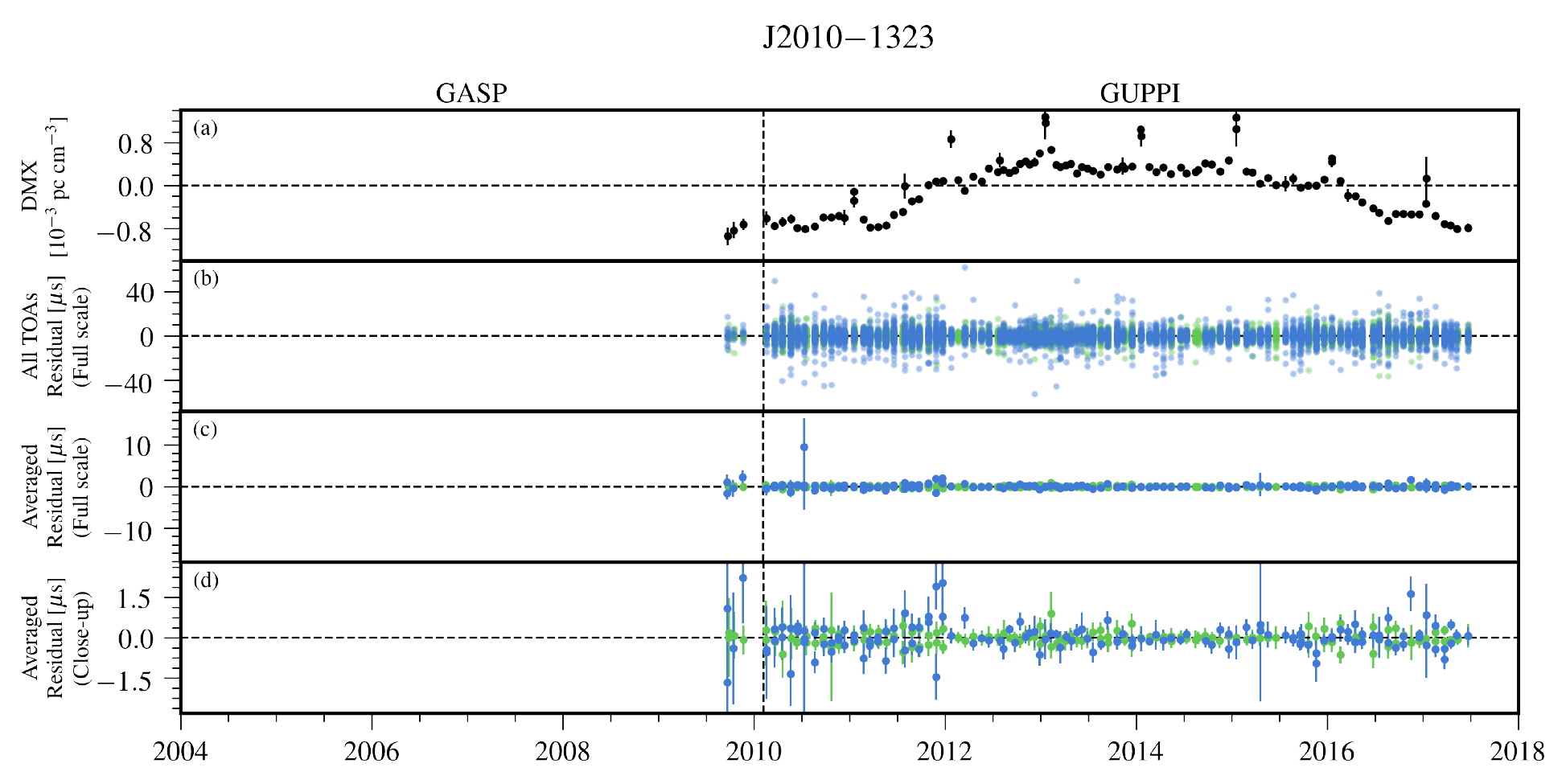}
\caption{Timing residuals and DM variations for PSR J2010$-$1323. See appendix~\ref{sec:resid} text for details.  In residual plots, colored points indicate the receiver of each observation: 820~MHz (Green) and 1.4~GHz (Dark blue).  (a) Variations in DMX.  (b) Residual arrival times for all TOAs.  Points are semi-transparent; dark regions arise from the overlap of many points.  (c,d) Average residual arrival times shown full scale (panel c) and close-up of low residuals (panel d). }
\label{fig:summary-J2010-1323}
\end{figure*}

\begin{figure*}[p]
\centering
\includegraphics[scale=0.76]{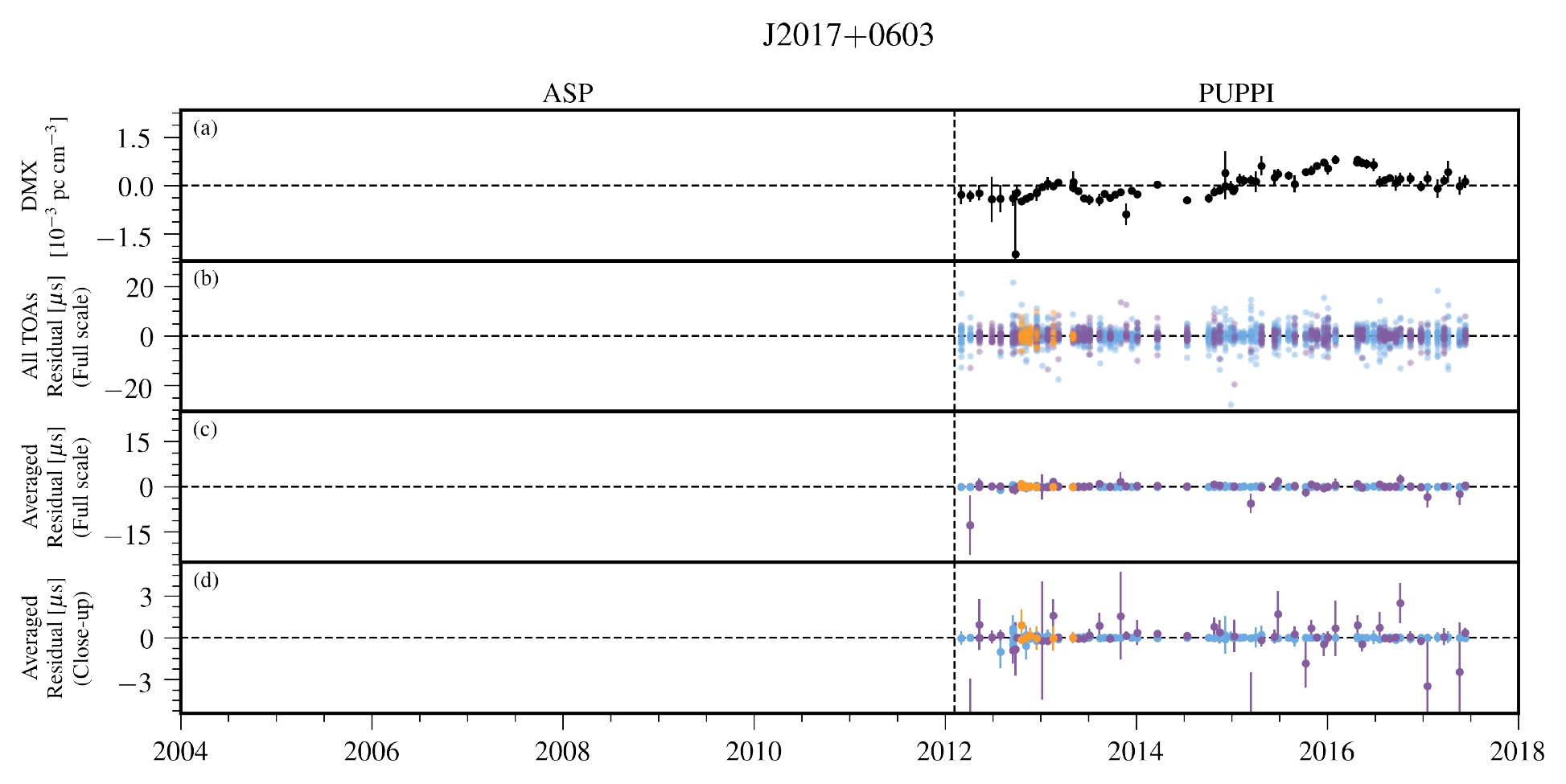}
\caption{Timing residuals and DM variations for PSR J2017$+$0603. See appendix~\ref{sec:resid} text for details.  In residual plots, colored points indicate the receiver of each observation: 430~MHz (Orange), 1.4~GHz (Light blue), and 2.1~GHz (Purple).  (a) Variations in DMX.  (b) Residual arrival times for all TOAs.  Points are semi-transparent; dark regions arise from the overlap of many points.  (c,d) Average residual arrival times shown full scale (panel c) and close-up of low residuals (panel d). }
\label{fig:summary-J2017+0603}
\end{figure*}

\begin{figure*}[p]
\centering
\includegraphics[scale=0.76]{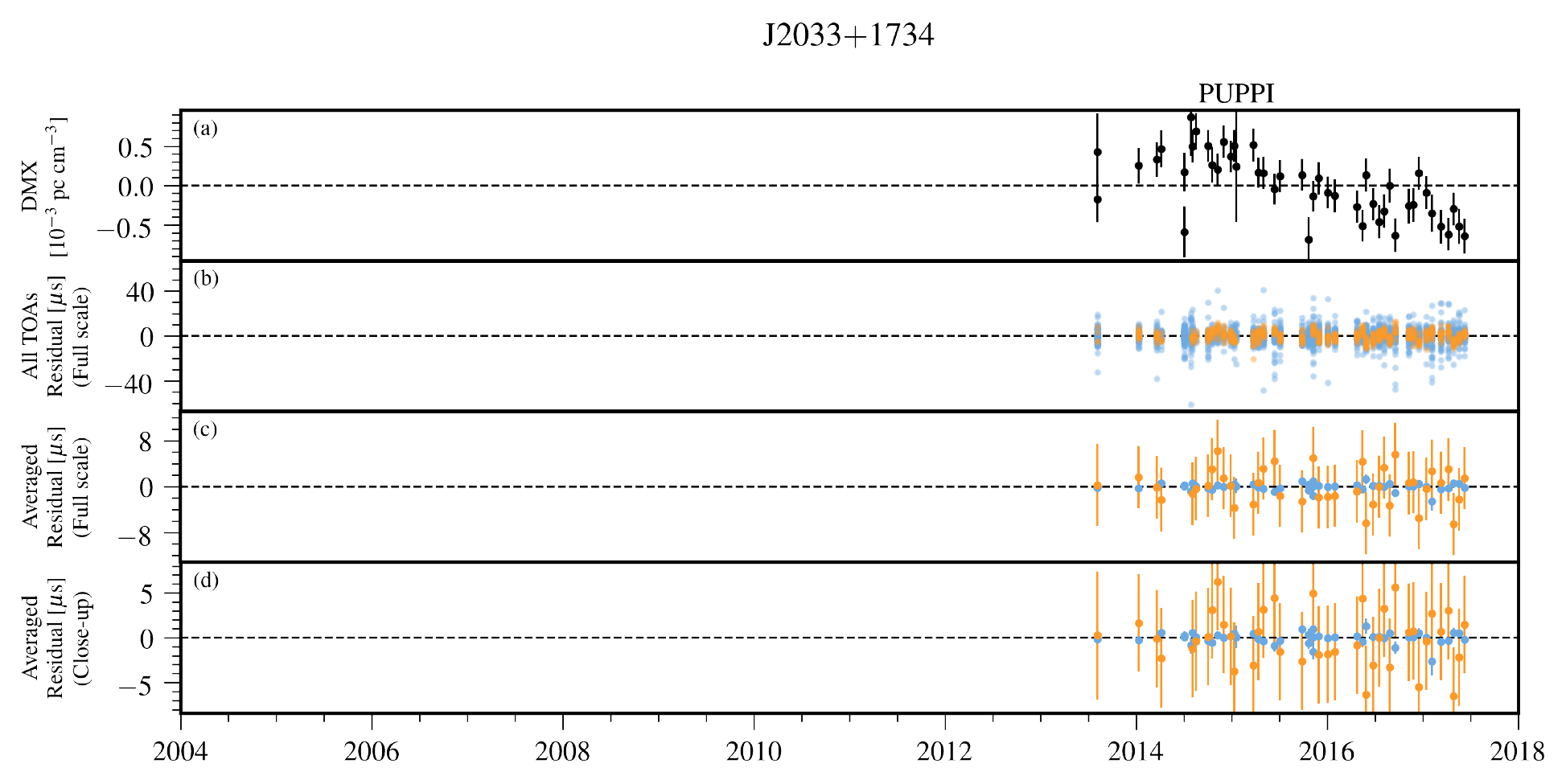}
\caption{Timing residuals and DM variations for PSR J2033$+$1734. See appendix~\ref{sec:resid} text for details.  In residual plots, colored points indicate the receiver of each observation: 430~MHz (Orange) and 1.4~GHz (Light blue).  (a) Variations in DMX.  (b) Residual arrival times for all TOAs.  Points are semi-transparent; dark regions arise from the overlap of many points.  (c,d) Average residual arrival times shown full scale (panel c) and close-up of low residuals (panel d). }
\label{fig:summary-J2033+1734}
\end{figure*}

\begin{figure*}[p]
\centering
\includegraphics[scale=0.76]{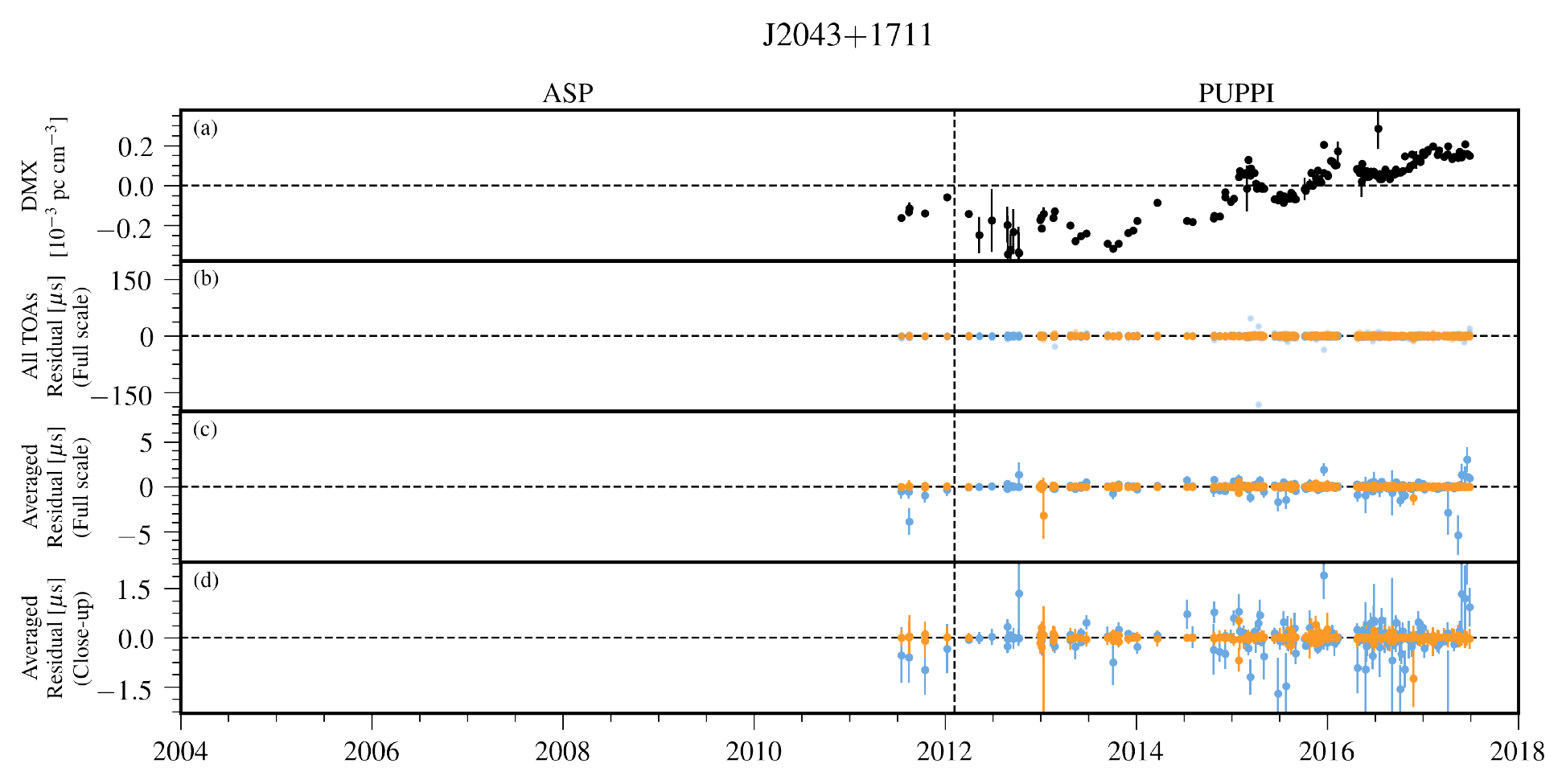}
\caption{Timing residuals and DM variations for PSR J2043$+$1711. See appendix~\ref{sec:resid} text for details.  In residual plots, colored points indicate the receiver of each observation: 430~MHz (Orange) and 1.4~GHz (Light blue).  (a) Variations in DMX.  (b) Residual arrival times for all TOAs.  Points are semi-transparent; dark regions arise from the overlap of many points.  (c,d) Average residual arrival times shown full scale (panel c) and close-up of low residuals (panel d). }
\label{fig:summary-J2043+1711}
\end{figure*}

\begin{figure*}[p]
\centering
\includegraphics[scale=0.76]{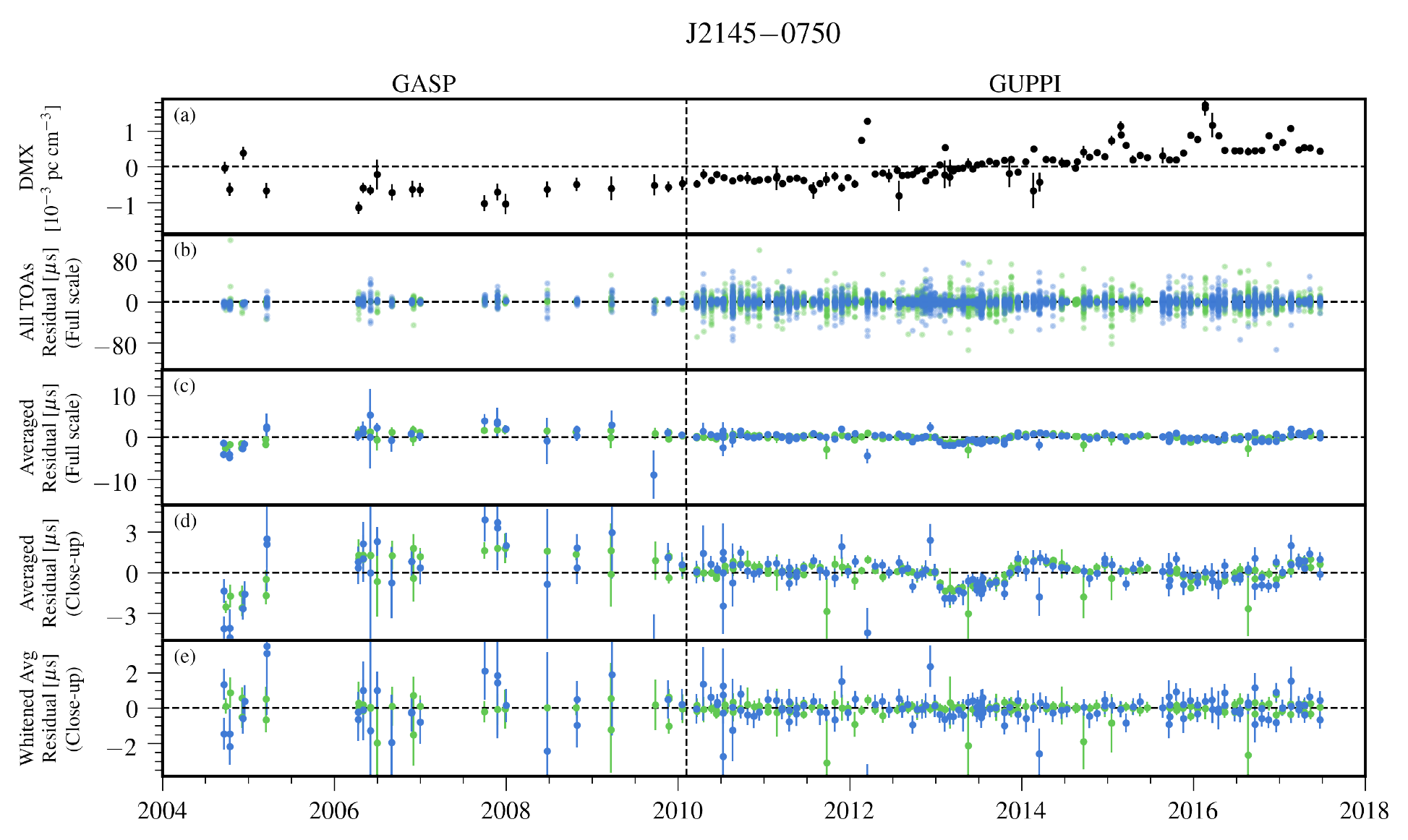}
\caption{Timing residuals and DM variations for PSR J2145$-$0750. See appendix~\ref{sec:resid} text for details.  In residual plots, colored points indicate the receiver of each observation: 820~MHz (Green) and 1.4~GHz (Dark blue).  (a) Variations in DMX.  (b) Residual arrival times for all TOAs.  Points are semi-transparent; dark regions arise from the overlap of many points.  (c,d) Average residual arrival times shown full scale (panel c) and close-up of low residuals (panel d).  (e) Whitened average residual arrival times after removing the red noise model (close-up of low residuals).}
\label{fig:summary-J2145-0750}
\end{figure*}

\begin{figure*}[p]
\centering
\includegraphics[scale=0.76]{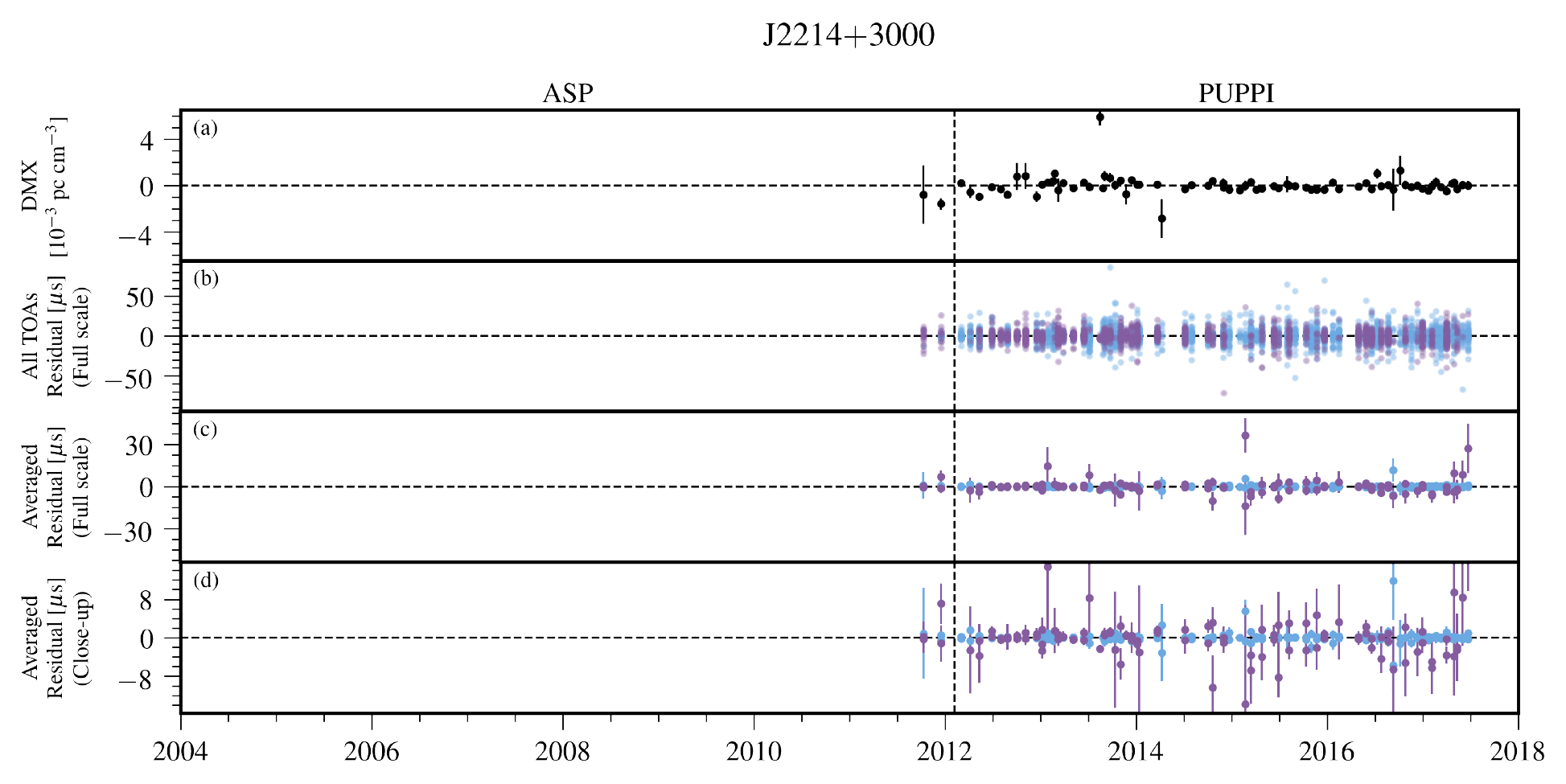}
\caption{Timing residuals and DM variations for PSR J2214$+$3000. See appendix~\ref{sec:resid} text for details.  In residual plots, colored points indicate the receiver of each observation: 1.4~GHz (Light blue) and 2.1~GHz (Purple).  (a) Variations in DMX.  (b) Residual arrival times for all TOAs.  Points are semi-transparent; dark regions arise from the overlap of many points.  (c,d) Average residual arrival times shown full scale (panel c) and close-up of low residuals (panel d). }
\label{fig:summary-J2214+3000}
\end{figure*}

\begin{figure*}[p]
\centering
\includegraphics[scale=0.76]{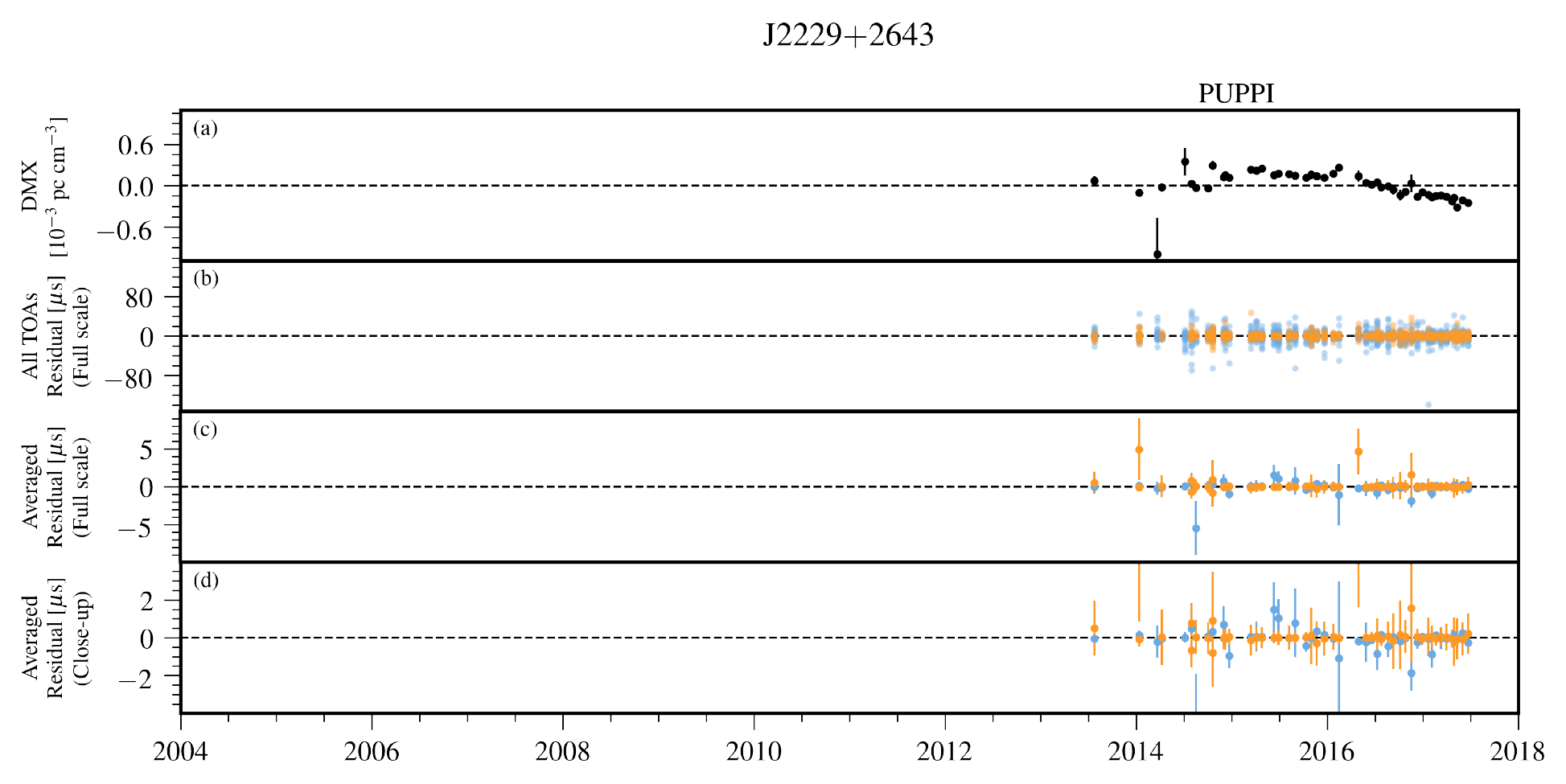}
\caption{Timing residuals and DM variations for PSR J2229$+$2643. See appendix~\ref{sec:resid} text for details.  In residual plots, colored points indicate the receiver of each observation: 430~MHz (Orange) and 1.4~GHz (Light blue).  (a) Variations in DMX.  (b) Residual arrival times for all TOAs.  Points are semi-transparent; dark regions arise from the overlap of many points.  (c,d) Average residual arrival times shown full scale (panel c) and close-up of low residuals (panel d). }
\label{fig:summary-J2229+2643}
\end{figure*}

\begin{figure*}[p]
\centering
\includegraphics[scale=0.76]{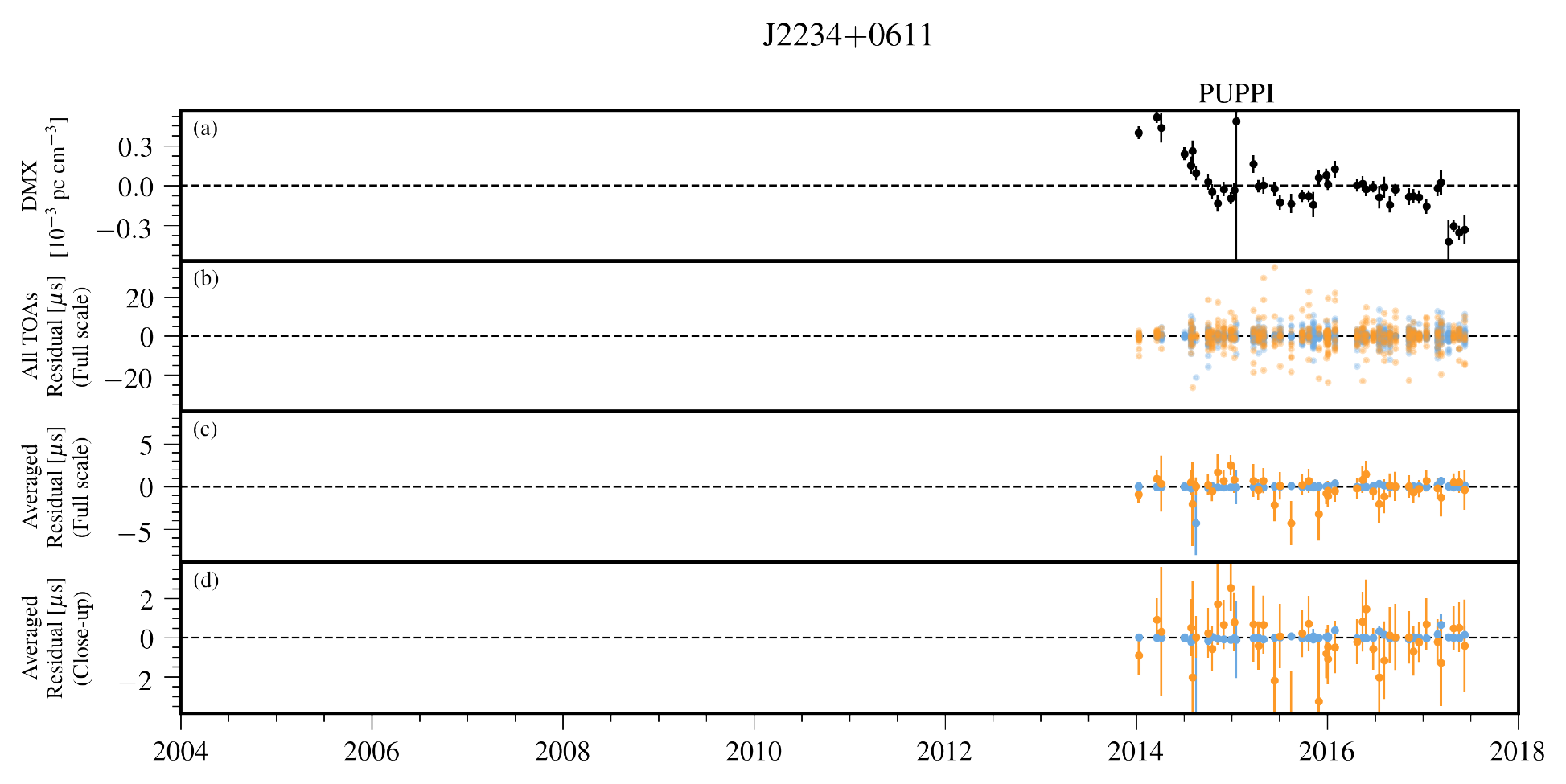}
\caption{Timing residuals and DM variations for PSR J2234$+$0611. See appendix~\ref{sec:resid} text for details.  In residual plots, colored points indicate the receiver of each observation: 430~MHz (Orange) and 1.4~GHz (Light blue).  (a) Variations in DMX.  (b) Residual arrival times for all TOAs.  Points are semi-transparent; dark regions arise from the overlap of many points.  (c,d) Average residual arrival times shown full scale (panel c) and close-up of low residuals (panel d). }
\label{fig:summary-J2234+0611}
\end{figure*}

\begin{figure*}[p]
\centering
\includegraphics[scale=0.76]{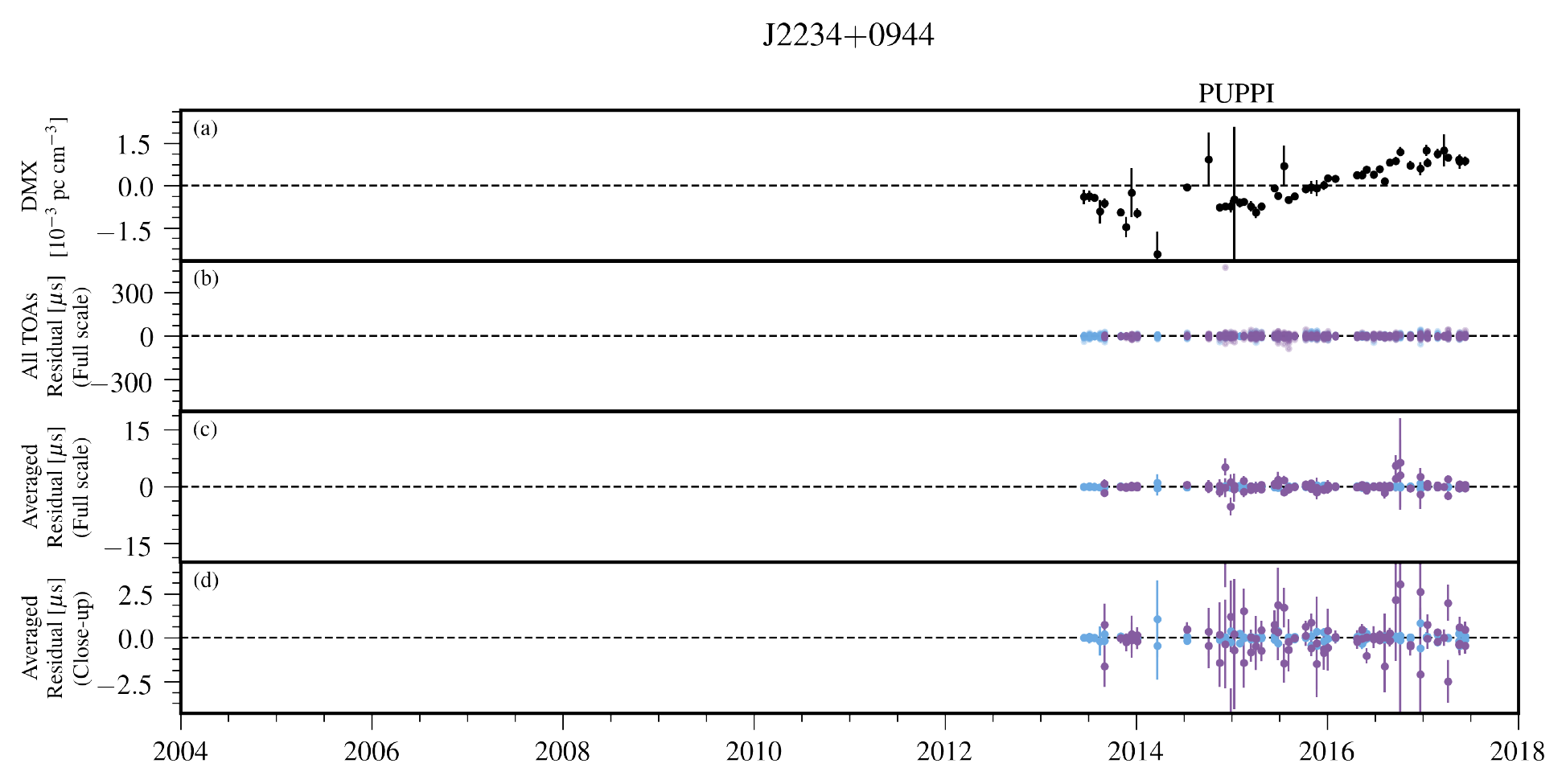}
\caption{Timing residuals and DM variations for PSR J2234$+$0944. See appendix~\ref{sec:resid} text for details.  In residual plots, colored points indicate the receiver of each observation: 1.4~GHz (Light blue) and 2.1~GHz (Purple).  (a) Variations in DMX.  (b) Residual arrival times for all TOAs.  Points are semi-transparent; dark regions arise from the overlap of many points.  (c,d) Average residual arrival times shown full scale (panel c) and close-up of low residuals (panel d). }
\label{fig:summary-J2234+0944}
\end{figure*}

\begin{figure*}[p]
\centering
\includegraphics[scale=0.76]{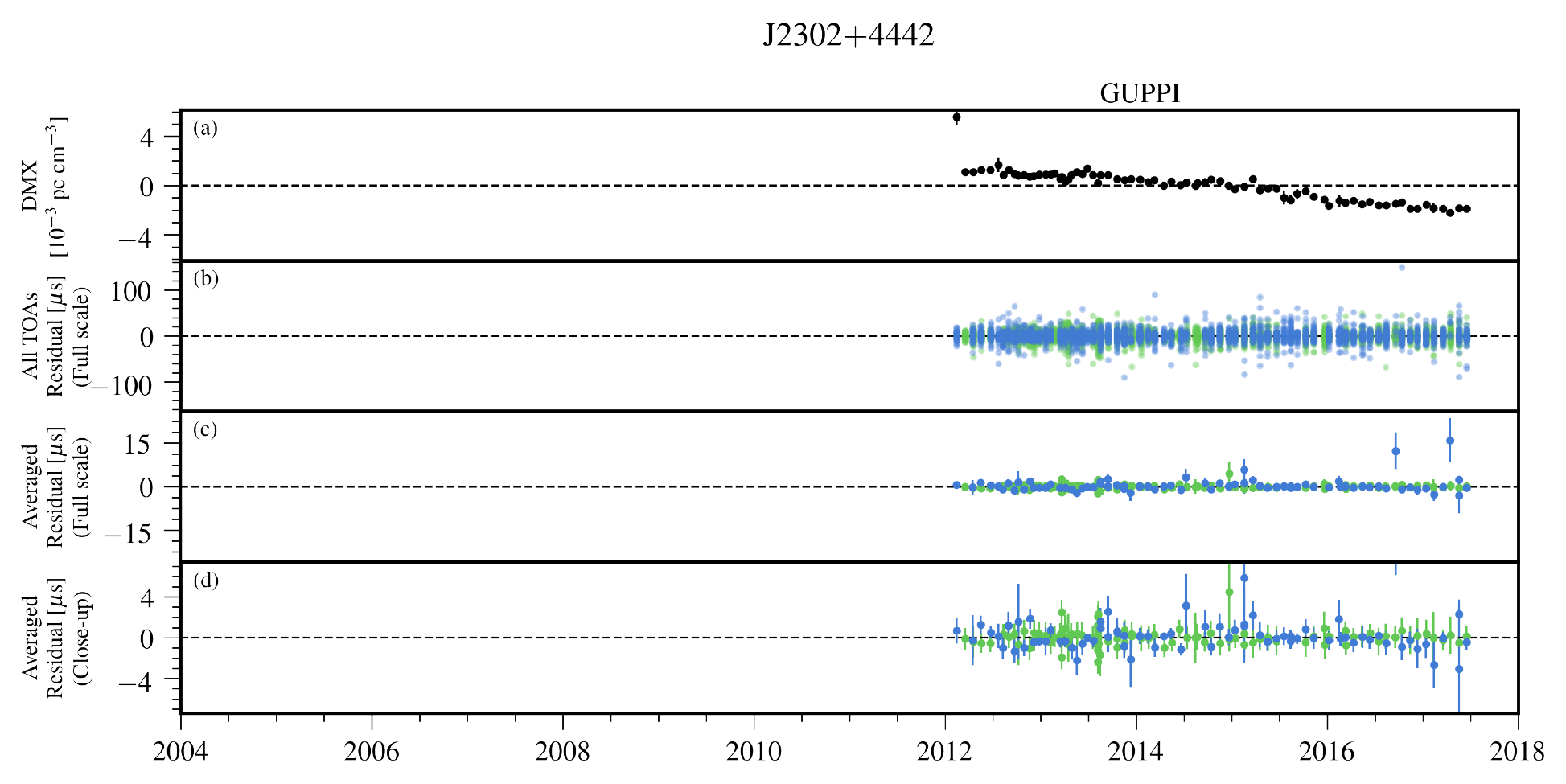}
\caption{Timing residuals and DM variations for PSR J2302$+$4442. See appendix~\ref{sec:resid} text for details.  In residual plots, colored points indicate the receiver of each observation: 820~MHz (Green) and 1.4~GHz (Dark blue).  (a) Variations in DMX.  (b) Residual arrival times for all TOAs.  Points are semi-transparent; dark regions arise from the overlap of many points.  (c,d) Average residual arrival times shown full scale (panel c) and close-up of low residuals (panel d). }
\label{fig:summary-J2302+4442}
\end{figure*}

\begin{figure*}[p]
\centering
\includegraphics[scale=0.76]{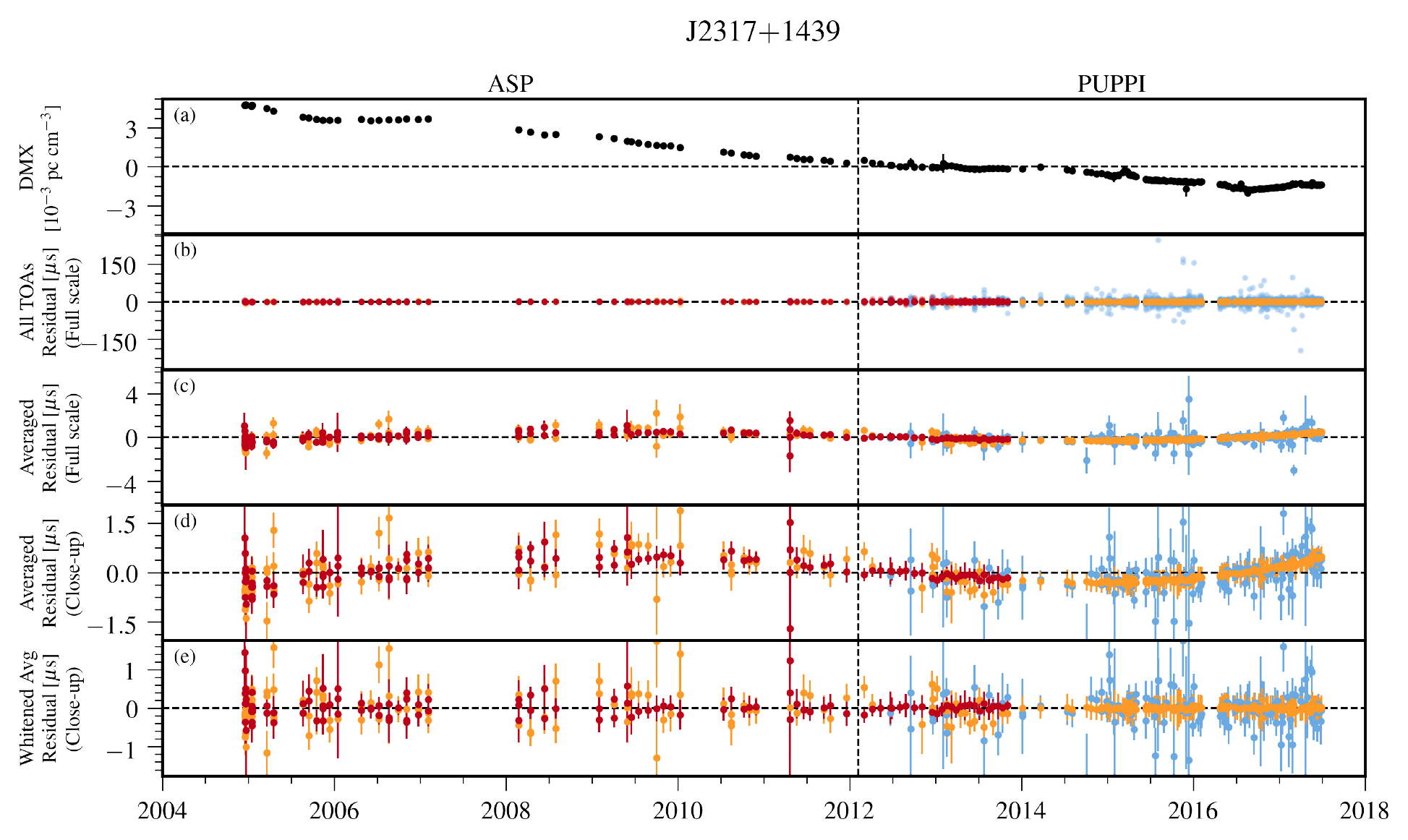}
\caption{Timing residuals and DM variations for PSR J2317$+$1439. See appendix~\ref{sec:resid} text for details.  In residual plots, colored points indicate the receiver of each observation: 327~MHz (Red), 430~MHz (Orange), and 1.4~GHz (Light blue).  (a) Variations in DMX.  (b) Residual arrival times for all TOAs.  Points are semi-transparent; dark regions arise from the overlap of many points.  (c,d) Average residual arrival times shown full scale (panel c) and close-up of low residuals (panel d).  (e) Whitened average residual arrival times after removing the red noise model (close-up of low residuals).}
\label{fig:summary-J2317+1439}
\end{figure*}

\begin{figure*}[p]
\centering
\includegraphics[scale=0.76]{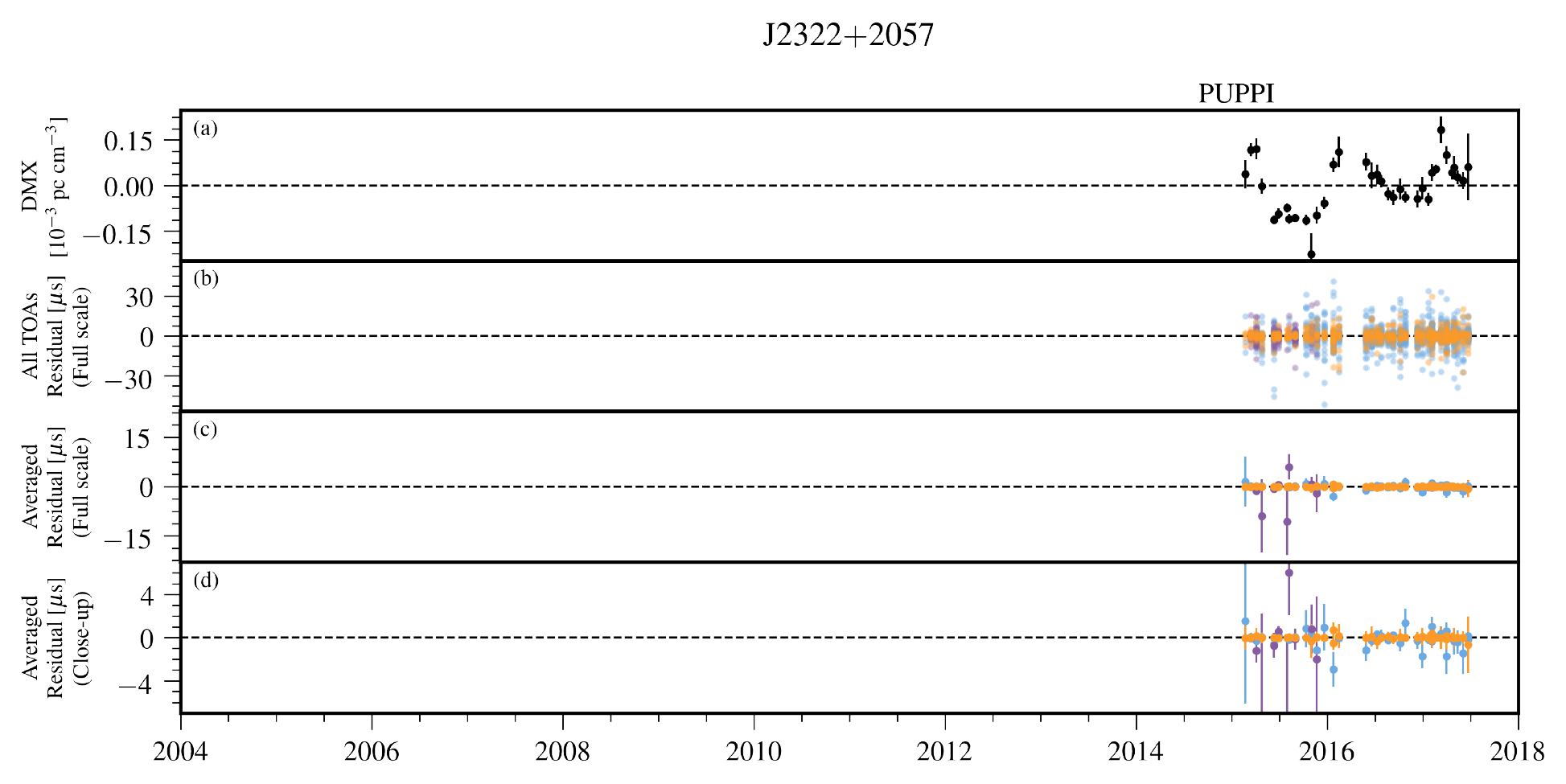}
\caption{Timing residuals and DM variations for PSR J2322$+$2057. See appendix~\ref{sec:resid} text for details.  In residual plots, colored points indicate the receiver of each observation: 430~MHz (Orange), 1.4~GHz (Light blue), and 2.1~GHz (Purple).  (a) Variations in DMX.  (b) Residual arrival times for all TOAs.  Points are semi-transparent; dark regions arise from the overlap of many points.  (c,d) Average residual arrival times shown full scale (panel c) and close-up of low residuals (panel d). }
\label{fig:summary-J2322+2057}
\end{figure*}

This appendix includes figures~\ref{fig:summary-J0023+0923}--\ref{fig:summary-J2322+2057} showing timing residuals and DM variations for each pulsar in our data set.

{\it DM variations.}  In panel (a) of each of the appendix figures, we plot the mean-subtracted DM time-series, where each point represents a DM parameter in the DMX model (Section~\ref{subsec:dm_variation_msmts}).
The division of each timing baseline into DMX epochs is described in Section~\ref{subsec:dm_variation_msmts}, and the epochs are typically six days in length or less (usually one day for Arecibo observations), except in the earliest data.
The mean-subtracted DM values are plotted in part because it allows us to segregate the uncertainty in the average DM, which arises due to covariance with the FD parameters and template pulse profiles, from the uncertainties in the DMX parameters that are shown in the figures.

{\it Timing residuals.}  As described in Section~\ref{sec:obs}, each observation is
comprised of many simultaneously-obtained narrowband TOAs.  We plot the residual arrival times
(observed minus that predicted from the timing model) for each pulsar in multiple ways described below.
In each residual plot, linear and quadratic trends have been removed, as they are completely covariant with the pulsar's rotation frequency and frequency derivative in the timing model, and hence would be absorbed.

Residuals from every TOA measurement are plotted in panel (b) of each figure.
The color of each point encodes the receiver as in Figure~\ref{fig:obs:epochs}: 327~MHz (red), 430~MHz (orange), 820~MHz (green), 1400~MHz (lighter blue for AO, darker blue for the GBT), 2100~MHz (purple).
The predominant data acquisition backend instrument over any given time period is indicated at the top of each figure, and vertical dashed lines indicate the times at which instruments changed.
Averaged residuals of simultaneously measured TOAs are shown in panels (c) and (d); these were computed using the procedure described in Appendix~D of \nineyr. The full vertical range of these residuals is shown in panel (c), while panel (d) shows a close-up of the low residuals.
For pulsars with red noise above our defined threshold, red noise Bayes factor $B>100$ (Table~\ref{tab:timingpars}), panel (e) shows whitened timing residuals, which have the red noise contribution subtracted from the averaged residuals.

\clearpage

\bibliographystyle{aasjournal}

\end{document}